\documentclass[11pt,a4paper]{article}
\usepackage{jcappub}
\pdfoutput=1
\usepackage[T1]{fontenc}
\usepackage{subcaption}
\usepackage{caption}
\usepackage{tikz,xcolor,hyperref}
\usepackage{amsmath,amssymb}
\usepackage{braket}
\usepackage{epsfig}
\usepackage{graphicx}
\usepackage{slashed}
\usepackage{comment}
\usepackage{orcidlink}
\begin{document}
%\preprint{IP/BBSR/2015-4}
%\title{Imprints of ultralight PBH formed during phase transitions on gravitational waves}

%\title{Probing ultra-light PBH formation mechanism via unique doubly-peaked gravitational waves spectrum}
%\preprint{IP/BBSR/2015-4}
%\title{Imprints of ultralight PBH formed during phase transitions on gravitational waves}

%\title{Probing ultra-light PBH formation mechanism via unique doubly-peaked gravitational waves spectrum}

\title{QCD axion misalignment during reheating}
%%%%%%%%%%%%%%%%%%
\author[a]{Disha Bandyopadhyay\orcidlink{0009-0002-0871-8538},}
\author[a]{Debasish Borah\orcidlink{0000-0001-8375-282X},}
\author[a, b]{and Nayan Das\orcidlink{0000-0002-7627-0889}}
\affiliation[a]{\,\,Department of Physics, Indian Institute of Technology Guwahati, Assam 781039, India}
\affiliation[b]{\,\,Centre for High Energy Physics, Indian Institute of Science, C. V. Raman Avenue, Bengaluru 560012, India}
\emailAdd{b.disha@iitg.ac.in}
\emailAdd{dborah@iitg.ac.in}
\emailAdd{nayan.das@iitg.ac.in }

\abstract{We study the production of QCD axion dark matter (DM) via vacuum and kinetic misalignment mechanisms during an extended reheating period after the end of slow-roll inflation. With pre-inflationary Peccei-Quinn breaking and considering a monomial inflaton potential during reheating, we classify our study into three different perturbative reheating scenarios in which the inflaton decays into (i) a pair of Standard Model (SM)-like bosons, (ii) a pair of SM-like fermions, or (iii) exclusively into a pair of heavy right-handed neutrinos, which eventually decays into the SM final states after dominating the energy density of the Universe for an intermediate epoch. For each of these reheating scenarios, we study the details of QCD axion relic separately via vacuum and kinetic misalignment mechanisms. For different combinations of reheating, monomial inflaton potential and misalignment mechanisms, we identify the parameter space consistent with the observed DM relic density while satisfying other cosmological and laboratory constraints. We also check the scope of probing the currently allowed parameter space at axion-detection and gravitational wave (GW) experiments.}
%\pacs{}
\maketitle
%%%%%%%%%%%%%%%%%%%%%%%%%%%%%%%%%%%%%%%%%%%%%%%%%%%%%%%%%%%%%%%%%

\section{Introduction}
Experimental observations related to neutron electric dipole moment (EDM) put stringent upper bounds on the strong sector CP violation or the $\bar{\theta}$ parameter of quantum chromodynamics (QCD) as $\bar{\theta} < 10^{-10}$ \cite{Abel:2020pzs}. The requirement for such a fine-tuned $\bar{\theta}$ parameter gives rise to the strong CP problem. The most popular solution is the Peccei-Quinn (PQ) mechanism~\cite{Peccei:1977hh, Peccei:1977ur, Wilczek:1977pj, Weinberg:1977ma, Preskill:1982cy, Kawasaki:2013ae,DiLuzio:2020wdo} where a pseudoscalar field, known as axion, determines the $\bar{\theta}$ parameter dynamically. When the axion acquires a potential from non-perturbative QCD effects, $\bar{\theta}$ dynamically relaxes to zero, the ground state of the potential thereby solving the strong CP problem. In addition to providing an elegant solution to the strong CP problem, axions can also solve other longstanding puzzles in particle physics like the origin of dark matter (DM) \cite{Planck:2018vyg, ParticleDataGroup:2024cfk} in the Universe. Another appealing feature of axions is their promising detection prospects. Axion couplings to the standard model (SM) particles, decided by the PQ symmetry breaking scale or the axion decay constant $f_a$, remain tightly constrained. For QCD axion models with sub-eV axion mass, there exists a stringent astrophysical bound $f_a\gtrsim 10^{8\div9}$ GeV \cite{Raffelt:2006cw, Caputo:2024oqc}. Depending on ths specific production mechanism, ultra-light axion with sub-eV mass can constitute the entire cold dark matter (CDM) in the Universe. In the vacuum misalignment mechanism \cite{Preskill:1982cy, Abbott:1982af, Dine:1982ah}, coherent oscillations of the axion field in an expanding Universe lead the production of CDM. In order to produce the entire CDM observed in the present Universe, the axion decay constant is constrained to be around $f_a\sim 10^{11}$ GeV \cite{Kawasaki:2013ae}. Depending on the details of the QCD axion model namely, Dine-Fischler-Srednicki-Zhitnitsky (DFSZ) \cite{Dine:1981rt, Zhitnitsky:1980tq} or Kim-Shifman-Vainshtein-Zakharov (KSVZ) \cite{Kim:1979if, Shifman:1979if}, additional contributions from the topological defects can alter this window by less than an order of magnitude \cite{Kawasaki:2013ae}. If the axion field has an initial non-zero velocity, as considered in the kinetic misalignment mechanism \cite{Co:2019jts, Chang:2019tvx}, the decay constant $f_a$ can be much lower while being consistent with the observed CDM relic.

Conventional estimates of axion DM relic abundance assume a standard radiation dominated prior to the big bang nucleosynthesis (BBN) epoch. Deviations from standard cosmology can change these predictions leading to a different parameter space in terms of the axion decay constant $f_a$. Since the same $f_a$ also dictates the couplings of QCD axion with the SM particles affecting the detection aspects, it is important to exhaust possible non-standard cosmological histories in the context of axion DM production. Several earlier works \cite{Arias:2021rer, Visinelli:2017imh} considered axion production via vacuum misalignment in a non-standard cosmological epoch with general equation of state. Early matter dominated phase either due to long-lived field \cite{Visinelli:2009kt,Nelson:2018via} or by ultra-light primordial black holes (PBH) \cite{Bernal:2021yyb, Mazde:2022sdx, Borah:2024qyo, Bandyopadhyay:2025ast} was also studied in the context of axion DM. Motivated by this, here we consider a scenario with an extended reheating period after inflation. If the reheating temperature is sufficiently low, the extended reheating period can alter the axion DM abundance by changing the oscillation temperature. A related earlier work \cite{Xu:2023lxw} considered axion DM production for post-inflationary PQ breaking via vacuum misalignment in an extended reheating epoch, during which the inflaton oscillates around the minimum of a generic monomial potential. Another work \cite{Barman:2023icn} considered a similar inflaton dominated epoch but with gravitational reheating only. Here we generalize these earlier works to both vacuum and kinetic misalignment mechanisms by considering an extended reheating era following which the inflaton decays perturbatively into (i) bosonic or fermionic SM particles, and (ii) heavy right-handed neutrinos (RHNs) exclusively, which subsequently decay into the SM final states. In addition, we focus on pre-inflationary PQ breaking such the axion production from topological defects can be ignored. Unlike the earlier works, here we also investigate the complementary observational prospects of this framework through its imprint on the primordial gravitational wave (GW) spectrum, with an inflationary origin. The prolonged reheating phase can significantly modify the shape of the primordial GW spectrum, potentially bringing the resulting signal within the reach of future GW observatories. For different monomial potentials and decay mode of the inflaton field, we find the allowed parameter space consistent with axion DM relic in both the misalignment mechanisms and show the complementary detection prospects at axion and GW experiments. 

This paper is organised as follows. In section \ref{sec:axions} we briefly review the axion misalignment mechanisms in standard cosmology followed by the details of the reheating dynamics in section \ref{sec:reheating}. In section \ref{sec:axionreheating} we discuss axion misalignment during an extended reheating period. In section \ref{sec:detection} we discuss the detection prospects of such scenarios in GW and axion experiments. Finally, in section \ref{sec:conclude}, we summarize our results and conclude. In Appendix \ref{appen1} and Appendix \ref{appen2}, we provide approximate analytical expressions of axion relic abundance for vacuum and kinetic misalignment respectively considering all possible reheating schemes.

\section{Axion Misalignment in Standard Cosmology}
\label{sec:axions}
The axion arises as the angular component of a complex scalar field,
\begin{equation}
\Phi_a=\frac{1}{\sqrt{2}}\,S\,
e^{i\phi_a/f_a},
\end{equation}
charged under the global Peccei-Quinn symmetry $U(1)_{\rm PQ}$. Here $S$ and $\phi_a$ denote radial and angular degrees of freedom, respectively. The axion decay constant $f_a$ is identified with the vacuum expectation value (VEV) of the real part of the singlet scalar charged under $U(1)_{\rm PQ}$, responsible for spontaneous breaking of the latter. The mass of the QCD axion is given as
\begin{equation}
 m_{a} (T)
=
m_{a} \left\{
 \begin{array}{lll}
 1
 & \mbox{for $T \leq T_{\rm QCD} $} 

 \\
 \left(\frac{T_{\rm QCD}}{T}\right)^{4} & \mbox{for $T>T_{\rm QCD}$} 
 \end{array}
 \right.
 \label{axion_mass}
 \end{equation}
 where, $T_{\rm QCD} \simeq$ 160 MeV and 
$   m_{a} \simeq 5.7 \left(\frac{10^{12}~\rm GeV}{f_{a}}\right) \rm \mu eV$ is the zero-temperature mass of the QCD axion. As indicated by lattice simulation, the power of 4 used in the above expression for $T>T_{\rm QCD}$ is not precise \cite{DiLuzio:2020wdo}, but it does not affect our overall results significantly. The corresponding axion potential is given as \begin{eqnarray}
    V(\phi_a, T) = f^2_{a} m^2_{a}(T) \left(1-\cos{\left(\frac{\phi_a}{f_{a}}\right)}\right).
    \label{potential}
\end{eqnarray}
For a large decay constant $f_a$ as required from astrophysical bounds \cite{Raffelt:2006cw, Caputo:2024oqc}, the axion couplings to the SM particles are highly suppressed, keeping production of axions from the thermal bath highly suppressed. However, in the ultra-light mass window, the axion field can behave coherently oscillating scalar field. This can lead to production of cold axions within the purview of misalignment mechanisms which we briefly summarize below. Throughout this study, we assume the PQ symmetry to be broken before inflation such that the production of axions from topological defects \cite{Kawasaki:2014sqa} remains sub-dominant.
%\textcolor{Red}{Check if notation are consistent in paper and if any additional detail needed}
\subsection{Vacuum Misalignment}\label{VM}
The evolution of the homogeneous axion field $\phi_a$ in an expanding Universe is governed by 
\begin{eqnarray}
    \Ddot{\phi_a} + 3\mathcal{H} \Dot{\phi_a} + \frac{\partial V(\phi_a, T)}{\partial\phi_a} = 0,
\label{axevo}
\end{eqnarray}
where $\mathcal{H}$ denotes the Hubble expansion parameter. For our convenience, we parametrize the axion field in terms of the dimensionless quantity $\theta \equiv \frac{\phi_a}{f_{a}}$. In terms of $\theta$, Eq. \eqref{axevo} can be rewritten as
\begin{eqnarray}
    \Ddot{\theta} + 3\mathcal{H} \Dot{\theta} + m_a^2(T)\,\rm sin\theta = 0.
\label{thetaEvo}
\end{eqnarray}
In the vacuum misalignment mechanism \cite{Preskill:1982cy, Abbott:1982af, Dine:1982ah}, $\theta$ is initially displaced from its minima and remains fixed at a constant value $ \theta_i \, \in \, (-\pi, \pi)$ with $\dot{\theta}_i \, = \,0$ due to Hubble friction. As the temperature drops in an expanding Universe, at some point the mass of the axion becomes comparable to the Hubble term and this is when the axion starts oscillating. This oscillation temperature $T_{\rm osc}$ is defined as
\begin{eqnarray}
  3  \mathcal{H} (T_{\rm osc}) = m_{a} (T_{\rm osc}).
\label{Tosc}
\end{eqnarray}
The energy density in the axion field at any given temperature is, \begin{equation}\label{energyden}
    \rho_a \, = \, \dfrac{1}{2} \, \dot{\phi_a}^2 + V(\phi_a).
\end{equation}
Since the axion behaves as non-relativistic matter after the onset of oscillations, its energy density is related to the number density by
\begin{equation}
    \rho_a(T)=m_a(T)\,n_a(T).
\end{equation} 
From the conservation of comoving number density, the axion number density at a later epoch can be written as 
\begin{align}
    n_{a}(T) = n_{a}(T_{\rm osc}) \frac{s(T)}{s(T_{\rm osc})}.
    \label{num den}
\end{align}
Here $s(T)$ denotes the entropy density at a temperature $T$. 
The present axion dark matter relic abundance can be found to be
\begin{align}
    \Omega_{a}h^2 = \frac{\rho_{a}(T_{0})}{\rho_{c}}h^2 = \frac{m_{a}}{\rho_{c}} \left(\frac{\rho_{a}(T_{\rm osc})}{m_{a}(T_{\rm osc})} \frac{s(T_0)}{s(T_{\rm osc})} \right) h^2.
    \label{Axion_DM}
\end{align}
where $\rho_c$ is the critical density of the present Universe given by $\rho_c=3M_{\rm P}^2\mathcal{H}_0^2
    \simeq 8.10\times10^{-11}\,h^2~{\rm eV}^4$.
% with $\mathcal{H}_0=100\,h~{\rm km\,s^{-1}\,Mpc^{-1}}$ being the present Hubble parameter. For sufficiently small initial misalignment angles, $\theta \ll 1$, the potential can be approximated by its harmonic form,
% \begin{equation}
% V(\theta,T)\simeq \frac{1}{2}m_a^2(T)f_a^2\theta^2,
% \end{equation}
% such that Eq.~\eqref{thetaEvo} can be solved analytically. Near the onset of oscillations the field velocity is still small, so the energy density is dominated by the potential term is approximately
% \begin{equation}
% \rho_a(T_{\rm osc}) \simeq \frac{1}{2}m_a^2(T_{\rm osc})f_a^2\theta_i^2 \, \, \,\, {\rm or} \,\,\,\, \Omega_a h^2 \propto
% \theta_i^2
% \label{rhoosc}
% \end{equation}

%  Hence, it is evident that the relic abundance of the axion remains sensitive to the initial displacement of the field from its minimum. This initial misalignment angle depends on the PQ symmetry breaking history. 

% In our work, we consider the PQ symmetry is broken after inflation, in such scenario different Hubble patches acquire random values of $\theta_i$, and the physically relevant quantity is the root mean square over all patches, which gives
% $\theta_i=\sqrt{\langle\theta_i^2\rangle}=\frac{\pi}{\sqrt{3}}\simeq1.81$. The relic abundance thus computed  numerically by using Eqs. \eqref{potential}-\eqref{Axion_DM} for the reference value $\theta_i=\pi/\sqrt{3}$, corresponding to $f_a\simeq2.4\times10^{11}\,\mathrm{GeV}$. 
If the PQ symmetry is broken before or during inflation and is never restored afterwards, the observable Universe inherits a single homogeneous value of $\theta_i$, which remains a free parameter. The analytic dependence of the relic abundance on $\theta_i$ and $f_a$ is 
\begin{equation}
\Omega_a h^2 \propto \theta_i^2
\begin{cases}
f_a^{3/2}, & T_{\rm osc} \gtrsim T_{\rm QCD}, \\
f_a^{7/6}, & T_{\rm osc} < T_{\rm QCD}.
\end{cases}
\end{equation} 
It is evident that smaller values of  $\theta_i$ allow larger values of $f_a$, corresponding to lighter axions, while still reproducing the observed dark matter abundance. However, choosing an extremely small initial misalignment angle requires significant fine-tuning of the initial condition and diminishes one of the primary motivations of the Peccei--Quinn mechanism as a natural solution to the strong CP problem. Hence $\theta_i$ in this case is typically taken as $\mathcal{O}(1)$ and the correct relic abundance is obtained typically for $f_a \approx \mathcal{O}(\,10^{11} - 10^{12} \,)$ GeV.

% In the remainder of this work, we focus on the pre-inflationary PQ symmetry breaking scenario, where $\theta_i$ is treated as a free parameter. The relic abundance computed numerically for the reference value $\theta_i=\pi/\sqrt{3}$ can subsequently map the results to arbitrary homogeneous values of $\theta_i$ using the analytic formula for abundance in Eq. \eqref{rhoosc} to obtain the corresponding $f_a$ and $m_a$ whenever required for the pre-inflationary PQ symmetry breaking scenario. However, the analytic and numerical results may differ due to approximations such as the small-angle limit, neglecting kinetic energy at $T_{\rm osc}$, and the assumption that axion starts behaving like cold dark matter of the onset of oscillations itself and temperature-dependent quantities, but both predict the correct dark matter abundance for $f_a$ of the same order.

\subsection{Kinetic Misalignment}
In the kinetic misalignment mechanism \cite{Co:2019jts, Chang:2019tvx}, the axion field possesses a non-zero initial velocity, $\dot{\theta}_i \neq0$, while initially remaining nearly homogeneous. The conserved Noether charge associated with the shift symmetry is defined in terms of number density and its comoving counterpart as
\begin{equation}
n_\theta \equiv \dot{\theta}f_a^2,
\qquad
Y_\theta \equiv \frac{n_\theta}{s}.
\label{eq:Ytheta}
\end{equation}
Since both $n_\theta$ and $s$ decrease with the scale factor $a$ as $a^{-3}$, the yield $Y_\theta$ remains constant during the expansion. 
For sufficiently small initial velocity, oscillations begin at the usual temperature determined by Eq. \eqref{Tosc}. If the kinetic energy dominates over the potential barrier at $T_{\rm osc}$, the field continues to roll until the kinetic energy becomes insufficient to cross the potential barrier which is given by the following condition
\begin{equation}
\dot{\theta}(T')=2m_a(T'),
\label{eq:Tprime}
\end{equation}
where $T'$ denotes the delayed onset of oscillations. The kinetic misalignment regime requires
\begin{equation}
m_a(T')\ge 3\mathcal{H}(T').
\label{eq:KMcondition}
\end{equation}
After the onset of oscillations, the comoving axion number density is conserved resulting in the comoving axion energy density as
\begin{equation}
\frac{\rho_a}{s}
= m_a(0) \frac{n_a(T')}{s(T')} =
\mathcal{C}_{\rm KM}\,m_a(0)\,Y_\theta ,
\end{equation}
where $
n_a(T')=
\frac{V}{m_a(T')}
\simeq
2m_a(T')f_a^2
\simeq
\dot{\theta}(T')f_a^2
=
n_\theta,
$ and $ \mathcal C_{\rm KM}$ is a numerical correction factor accounting for the delayed onset of oscillations due to anharmonic effects. In the simple analytic estimate $ \mathcal C_{\rm KM} =1$, while a full numerical solution obtained by solving Eqs.~\eqref{thetaEvo}-\eqref{num den}, numerically, averaging the number density over a few oscillations, and taking anharmonicity effects into account yields $ \mathcal C_{\rm KM} \simeq 2$ which is consistent with~\cite{Co:2019jts}. In the regime where kinetic misalignment remains valid, the axion relic abundance becomes
\begin{equation}
\Omega_a h^2
\simeq
0.12
\left(
\frac{10^9~{\rm GeV}}{f_a}
\right)
\left(
\frac{Y_\theta}{40}
\right),
\label{eq:OmegaKM}
\end{equation}
which is independent of the temperature evolution of the axion mass and initial value of $\theta_i$.

\section{Reheating Dynamics}
\label{sec:reheating}
At the end of slow-roll inflation, the inflaton field $\phi$ begins coherent oscillations around the minimum of its potential. The characteristic inflationary energy scale is related to the tensor-to-scalar ratio $r$ as $\Lambda_I \equiv V^{1/4} = M_P\left(\frac{3\pi^2}{2}A_s r\right)^{1/4}
\simeq 3.23\times10^{16}\, r^{1/4}\ {\rm GeV}$, where $A_s\simeq2.1\times10^{-9}$ \cite{Planck:2018jri}. Adopting the bound $r<0.036$ \cite{BICEPKeck:2024stm} gives
$\Lambda_I\lesssim1.41\times10^{16}\,{\rm GeV}$. The initial inflaton energy density is therefore $\rho_{\phi,{I}} \simeq V_{\rm I}\sim \Lambda_I^4 $ and the corresponding initial Hubble rate is
\begin{equation}
    \mathcal{H}_{\rm I}
    \simeq \sqrt{\frac{\rho_{\phi,{I}}}{3M_{\rm P}^2}}
    \sim \frac{\Lambda_I^2}{\sqrt{3}\,M_{\rm P}} \leq 4.7 \times 10^{13}\, \rm GeV 
    \label{Hmax}
\end{equation}
%\textcolor{Red}{Check this and add upper bound on $H_I$}
where $M_P$ is the reduced Planck scale. We study the post-inflationary oscillations of the inflaton $\phi$ around the minimum of a monomial potential
\begin{equation}\label{eq:inf-pot}
V(\phi) = \lambda\, \frac{\phi^n}{\Lambda^{n-4}}\,,
\end{equation}
where the dimensionless coupling $\lambda$ decides the overall scale of the potential and the dimensionful parameter $\Lambda$ depends on the form of the full inflationary potential, details of which can be found in \cite{Barman:2026yhc}. Potential of the form in Eq.~\eqref{eq:inf-pot} arise in several inflationary frameworks, such as $\alpha$-attractor T- or E-models~\cite{Kallosh:2013hoa, Kallosh:2013yoa, Kallosh:2013maa, Kallosh:2015lwa} and the Starobinsky model~\cite{Starobinsky:1980te, Starobinsky:1983zz, Kofman:1985aw}. Averaging over oscillations, we can find the equation of state of the inflaton as $w_\phi = \frac{n-2}{n+2}$.
In the absence of decays the energy density of the inflaton condensate scales as
\begin{equation}
    \rho_\phi(a)
    = \rho_{\phi,{\rm rh}}
      \left(\frac{a}{a_{\rm RH}}\right)^{-3(1+w_\phi)}
    = \rho_{\phi,{\rm rh}}
      \left(\frac{a}{a_{\rm RH}}\right)^{-\frac{6n}{n+2}}.
\end{equation}
For $n=2$ this reproduces matter-like behaviour ($\rho_\phi\propto a^{-3}$),
while for $n=4$ one obtains radiation-like scaling. If we keep increasing n, then in the limit $n \to \infty$ we can obtain kination-like scaling $(\omega_\phi \rightarrow 1)$. The effective inflaton mass during oscillations is
\begin{equation}
    m_\phi^2(a)
    = \frac{\partial^2 V}{\partial\phi^2}
    = n(n-1)\lambda\,\frac{\phi^{\,n-2}}{\Lambda^{\,n-4}}.
\end{equation}
Using $\rho_\phi\simeq V(\phi)$ during oscillations,
\begin{equation}
    m_\phi^2(a)
    \simeq n(n-1)\lambda^{2/n}
    \Lambda^{2(4-n)/n}
    \rho_\phi(a)^{(n-2)/n},
    \label{eq:inflatonmass}
\end{equation}
which shows that for $n\neq2$ the effective mass evolves with time
through $\rho_\phi(a)$.

% \begin{figure*}
%     \centering
%    \includegraphics[scale=0.35]{Figures/energy density evo.pdf}
%     \caption{Evolution of radiation and inflaton energy densities for two different value of n = 2,4  for bosonic reheating when $\lambda =\frac{1}{10^{12}},\Lambda_I =10^{13} \, , \rm and \,\mu_{eff}=0.5 \times{10^5}$}
%     \label{fig:energy}
% \end{figure*}
%\subsection{Boltzmann system during reheating}

\subsection{Case I: Direct Reheating}
Following inflation, the inflaton $\phi$ oscillates about the minimum of its potential while decaying into the radiation. The coupled evolution equations for the inflaton and radiation energy densities are
\begin{align}
\dot{\rho}_{\phi}
+\frac{6n}{n+2}\mathcal{H}\rho_{\phi}
&=
-\frac{2n}{n+2}\Gamma_{\phi}(a)\rho_{\phi},
\\[2mm]
\dot{\rho}_{R}
+4\mathcal{H}\rho_{R}
&=
\frac{2n}{n+2}\Gamma_{\phi}(a)\rho_{\phi},
\\[2mm]
3M_{\rm P}^{2}\mathcal{H}^{2}
&=
\rho_{\phi}+\rho_{R},
\end{align}
where $\Gamma_\phi$ is the inflaton decay rate. During the early reheating phase when $\rho_\phi \gg \rho_R$, $\rho_R$ is negligible and the expansion is dominated by the inflaton. Therefore, the Hubble expansion parameter can be approximated as
\begin{equation}
    \mathcal{H}(a) \simeq
    \sqrt{\frac{\rho_\phi(a)}{3M_{\rm P}^2}}
    \propto a^{-\frac{3}{2}(1+w_\phi)}
    = a^{-\frac{3n}{n+2}}.
\end{equation}
or
\begin{equation}
 \mathcal{H}(a) \simeq \mathcal{H}(a_{\rm RH}) \times 
 \left\{
 \begin{array}{lll}
 \left (\dfrac{a_{\rm RH}}{a}\right)^{3n/(n+2)}
 & \mbox{for $a_I \leq a \leq a_{\rm RH}$} 
% & \mbox{(radiation-like)}
 \\
 \left (\dfrac{a_{\rm RH}}{a}\right)^{2}
 & \mbox{for $a_{\rm RH} \leq a$} 
\end{array}
 \right.
 \label{hub_case1}
 \end{equation}
where, $a_I$ and $a_{\rm RH}$ denote the scale factors at the end of inflation and reheating. More precisely, $a_{\rm RH}$ is scale factor when radiation energy density starts dominating the Universe $\rho_R (a_{\rm RH})=\rho_\phi (a_{\rm RH})$. The temperature of the bath at $a=a_{\rm RH}$ is identified as the reheating temperature $T_{\rm RH}$ such that the Universe becomes radiation dominated for $T< T_{\rm RH}$. Successful BBN requires $T_{\rm RH} > T_\text{BBN} \simeq 4$~MeV~\cite{Sarkar:1995dd, Kawasaki:2000en, Hannestad:2004px, DeBernardis:2008zz, deSalas:2015glj,Hasegawa:2019jsa}.

% A useful estimate is obtained
% by setting
% \begin{equation}
%     \mathcal{H}(a_{\rm rh}) \simeq \Gamma(a_{\rm rh}).
% \end{equation}
For direct reheating, we consider two separate scenarios. In the first scenario known as bosonic reheating, the inflaton decays into a pair of bosonic final states through interactions of the type $\mu\,\phi\,|H|^2$, where $H$ is the SM Higgs doublet. In the second scenario known as fermionic reheating, the reheating proceeds via inflaton's energy dissipation into fermion final states via Yukawa interaction of the type $y_\psi\,\bar{\psi}\,\psi\,\phi$.

For bosonic reheating ($\phi\to H^\dagger H$) with inflaton–scalar trilinear coupling, the effective decay rate is
\begin{equation}
\Gamma^H_\phi(a)
= \frac{\mu_{\rm eff}^2}{8\pi\,m_\phi(a)},
\end{equation}
where $\mu_{\rm eff}$ is the effective coupling after averaging over oscillations \cite{Shtanov:1994ce, Ichikawa:2008ne, Garcia:2020wiy} and $\mu_{\rm eff} \neq \mu$ for $n \neq 2$. Similarly, for fermionic reheating ($\phi\to\psi\overline{\psi}$) the effective decay rate averaged over oscillations is 
\begin{equation}
\Gamma^\psi_\phi(a)
= \frac{y_{\rm eff}^2}{8\pi\,}m_\phi(a),
\end{equation}
where $y_{\rm eff}\neq y_\psi$ for $n\neq2$ arises after averaging \cite{Shtanov:1994ce, Ichikawa:2008ne, Garcia:2020wiy}. 
During reheating (i.e., for $a_I\ll a\ll a_{\rm RH}$) when the inflaton dominates the total energy density, one can solve the continuity equations analytically to obtain the radiation energy density as a function of scale factor as \cite{Bernal:2022wck}
\begin{equation}
\rho_R(a)\simeq 
    \begin{cases}
    \frac{3n}{7-n}\,M_P^2\,\Gamma^\psi_\phi(a_{\rm RH})\,\mathcal{H}(a_{\rm RH})
    \left(\frac{a_{\rm RH}}{a}\right)^{\frac{6(n-1)}{n+2}}\,
    \left[\,1-\left(\frac{a_I}{a}\right)^{\frac{2(7-n)}{n+2}}\right],
    & \text{fermionic reheating (n < 7)},\\[0.7em]
    \frac{3n}{1+2n}\,M_P^2\,\Gamma^H_\phi(a_{\rm RH})\,\mathcal{H}(a_{\rm RH})
    \left(\frac{a_{\rm RH}}{a}\right)^{\frac{6}{n+2}}\,
    \left[\,1-\left(\frac{a_I}{a}\right)^{\frac{2(1+2n)}{n+2}}\right],
    & \text{bosonic reheating}.
    \end{cases} \nonumber \\
\end{equation}
Once the decay products thermalize, the radiation energy density is related to the temperature as $ \rho_R(a)
    = \frac{\pi^2}{30}\,g_*(T)\,T^4.$
Hence, using $ T(a)
    = \left(\frac{30}{\pi^2 g_*(T)}\rho_R(a)\right)^{1/4},$
the corresponding temperature of the thermal bath during reheating is
\begin{equation}
T(a)\simeq T_{\rm RH}\,
    \left(\frac{a_{\rm RH}}{a}\right)^\alpha
    \label{temp_case1}
\end{equation}

with
\begin{equation}
\alpha =
\begin{cases}
\dfrac{3}{2(n+2)}, 
& \text{bosonic reheating}, \\[0.8em]
\dfrac{3(n-1)}{2(n+2)}, 
& \text{fermionic reheating, } n<7, \\[0.8em]
1, 
& \text{fermionic reheating, } n>7.
\end{cases}
\label{alpha1}
\end{equation}
The Hubble parameter as a function of temperature during reheating can then be written as
\begin{equation}
\mathcal{H}(T)\simeq \mathcal{H}(T_{\rm RH})\,
    \left(\frac{T}{T_{\rm RH}}\right)^{\frac{3n}{(n+2)\alpha}},
\end{equation}
indicating nonstandard scaling during reheating before radiation domination sets in.

The temperature of thermal bath can exceed the eventual reheating temperature $T_{\rm RH}$ by a large factor before the onset of the radiation-dominated era:
\begin{equation}
T_{\rm max}\simeq T_{\rm RH}\,
    \begin{cases}
    \left(\frac{a_{\rm RH}}{a_I}\right)^{\frac{3(n-1)}{2(n+2)}},
    & \text{fermionic reheating},\\[0.7em]
    \left(\frac{a_{\rm RH}}{a_I}\right)^{\frac{3}{2(n+2)}},
    & \text{bosonic reheating},
    \end{cases}
    \label{eq:tmax}
\end{equation}
which can be substantially larger than $T_{\rm RH}$ for $a_{\rm RH}\gg a_I$.  The maximum value $T_{\rm max}$ can be reached during
the early reheating stage when radiation production is efficient, even
before reheating is complete. It is also worth noting that, for the same $T_{\rm RH}$, $T_{\rm max}$ corresponding to bosonic reheating is much lower compared to the fermionic scenario, for $n>2$.

\subsection{Case II: Reheating via RHN Decay}
Instead of directly decaying into the SM bath, the inflaton condensate can also decay into a heavy particle which subsequently decays into the SM bath. Such a possibility naturally emerges in Type-I seesaw mechanism~\cite{Minkowski:1977sc,Yanagida:1979as,Gell-Mann:1979ijt, GellMann:1980vs,Mohapatra:1979ia,Schechter:1980gr,Schechter:1981cv}, where heavy gauge singlet right-handed neutrinos $N_i$ are introduced to explain the origin of light neutrino masses. The relevant interaction Lagrangian is given by,
\begin{align}
-\mathcal{L}\supset\overline{\ell}_{L_\alpha}\,(y_{\nu})_{\alpha i}\,\widetilde{H}\,N_i
+\frac{1}{2}\,\overline{N_i^c}\,(M_N)_{ii}\,N_i +\left(y_{\phi N}\right)_i\,\phi\,\overline{N_i^c}\,N_i
+\text{h.c.},\qquad (i=1,2,3)\,,
\end{align}
with $\ell, H$ being the SM lepton and Higgs doublets respectively. For simplicity, we consider all the three RHNs to be mass-degenerate with mass $M_N$ and estimate their Yukawa coupling with the SM leptons using the Casas--Ibarra (CI) parametrization~\cite{Casas:2001sr} which incorporates the constraints from light neutrino mass and mixing. Another advantage of introducing heavy RHNs is the possibility of generating the baryon asymmetry of the Universe via leptogenesis \cite{Fukugita:1986hr}. While we do not discuss the results for leptogenesis here, a few recent works \cite{Garcia:2026ulw, Barman:2026yhc} have studied non-thermal leptogenesis with low reheating temperatures, similar to what we have considered in this work. On the other hand, several earlier works \cite{Ballesteros:2016xej, Clarke:2015bea, Sopov:2022bog, Borah:2024qyo} studied similar setups with Type-I seesaw extended by $U(1)_{\rm PQ}$ symmetry. Here we consider $N_1$ to be sufficiently long-lived such that its decay can decide the onset and duration of the reheating era itself. The other two RHNs $N_{2,3}$ can account for dominant contributions to light neutrino masses and baryon asymmetry of the Universe via leptogenesis. As our goal is to study axion misalignment in extended reheating via RHN decay, we skip the phenomenology related to the other two RHNs.

Due to the direct coupling of inflaton to the RHNs, here the inflaton decays first into heavy RHNs followed by the decay of the latter into the SM bath thereby reheating the Universe. The relevant Boltzmann equations (BEs) for this seesaw reheating\footnote{The term "Seesaw Reheating" was recently coined by \cite{Mambrini:2026tla} for a similar setup of type-I seesaw and inflaton but with an additional long-lived scalar.} scenario can be written as
\begin{align}\label{eq:BEQ-phiNN}
& \dot\rho_\phi+3\,\mathcal{H}\,(1+w_\phi)\,\rho_\phi=-\Gamma_{\phi\to N_1N_1} \,(1+w_\phi)\,\rho_\phi,  
\nonumber\\&
\dot\rho_R + 4\,\mathcal{H}\,\rho_R = \langle \Gamma_N\rangle\,\left(n_{N_1}-n_{N_1}^{\rm eq}\right),
\nonumber\\&
\Dot{n}_{N_1}+ 3\,\mathcal{H}\,n_{N_1}= - \langle \Gamma_N\rangle\,\left(n_{N_1}-n_{N_1}^{\rm eq}\right)+\frac{\Gamma_{\phi\to N_1N_1}}{m_\phi}\,(1+w_\phi)\,\rho_\phi,
\nonumber\\&
3\,\mathcal{H}^2\,M_P^2= \rho_\phi+\rho_R+ n_{N_1}\,E_N,
\nonumber\\&
E_N^2=M_N^2 + \left(\frac{m_\phi(a)}{2}\,\frac{a'}{a}\right)^2\,.
\end{align}
The thermally averaged $N_1\to\ell\,H$ decay width is defined as,
\begin{align}
\langle \Gamma_N\rangle= \frac{K_1(M_1/T)}{K_2(M_1/T)}\frac{M_1}{8\pi} \left(y_\nu^\dagger y_\nu\right)_{11}\,,  
\end{align}
with $K_i$'s being modified Bessel functions of $i$-th kind. The equilibrium number density of any species $j$ is given by
\begin{align}
    n_j^{\rm eq}= \frac{g_j\,T^3}{2 \pi^2} \left(\frac{M_j}{T}\right)^2 K_2(M_j/T),
\end{align}
with $g_j$ representing the degrees of freedom of the $j$-species. In the last equation of \eqref{eq:BEQ-phiNN}, $a'$ denotes the scale factor at which $\phi$ decays and produces RHNs. In reality, RHN production occurs continuously throughout the $\phi$-dominated era, implying that the corresponding value of $a'$ varies for RHNs produced at different times \cite{Bernal:2026lpr}. However, for solving the BEs,
%, we adopt an approximate value of $a_D$, determined by the condition
%$\Gamma_{\phi\to N_1N_1} \simeq \mathcal{H}$.
%We have explicitly verified that this approximation does not lead to any significant deviation from the results presented in Fig. 1 of \cite{Bernal:2026lpr}.  Correction by DB':
we have used the following approximation for the RHN energy
\begin{align}
E_N(a) =
\begin{cases}
\displaystyle \sqrt{M_N^2+\left(\frac{m_\phi(a)}{2}\right)^2}, & a<a_D,\\
\displaystyle \sqrt{M_N^2+\left[\frac{m_\phi(a_D)}{2}\frac{a_D}{a}\right]^2}, & a\geq a_D.
\end{cases}
\end{align}
Here, $a_D$ denotes the scale factor corresponding to the characteristic time of $\phi$ decay into RHNs, determined by the condition $\Gamma_{\phi\to N_1N_1}\simeq\mathcal{H}$. For $a<a_D$, RHN is produced relativistically, and the momentum of newly produced RHNs is approximated by $p_N\simeq m_\phi(a)/2$. For $a\geq a_D$, we approximate the produced RHNs as being generated around $a_D$, after which their momentum redshifts as $p_N(a)= [m_\phi(a_D)/2](a_D/a)$.

Our approximation works well as most of the RHNs are produced at around $a_{ D}$ and we have explicitly verified that our approximation does not lead to any significant deviation from the numerical results presented in \cite{Bernal:2026lpr}.
Unlike Case I, here the inflaton decays into massive RHNs bringing additional kinematical conditions. The effective mass of inflaton given in Eq. \eqref{eq:inflatonmass} can be rewritten as
\begin{equation}\label{eq:inf-mass}
m_\phi(a) \simeq m_I \left(\frac{a_I}{a}\right)^\frac{3 (n-2)}{n+2}\,,
\end{equation}
with $m_I\equiv m_\phi(a_I)$. For $n \neq 2$, $m_\phi$ is field-dependent and the decay into RHNs are allowed as long as $m_\phi(a)>2M_N$. For $n>2$, this decay channel gets kinematically forbidden beyond a critical scale factor,  
\begin{align}\label{eq:ast}
a_\star= a_I\,\left(\frac{M_N}{2\,m_I}\right)^\frac{n+2}{3(2-n)}.
\end{align}
For $n=2$, as the inflaton mass is independent of the field and the decay into RHNs continues forever if $m_\phi > 2 M_N$.

The evolution of the Hubble parameter and the radiation temperature can be approximated by the following piecewise expressions:
\begin{equation}
\mathcal{H}(a) \simeq
\begin{cases}
\displaystyle
\mathcal{H}_I\left(\frac{a_I}{a}\right)^{\frac{3n}{n+2}},
& \text{Inflaton dominated: } a_I \leq a \leq a_{\rm R},
\\[1.0em]
\displaystyle
\mathcal{H}_{\rm R}\left(\frac{a_{\rm R}}{a}\right)^2,
& \text{Relativistic RHN dominated: } a_{\rm R} \leq a \leq a_{\rm NR},
\\[1.0em]
\displaystyle
\mathcal{H}_{\rm NR}\left(\frac{a_{\rm NR}}{a}\right)^{\frac{3}{2}},
& \text{Non-relativistic RHN dominated: } a_{\rm NR} \leq a \leq a_{\rm RH},
\\[1.0em]
\displaystyle
\mathcal{H}_{\rm RH}\left(\frac{a_{\rm RH}}{a}\right)^2,
& \text{Radiation dominated: } a_{\rm RH} \leq a.
\end{cases}
\label{eq:Hubble_RHN}
\end{equation}
Here, \(\mathcal{H}_I\) denotes the Hubble parameter at the beginning of the reheating evolution, while \(\mathcal{H}_{\rm R}\), \(\mathcal{H}_{\rm NR}\), and \(\mathcal{H}_{\rm RH}\) denote its values at the corresponding characteristic transitions. The temperature evolution is given by
\begin{equation}
T(a) \simeq
\begin{cases}
\displaystyle
T_{\rm max},
& \text{Inflaton dominated: } a_I \leq a \leq a_{\rm R},
\\[1.0em]
\displaystyle
T_{\rm max}
\left(\frac{a_{\rm R}}{a}\right)^{\frac{1}{4}},
& \text{Relativistic RHN dominated: } a_{\rm R} \leq a \leq a_{\rm NR},
\\[1.0em]
\displaystyle
T_{\rm NR}
\left(\frac{a_{\rm NR}}{a}\right)^{\frac{3}{8}},
& \text{Non-relativistic RHN dominated: } a_{\rm NR} \leq a \leq a_{\rm RH},
\\[1.0em]
\displaystyle
T_{\rm RH}
\left(\frac{a_{\rm RH}}{a}\right),
& \text{Radiation dominated: } a_{\rm RH} \leq a.
\end{cases}
\label{eq:Temperature_RHN}
\end{equation}
Here, \(T_{\rm max}\) denotes the maximum temperature attained by the radiation bath during reheating, \(T_{\rm NR}\) is the temperature at the relativistic-to-non-relativistic transition of the RHNs, and \(T_{\rm RH}\) is the reheating temperature. The characteristic expansion rates and scale factors are related by
\begin{equation}
\mathcal{H}_{\rm RH}
=
\sqrt{\frac{\pi^2 g_{*}(T_{\rm RH})}{90}}
\frac{T_{\rm RH}^2}{M_{\rm P}},
\qquad
\mathcal{H}_{\rm NR}
=
\mathcal{H}_{\rm RH}
\left(\frac{T_{\rm NR}}{T_{\rm RH}}\right)^4,
\qquad
\mathcal{H}_{\rm R}
=
\mathcal{H}_{\rm RH}
\frac{T_{\rm max}^8}
{T_{\rm NR}^4 T_{\rm RH}^4},
\end{equation}
and
\begin{equation}
a_{\rm R}
=a_I\left(\frac{\mathcal{H}_I}{\mathcal{H}_{\rm R}}\right)^{\frac{n+2}{3n}},
\qquad
a_{\rm NR}
=a_{\rm R}\left(\frac{T_{\max}}{T_{\rm NR}}\right)^4,
\qquad
a_{\rm RH}
=a_{\rm NR}\left(\frac{T_{\rm NR}}{T_{\rm RH}}\right)^{8/3}.
\label{scale_fac}
\end{equation}
\\
The characteristic scale factors appearing above are summarized in Table~\ref{tab:scale_factors_RHN}.
\begin{table}[t]
\centering
\caption{Characteristic scale factors for different stages of indirect reheating via RHN decay.}
\label{tab:scale_factors_RHN}
\begin{tabular}{|c|l|}
\hline
Scale factor & Physical meaning \\
\hline
\(a_I\) & Beginning of the reheating stage \\
\(a_{\rm R}\) & Onset of the relativistic-RHN dominated regime \\
\(a_{\rm NR}\) & Relativistic-to-non-relativistic transition of the RHNs \\
\(a_{\rm RH}\) & End of reheating and onset of radiation domination \\
\hline
\end{tabular}
\end{table}

Thus, the reheating history in case II consists of an initial inflaton-dominated phase, followed by a relativistic-RHN dominated phase, a non-relativistic-RHN dominated phase, and finally the standard radiation-dominated era.

\section{Misalignment During Reheating}
\label{sec:axionreheating}

\subsection{Vacuum Misalignment During Reheating}
\label{sec:vacreheating}
In the presence of an extended reheating phase, the non-standard expansion history modifies the onset of axion oscillations, which is determined by $\mathcal{H}(T_{\rm osc}) \simeq m_a(T_{\rm osc})$ as given by Eq. \eqref{Tosc}. We can therefore consider two distinct cases depending on whether the axion starts oscillating before or after the completion of reheating.

\begin{itemize}
\item \textit{Case 1}: $T_{\rm osc}<T_{\rm RH}$: In this case, the axion starts oscillating after reheating, when the Universe is already radiation dominated. Hence, the standard cosmological result for vacuum misalignment is recovered. 
\item \textit{Case 2}: $T_{\rm osc}>T_{\rm RH}$: Here, the axion starts oscillating during the reheating phase. The modified Hubble expansion changes $T_{\rm osc}$ and hence the axion abundance produced through vacuum misalignment. In addition, the entropy produced during reheating dilutes the axion abundance. The present-day axion energy density is then given by
\begin{equation}
    \rho_a(T_0)
=
\rho_a(T_{\rm osc})
\frac{m_a}{m_a(T_{\rm osc})}
\frac{s(T_0)}{s(T_{\rm osc})}
\frac{S(T_{\rm osc})}{S(T_{\rm RH})},
\label{eq:rho_a_reheating}
\end{equation}\\ where $S=a^3s$ is the comoving entropy. Thus, when axion oscillations begin during reheating, both the modified expansion history and the entropy production affect the final axion relic abundance. We focus on this case in the remainder of this subsection.
\end{itemize}

Below we study vacuum misalignment of axions for different reheating scenarios, considering direct inflaton decay into bosonic and fermionic channels, as well as indirect decay through right-handed neutrinos, also referred to as the seesaw reheating.
\begin{figure}[htbp]
    \centering
    \captionsetup[subfigure]{labelformat=simple}
    \begin{subfigure}[b]{0.5\linewidth}
        \centering
        \includegraphics[width=\linewidth]{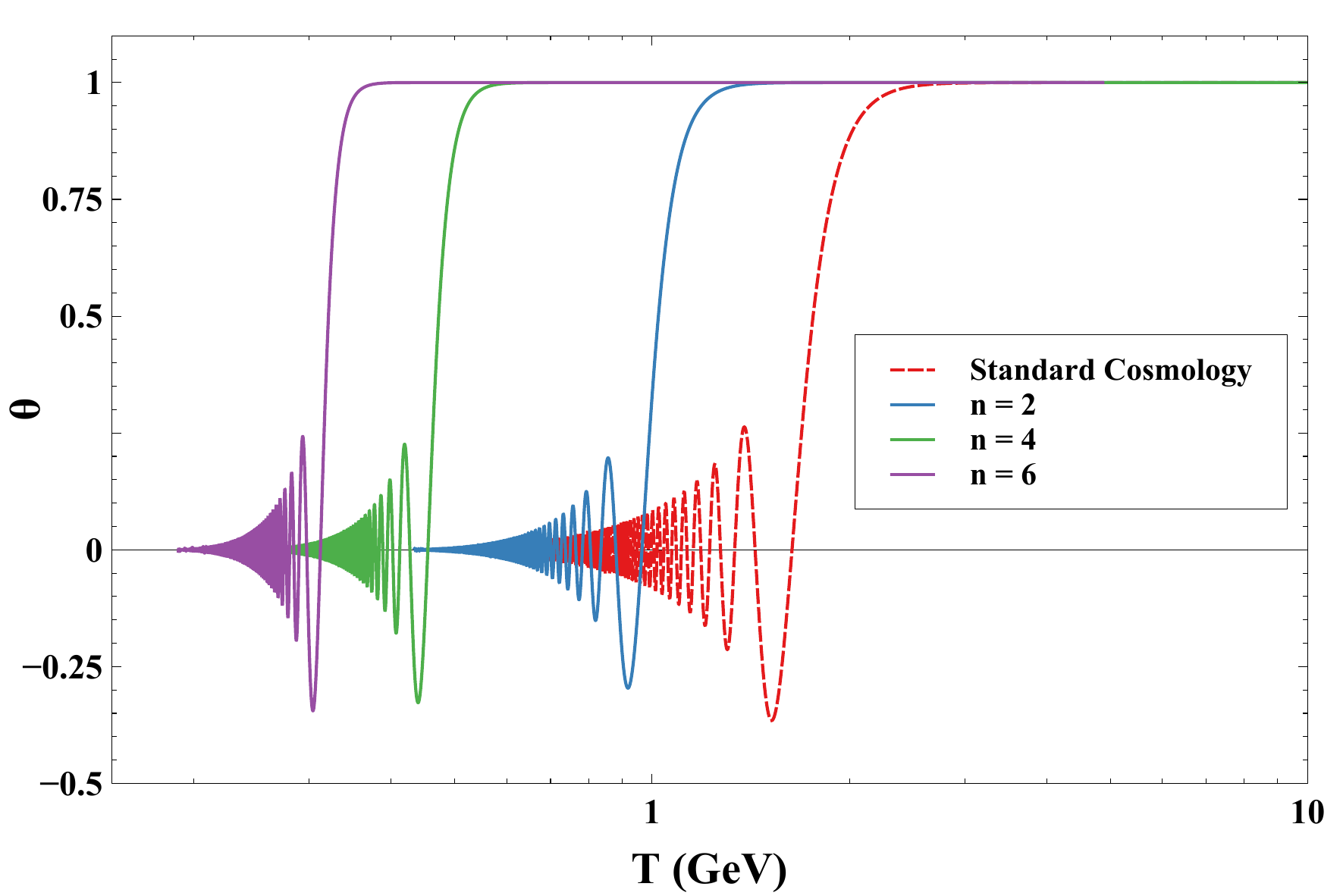}
        \caption{Bosonic reheating (vacuum)}
        \label{fig:axion_evo_bd}
    \end{subfigure}\hfill
    \begin{subfigure}[b]{0.5\linewidth}
        \centering
        \includegraphics[width=\linewidth]{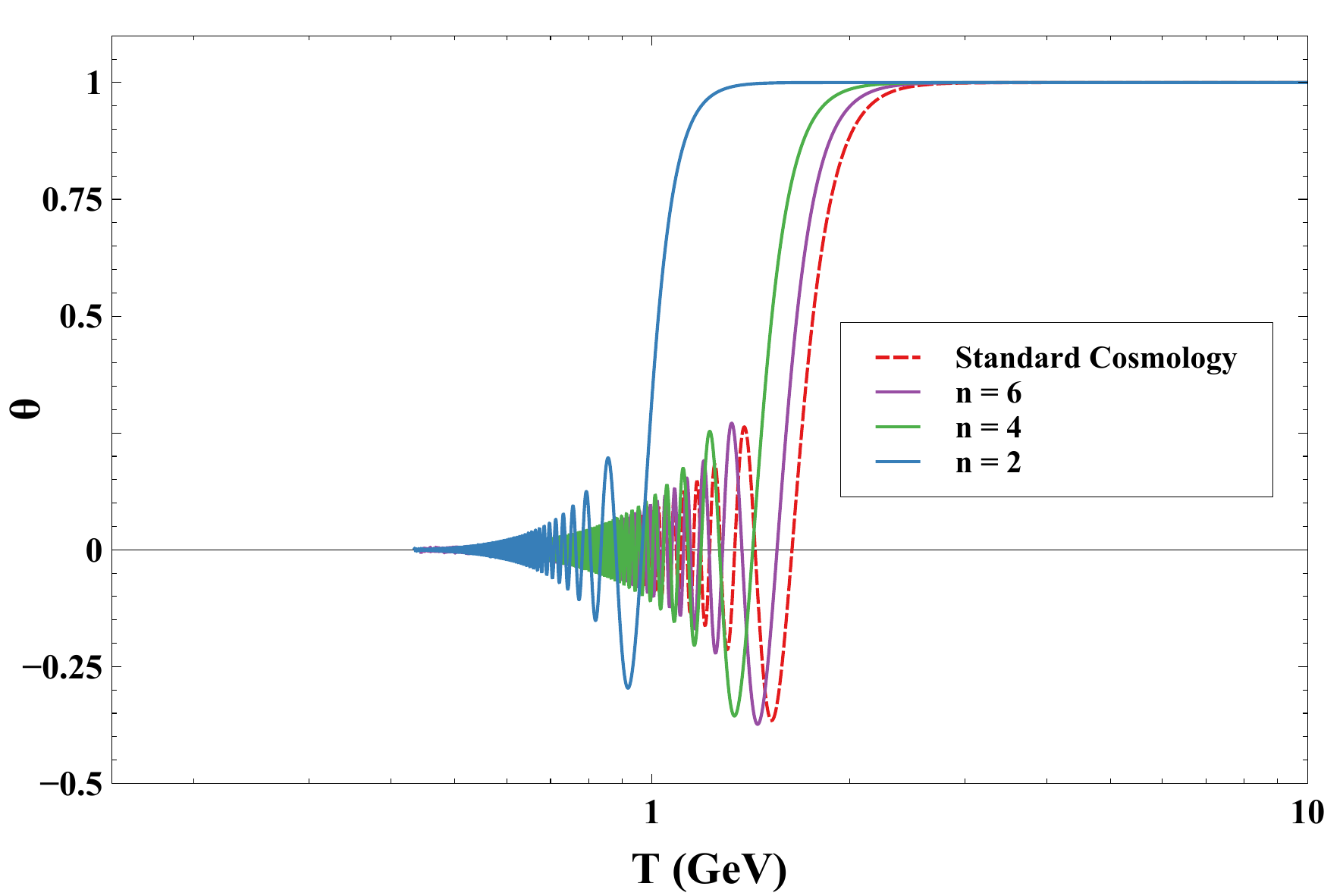}
        \caption{Fermionic reheating (vacuum)}
        \label{fig:axion_evo_fd}
    \end{subfigure}
    \caption{Direct Reheating (Vacuum Misalignment): Axion evolution for $f_a = 10^{10}\text{ GeV}$ and $T_{\text{RH}} = 0.1\text{ GeV}$. Left and right panels correspond to bosonic and fermionic reheating respectively.}
    \label{fig:axion_evolution_grid}
\end{figure}
We also note that for the vacuum misalignment mechanism when PQ symmetry is broken during inflation, primordial axion isocurvature perturbations generated during inflation place an upper bound on the inflationary Hubble scale \cite{Kobayashi:2013nva,DiLuzio:2020wdo},
\begin{equation}
\mathcal{H}_I \lesssim {\mathcal{H}}_I^{\rm iso}=
\frac{0.9 \times 10^{7}}
{F_a^{\rm CDM}(\theta_i,f_a,T_{\rm RH},n)}
\left(
\frac{\theta_i}{\pi}
\right)
\left(
\frac{f_a}{10^{11}\,\mathrm{GeV}}
\right)
\,\mathrm{GeV},
\label{bound}
\end{equation}
where
\begin{equation}
F_a^{\rm CDM}(\theta_i,f_a,T_{\rm RH},n)
=
\frac{\Omega_a^{\rm RH}(\theta_i,f_a,T_{\rm RH},n)}
{\Omega_{\rm CDM}}.
\end{equation}
Throughout this work, we consider the pre-inflationary Peccei--Quinn symmetry breaking scenario and therefore impose the consistency condition $\frac{\mathcal{H}_I}{2 \pi}<f_a.$
Accordingly from the above mentioned arguments and Eq. (\ref{Hmax}), the maximum allowed inflationary scale is taken to be $\mathcal{H}_I^{\rm max}
=
\min\left(\mathcal{H}_I^{\rm iso},2 \pi\,f_a, 4.7 \times 10^{13} \rm GeV \right).$
For a given $\mathcal{H}_I$, the condition in Eq. 
(\ref{Hmax}) translates into a corresponding maximum allowed value of $f_a$, which we denote by $f_a^{\rm max}$.

\subsubsection{Direct Reheating}
\begin{figure}
    \centering
    \captionsetup[subfigure]{labelformat=simple}
    \begin{subfigure}[b]{0.5\linewidth}
        \centering
        \includegraphics[width=\linewidth]{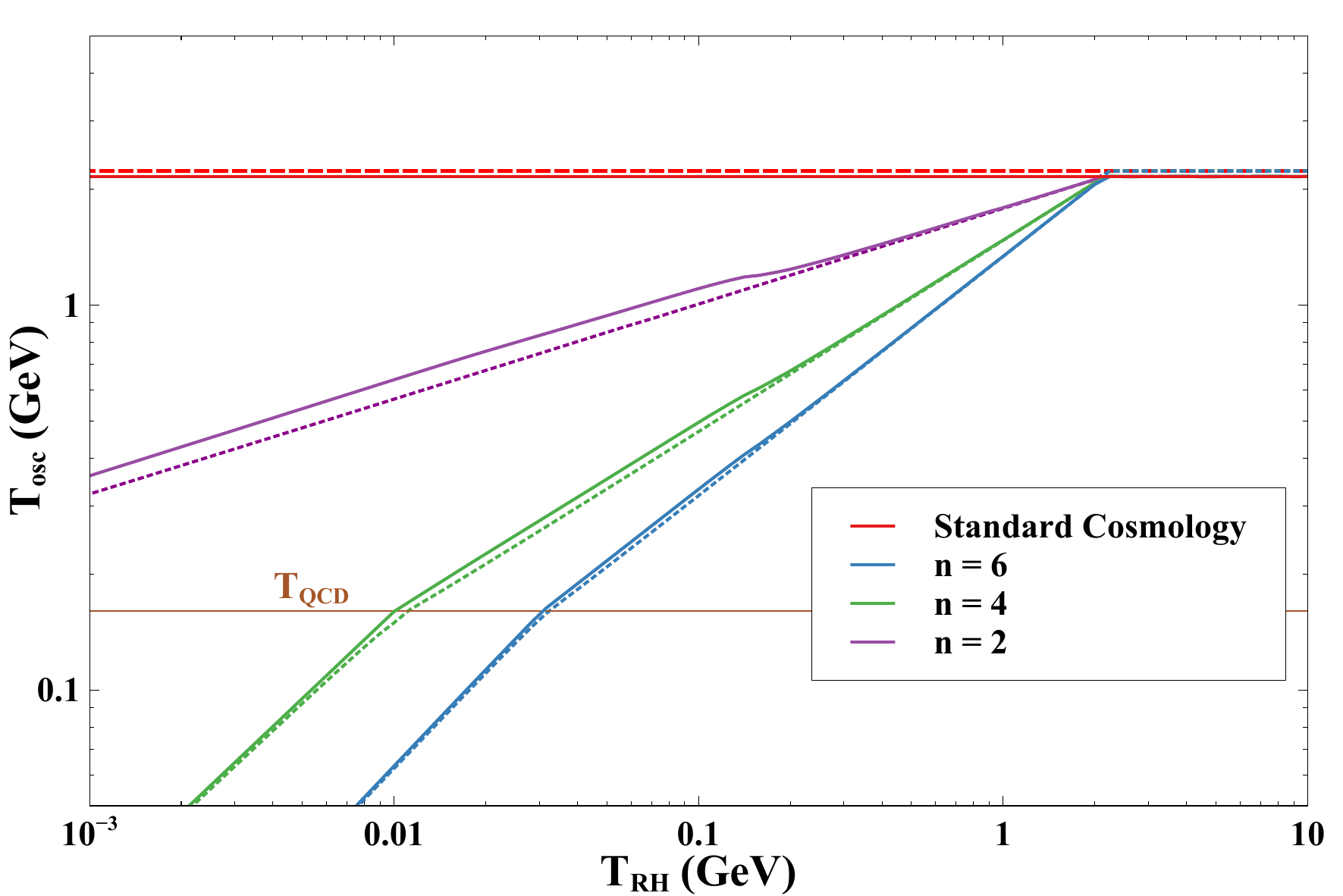}
        \caption{Bosonic reheating (vacuum)}
        \label{sfig:1}
    \end{subfigure}\hfill
    \begin{subfigure}[b]{0.5\linewidth}
        \centering
        \includegraphics[width=\linewidth]{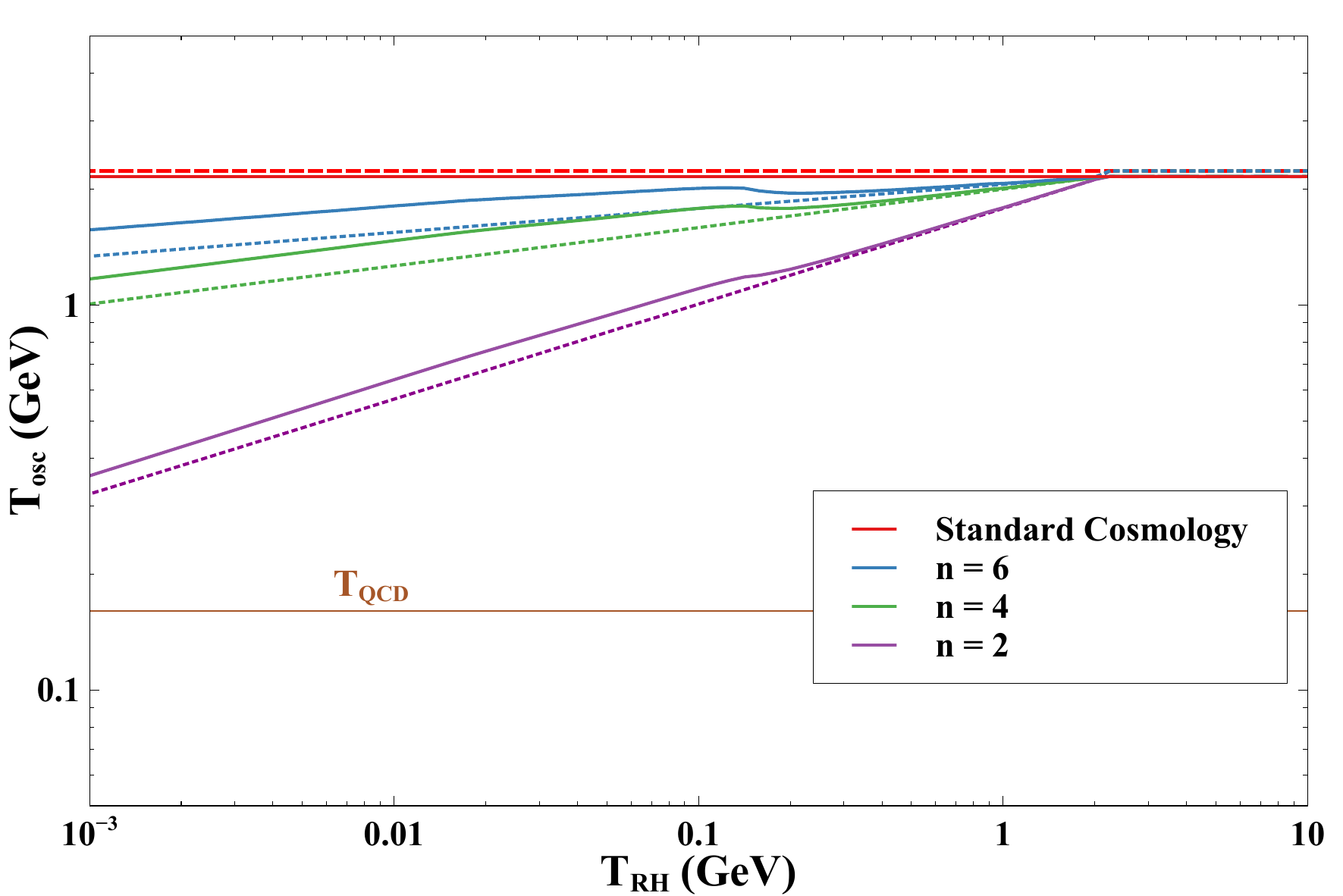}
        \caption{Fermionic reheating (vacuum)}
        \label{sfig:2}
    \end{subfigure}
    \caption{Direct Reheating (Vacuum Misalignment): Variation of $T_{\rm osc}$ with $T_{\rm RH}$ for bosonic (left) and fermionic (right) reheating with different $n$ and $f_a = 10^{10}$ GeV. The dotted and solid curves show the comparison between analytical and numerical results.}
    \label{fig:Tosc}
\end{figure}

We numerically solve Eq.~\eqref{thetaEvo} for axion evolution for different values of $n$ considering bosonic and fermionic reheating. We also use the corresponding Hubble evolution given in Eq.~\eqref{hub_case1} and the temperature evolution in Eq.~\eqref{temp_case1}. The resulting evolution is shown in Fig.~\ref{fig:axion_evolution_grid} for $f_a = 10^{10}\text{ GeV}$ and $T_{\text{RH}} = 0.1\text{ GeV}$. For the case in which the axion starts oscillating during reheating, $T_{\rm osc}>T_{\rm RH}$, the oscillation temperature depends on whether the oscillations begin above or below the QCD scale $T_{\rm QCD} \simeq 160$ MeV. Similar to the treatment of axion oscillations during a non-standard cosmological era in ~\cite{Xu:2023lxw}, we consider the following two cases.
\begin{itemize}
\item $T_{\rm osc}>T_{\rm QCD}$: For oscillations beginning above the QCD scale, the axion mass has a temperature dependence given by
$$
\tilde{m}_a(T)=m_a\left(\frac{T_{\rm QCD}}{T}\right)^{4},
$$
as noted in Eq. \eqref{axion_mass}. Using the oscillation condition in Eq.~\eqref{Tosc}, Hubble evolution given in Eq.~\eqref{hub_case1} and temperature evolution in Eq.~\eqref{temp_case1}

\begin{equation}
T_{\rm osc} = T_{\rm RH} \left[ \frac{1}{\pi}
\sqrt{\frac{10}{g_\star}}
\frac{m_a M_{\rm P} T_{\rm QCD}^{4}}
{T_{\rm RH}^{6}}
\right]^{\frac{\alpha(2+n)}
{3n+4\alpha(2+n)}} .
\label{eq:Tosc_above_QCD}
\end{equation}

\item $T_{\rm osc}\leq T_{\rm QCD}$: For oscillations starting below the QCD scale, the axion mass has reached its zero-temperature value, $\tilde{m}_a(T)=m_a$. In this case, the oscillation temperature is

\begin{equation}
T_{\rm osc} =
T_{\rm RH}
\left[
\frac{1}{\pi}
\sqrt{\frac{10}{g_\star}}
\frac{m_a M_{\rm P}}
{T_{\rm RH}^{2}}
\right]^{\frac{\alpha(2+n)}
{3n}} .
\label{eq:Tosc_below_QCD}
\end{equation}
\end{itemize}
The resulting $T_{\rm osc}$ as a function of $T_{\rm RH}$ is shown in Fig.~\ref{fig:Tosc} for bosonic and fermionic reheating in the left and right panels, respectively, for different values of $n$ and $f_a=10^9\,{\rm GeV}$. The dotted and solid curves correspond to the analytical and numerical results, respectively. The different behavior of $T_{\rm osc}$ in the two reheating modes originates from the different values of $\alpha$ for bosonic and fermionic reheating, as given in Eq. \eqref{alpha1}. In the bosonic reheating, $T_{\rm osc}$ decreases as $n$ increases, corresponding to a delayed onset of oscillations, whereas the opposite behavior is observed for fermionic reheating.

%\begin{figure}
%    \centering
%    \includegraphics[width=0.49\linewidth]{Figure/Tosc1.pdf}
%    \includegraphics[width=0.49\linewidth]{Figure/Tosc2.pdf}
%    \caption{Direct Reheating (Vacuum Misalignment): Variation of $T_{\rm osc}$ with $T_{\rm RH}$ for bosonic (left) and fermionic (right) reheating with different $n$ and $f_a = 10^9$ GeV. The dotted and solid curves show the comparison between analytical and numerical results.}
%    \label{fig:Tosc}
%\end{figure}
%

\begin{figure}[t]
    \centering
    \begin{tabular}{cc}
        \begin{subfigure}{0.49\linewidth}
            \centering
            \includegraphics[width=\linewidth]{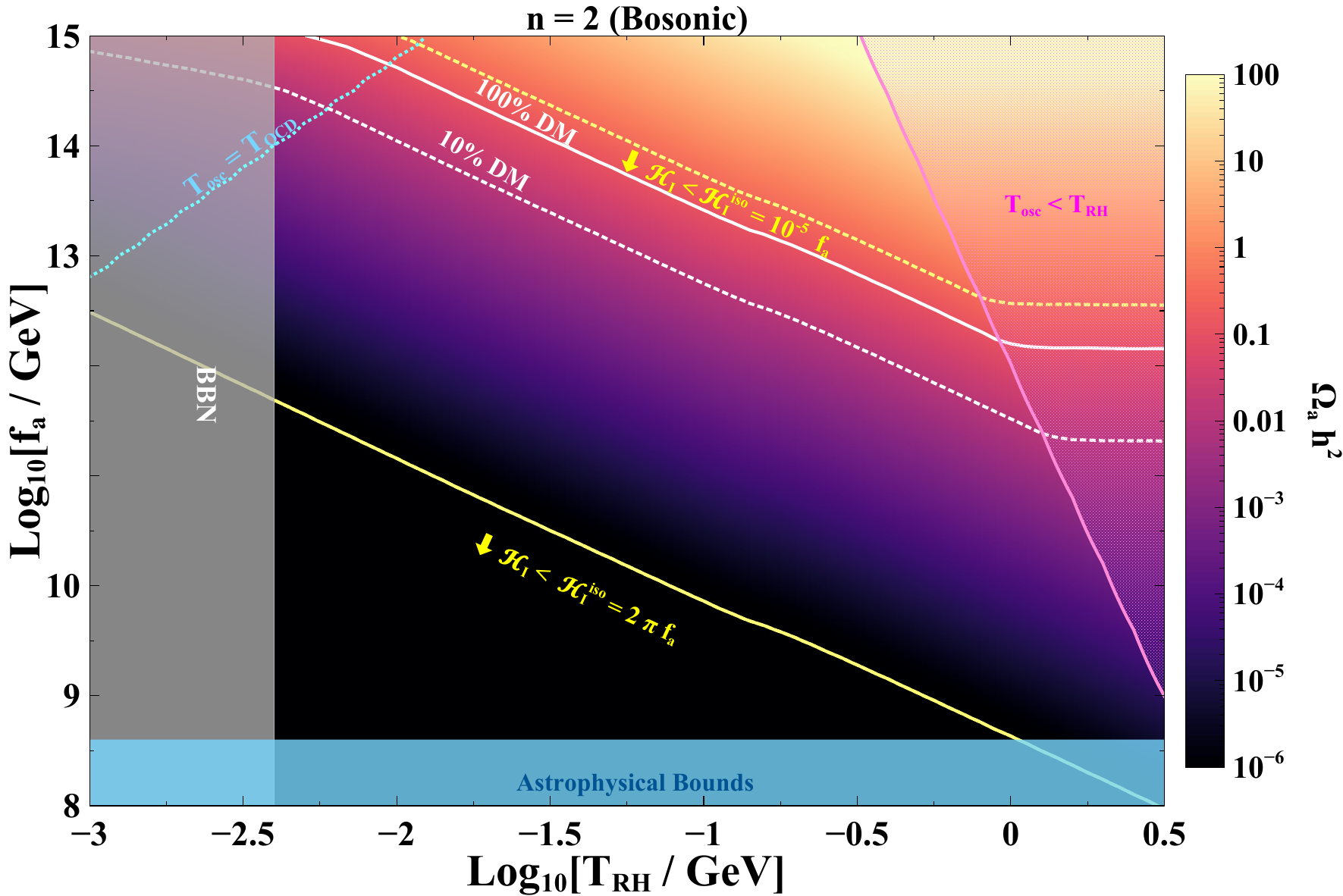}
        \end{subfigure}
        &
        \begin{subfigure}{0.49\linewidth}
            \centering
            \includegraphics[width=\linewidth]{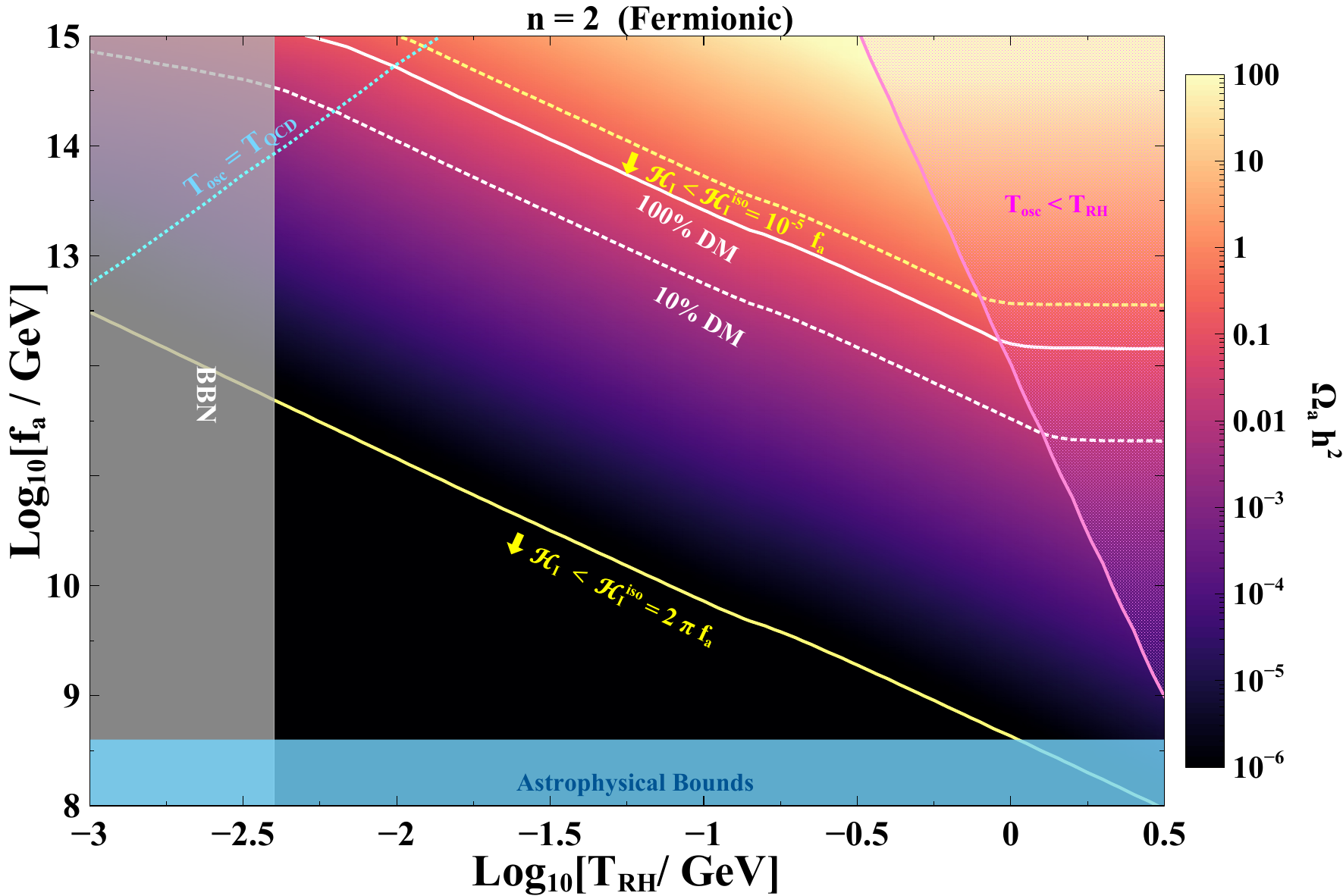}
        \end{subfigure}
        \\[0.5em]
        \begin{subfigure}{0.49\linewidth}
            \centering
            \includegraphics[width=\linewidth]{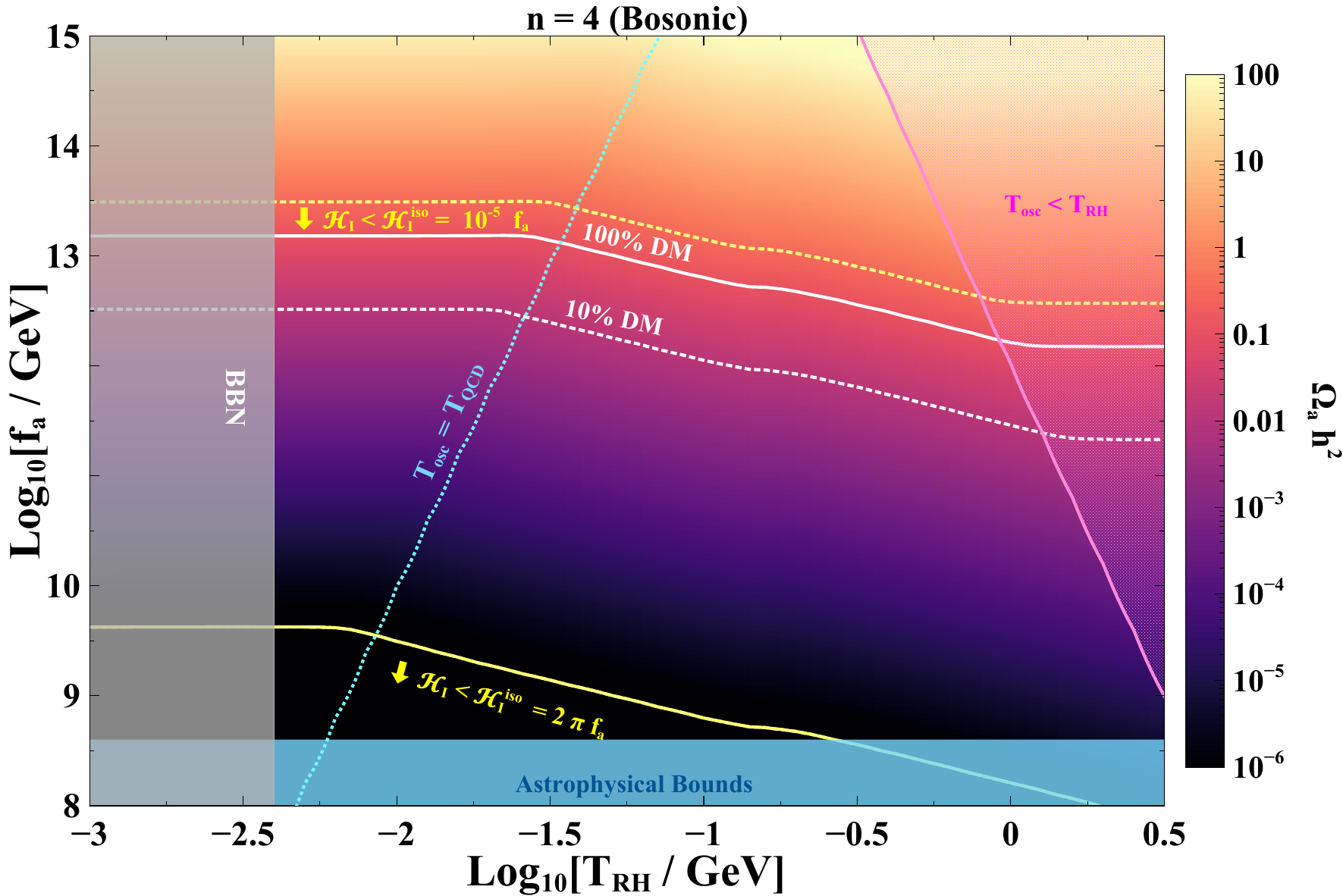}
        \end{subfigure}
        &
        \begin{subfigure}{0.49\linewidth}
            \centering
            \includegraphics[width=\linewidth]{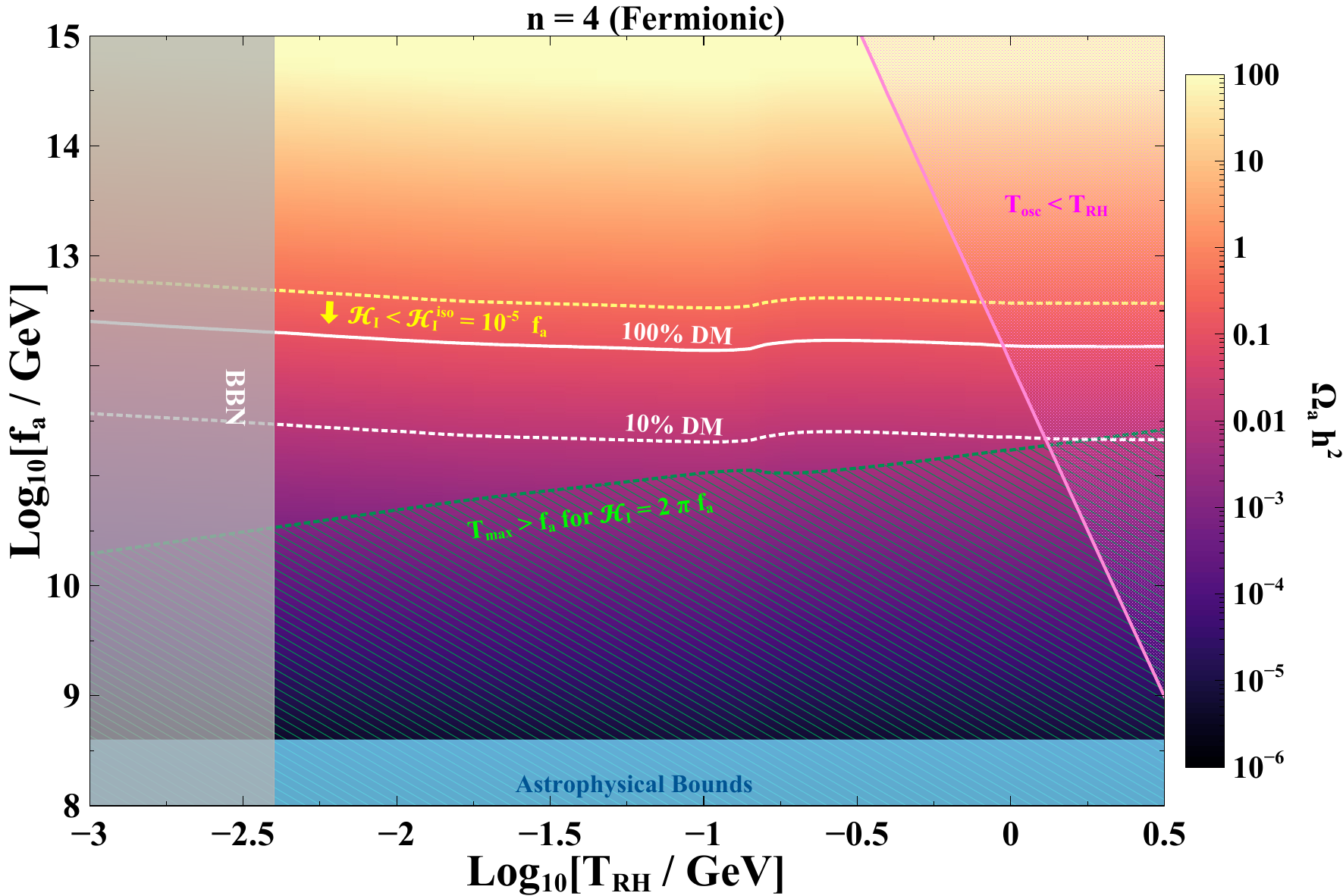}
        \end{subfigure}
        \\[0.5em]
        \begin{subfigure}{0.49\linewidth}
            \centering
            \includegraphics[width=\linewidth]{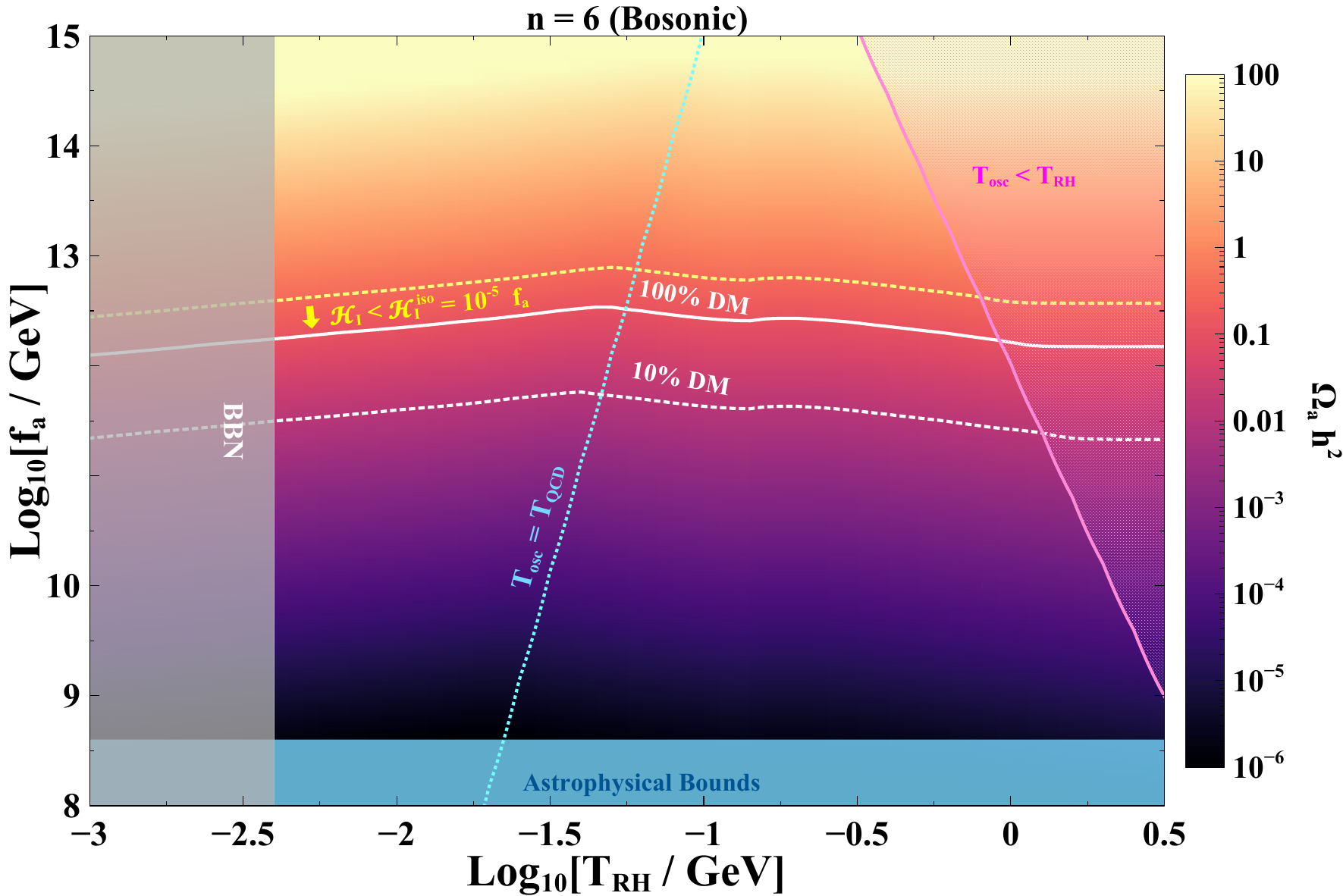}
        \end{subfigure}
        &
        \begin{subfigure}{0.49\linewidth}
            \centering
            \includegraphics[width=\linewidth]{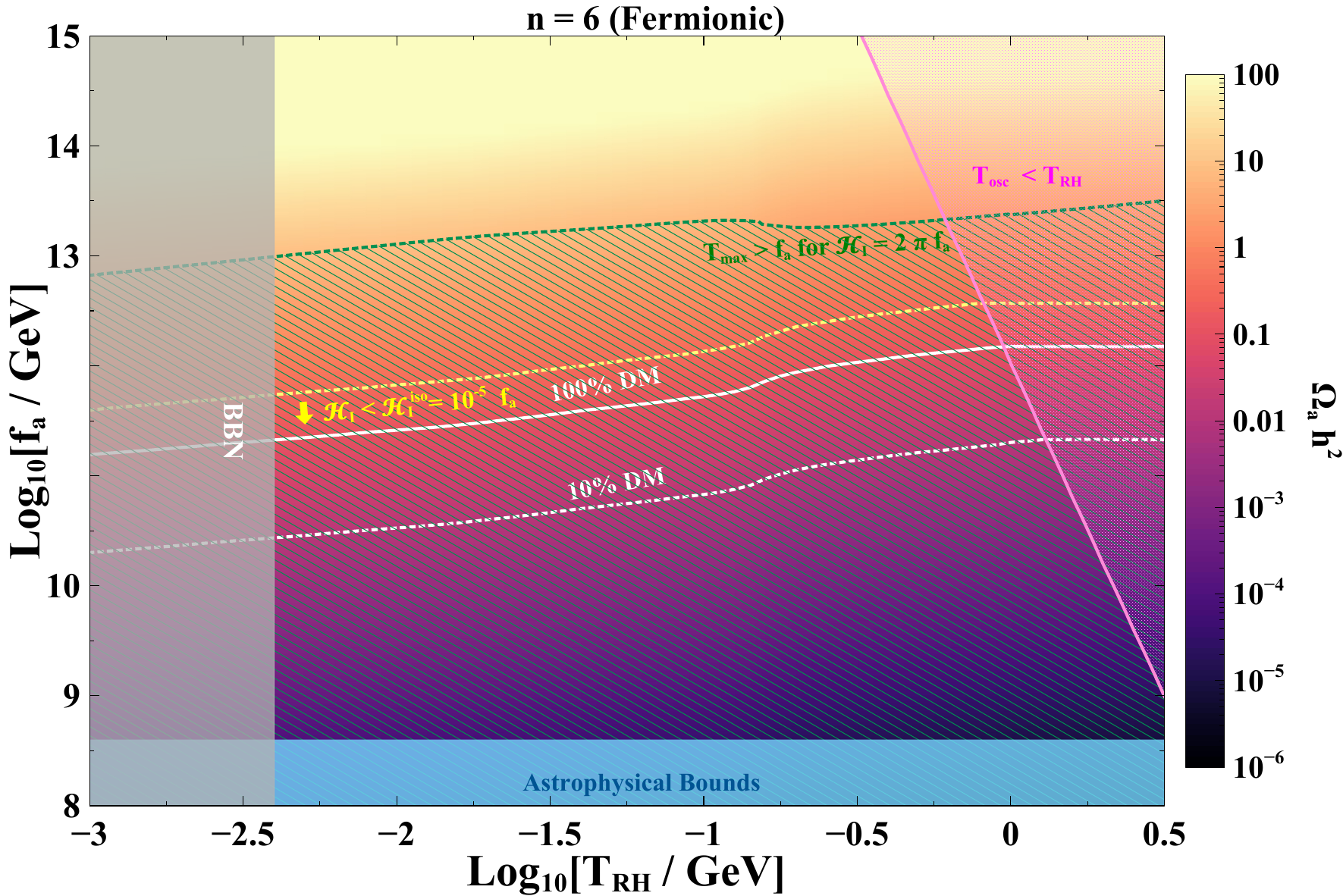}
        \end{subfigure}
    \end{tabular}

    \caption{Direct Reheating (Vacuum Misalignment):  Parameter space in the vacuum misalignment scenario for bosonic (left column) and fermionic (right column) reheating. The top, middle, and bottom rows correspond to $n=2$, $4$, and $6$, respectively. The solid white line in each panel corresponds to $100\%$ axion dark matter for $\theta_i=1$. See text for details related to the other bounds and contours.}
    \label{fig:vacuum_misalignment}
\end{figure}

To obtain the axion relic abundance, we solve the equations in Eqs.~\eqref{energyden}--\eqref{Axion_DM} numerically, using the modified Hubble and temperature evolutions given by Eqs.~\eqref{hub_case1} and \eqref{temp_case1} respectively. We also use the modified expression for $T_{\rm osc}$ obtained above in \eqref{eq:Tosc_above_QCD} - \eqref{eq:Tosc_below_QCD} and take into account the entropy dilution during reheating, as discussed in Eq.~\eqref{eq:rho_a_reheating}. The numerically obtained relic abundance is in close agreement
with the analytical expressions presented in
Appendix~\ref{Sec:Direct Reheating}. The resulting vacuum misalignment parameter space in the $T_{\rm RH}$--$f_a$ plane is shown in Fig.~\ref{fig:vacuum_misalignment} for bosonic (left panel) and fermionic (right panel) reheating. The top, middle, and bottom rows correspond to $n=2$, $4$, and $6$, respectively. The color bar shows the corresponding axion abundance $\Omega_a h^2$ for each point in the parameter space. The solid white contour in each panel corresponds to $100\%$ axion dark matter for $\theta_i=1$, while the dotted white line corresponds to $10\%$ of the observed dark matter abundance.

The cyan dotted line in Fig.~\ref{fig:vacuum_misalignment} denotes the boundary where $T_{\rm osc}=T_{\rm QCD}$. To the left of this line, the axion starts oscillating below the QCD scale, while to the right it starts oscillating above the QCD scale. The pink shaded region on the right corresponds to $T_{\rm osc}<T_{\rm RH}$, where oscillations begin after reheating and the standard radiation-dominated result is recovered. The green hatched region visible in some of the panels corresponds to $T_{\rm max}>f_a$ for $\mathcal{H}_I = 2 \pi f_a$, where thermal restoration of the Peccei--Quinn symmetry can occur. We therefore require $T_{\rm max}<f_a$ to safely avoid thermal restoration in the minimal scenario. The region shown in green is obtained for the maximal value $\mathcal{H}_I=2\pi f_a$, for which $T_{\rm max}$ is the largest and the resulting constraint is the strongest. For a lower value of the inflationary Hubble scale $\mathcal{H}_I$, the bound from $T_{\rm max}<f_a$ becomes weaker, leaving a larger fraction of the parameter space allowed. Since the axion abundance from vacuum misalignment is independent of the benchmark value chosen for $\mathcal{H}_I$, it can be varied while imposing this constraint without affecting the relic abundance contours shown in the figure. We therefore show the $T_{\rm max}<f_a$ constraint for $\mathcal{H}_I=2\pi f_a$ as a representative, conservative bound, while the corresponding isocurvature constraint for different values of $\mathcal{H}_I$ is discussed in Eq.~\eqref{bound} and shown in the form of yellow contours in Fig. \ref{fig:vacuum_misalignment}. The correct relic abundance is compatible with the isocurvature constraint for approximately $\mathcal H_I\lesssim10^{-5}f_a$. Since the $n=2$ case allows larger values of $f_a$ than the standard radiation-dominated scenario, it can accommodate inflationary Hubble scales of
order $10^{10}\,{\rm GeV}$, compared with order $10^7\,{\rm GeV}$ in the standard scenario. The thermal restoration constraint becomes stronger with increasing $n$ for fermionic reheating. In particular, for fermionic reheating with $n=6$, the region satisfying $100\%$ axion dark matter lies within the green hatched region for $\mathcal{H}_I=2\pi f_a$ and is therefore excluded by the thermal restoration constraint. However, for the lower inflationary scales compatible with both the observed axion abundance and the isocurvature constraint, $\mathcal H_I\lesssim10^{-5}f_a$, the maximum temperature is reduced and the thermal restoration constraint is substantially weakened even for fermionic reheating. On the other hand, the bosonic reheating scenario does not suffer from thermal restoration constraints as $T_{\rm max}$ remains much smaller compared to the fermionic counterpart for $n>2$, as noted in Eq. \eqref{eq:tmax}. For $\mathcal H_I=10^{-5}f_a$, the inflationary upper bound in Eq.~\eqref{Hmax} gives
$f_a^{\max}\simeq4.7\times10^{18}\,{\rm GeV}$,
following the prescription discussed above. This lies above the range of $f_a$ displayed, so the corresponding inflationary-scale constraint does not restrict the parameter space shown in Fig. \ref{fig:vacuum_misalignment}.

For $n=2$, the inflaton mass remains constant, and hence its decay rate $\Gamma_\phi$ has no time dependence. As a result, the reheating evolution is the same for bosonic and fermionic decay channels in this case. Consequently, the entropy dilution factor in Eq.~\eqref{eq:rho_a_reheating} and the oscillation temperature in Eq. \eqref{Tosc} take the same form for both bosonic and fermion reheating. This leads to an identical expression for axion relic $\Omega_a h^2$. For higher values of $n$, the time dependence of the inflaton mass leads to different reheating dynamics for the two decay channels, resulting in different parameter space for axions produced from vacuum misalignment mechanism. The resulting axion relic abundance shows the expected dependence on $m_a$, $T_{\rm RH}$, and $\theta_i$, in close agreement with the results of~\cite{Xu:2023lxw} and the analytical expressions presented in
Appendix~\ref{Sec:Direct Reheating}. While Fig. \ref{fig:vacuum_misalignment} is made for $\theta_i = 1$, in the small-angle regime, the abundance approximately scales as $\Omega_a h^2\propto\theta_i^2,$ so the results for other values of $\theta_i$ can be approximately obtained by the corresponding rescaling. Although we focus on the pre-inflationary PQ-breaking scenario, the results can also be applied to the post-inflationary case by using an appropriate effective initial angle, typically $\theta_i\simeq 1.81$, and rescaling the corresponding relic abundance.

\subsubsection{Reheating via RHN Decay: Seesaw Reheating}
\label{subc412}

\begin{figure}
    \centering
    \includegraphics[width=0.49\linewidth]{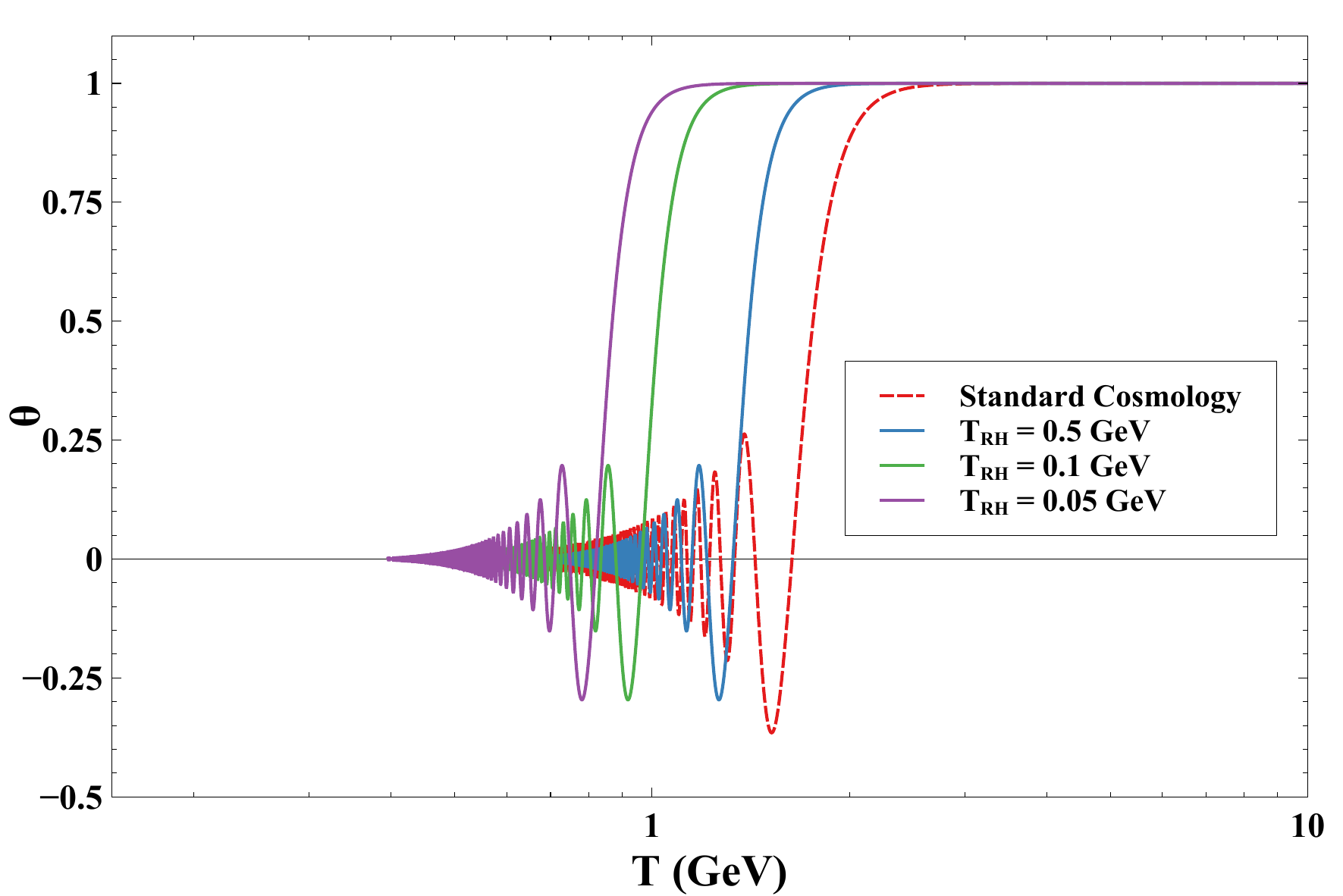}
    \includegraphics[width=0.49\linewidth]{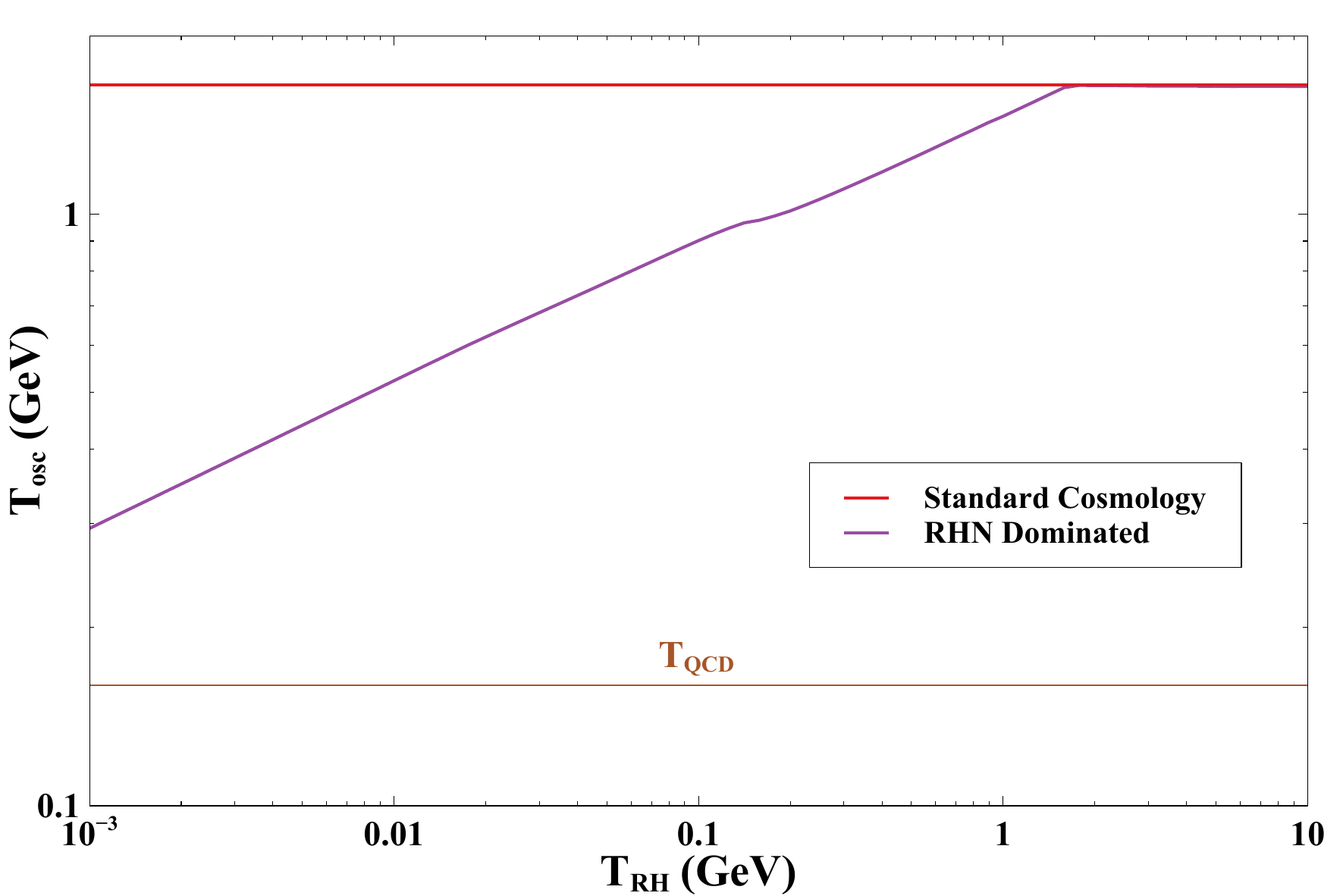}
   
    \caption{Seesaw Reheating (Vacuum Misalignment): Axion evolution (left) and variation of $T_{\rm osc}$ with $T_{\rm RH}$ (right) for n=2 with  $T_{\rm NR}=10^3\, T_{\rm RH}$ and $T_{\rm max}=10^4 \, T_{\rm RH}$. }
    \label{fig:RHN_evo}
\end{figure}

Similar to the previous subsection, we solve the axion evolution equation in Eq.~\eqref{thetaEvo} numerically for different values of $n$ for the indirect reheating scenario via RHN decay, referred to as the seesaw reheating. However, in the present case, the Hubble and temperature evolutions are different and are given by Eqs.~\eqref{eq:Hubble_RHN} and \eqref{eq:Temperature_RHN}, respectively. Unlike the direct reheating scenarios discussed above, the axion evolution is not determined only by $T_{\rm max}$ and $T_{\rm RH}$. Here, the inflaton first decays into right-handed neutrinos, which subsequently evolve from a relativistic to a non-relativistic regime while dominating the energy density of the Universe and eventually decaying into the thermal bath. Therefore, the transition scales $a_{\rm R}$ and $a_{\rm NR}$ (or equivalently $T_{\rm max}$ and $T_{\rm NR}$), together with the end of reheating at $a_{\rm RH}$ (or $T_{\rm RH}$), play an important role in determining the axion evolution.

  The axion evolution is shown in the left panel of Fig.~\ref{fig:RHN_evo}, where we take  $T_{\rm NR}=10^3\, T_{\rm RH}$ and $T_{\rm max}=10^4 \, T_{\rm RH}$. For this parameter choice, the axion starts oscillating during the non-relativistic RHN-dominated phase. The evolution is shown for different values of $T_{\rm RH}$, and we find that a lower $T_{\rm RH}$ leads to a delayed onset of axion oscillations. Since axion oscillations occur during RHN domination, the evolution relevant for vacuum misalignment is insensitive to the preceding reheating history and, in particular, to the value of $n$ in the inflaton potential $V(\phi)\propto\phi^n$. The oscillation temperature is therefore determined by the RHN-dominated evolution and varies with $T_{\rm RH}$, but not with $n$. This is also illustrated on the right panel of Fig.~\ref{fig:RHN_evo}, where the $T_{\rm RH}$--$T_{\rm osc}$ relation is shown for different values of $n$. Since the oscillations occur during RHN domination, the curves for different values of $n$ overlap, giving essentially a single relation between $T_{\rm RH}$ and $T_{\rm osc}$. This is also evident from the analytical expressions for $T_{\rm osc}$ obtained in Eqs.~\eqref{Tosc:seesaw-vac} and \eqref{eq:Tosc_preQCD1} of Appendix \ref{sec:seesaw-vac}. 
  %It is worth noting that this particular scenario of axion misalignment is qualitatively similar to the one in primordial black hole (PBH)-dominated scenarios discussed in \cite{Mazde:2022sdx, Bandyopadhyay:2025ast}. \textcolor{red}{Can the last line be changed to? - 
  The shift in the onset of the oscillation to lower temperatures relative to the ones in standard radiation domination is qualitatively similar to PBH-dominated scenarios discussed in \cite{Mazde:2022sdx, Bandyopadhyay:2025ast}.

\begin{figure}
    \centering
    \includegraphics[width=0.49\linewidth]{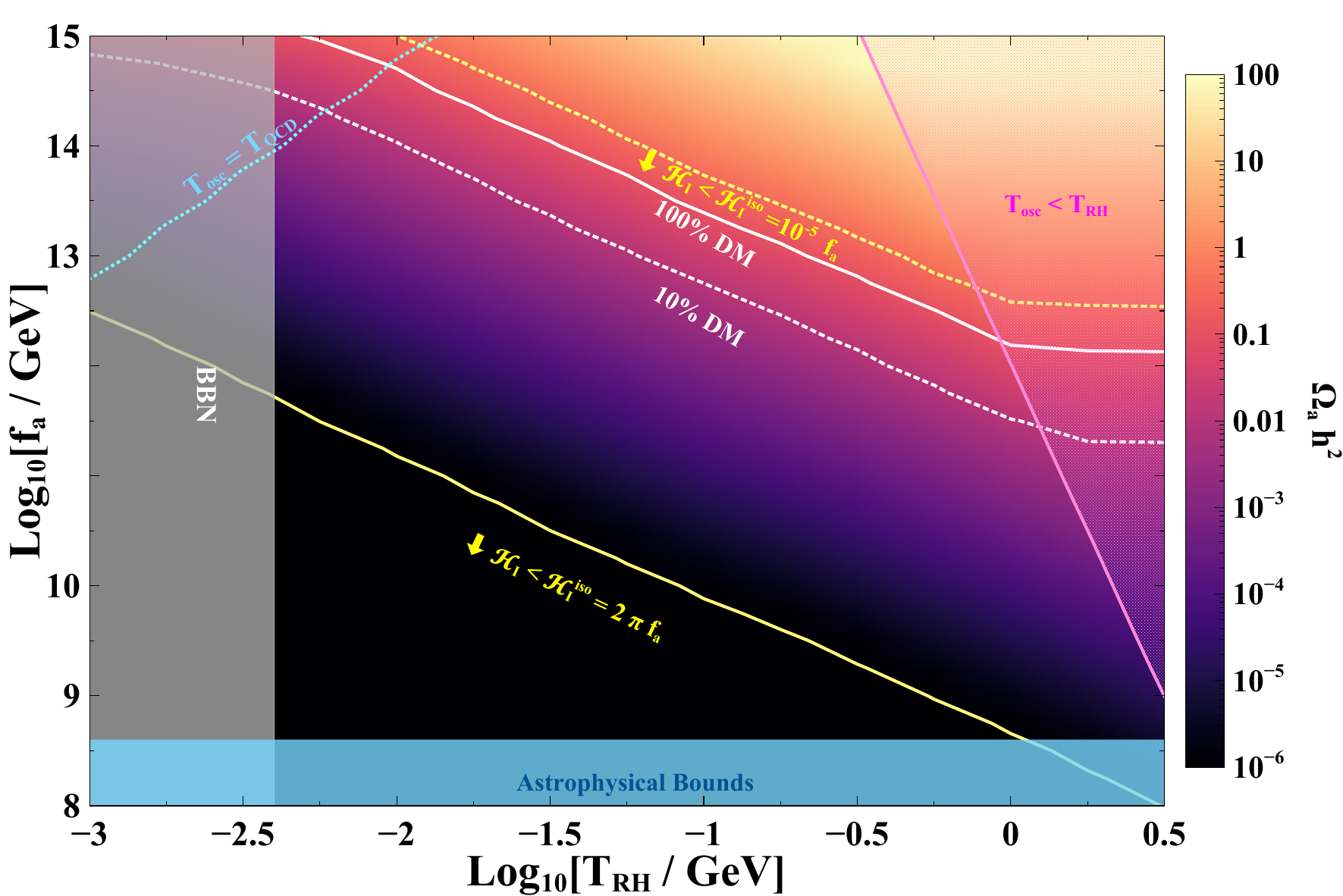}
    \includegraphics[width=0.49\linewidth]{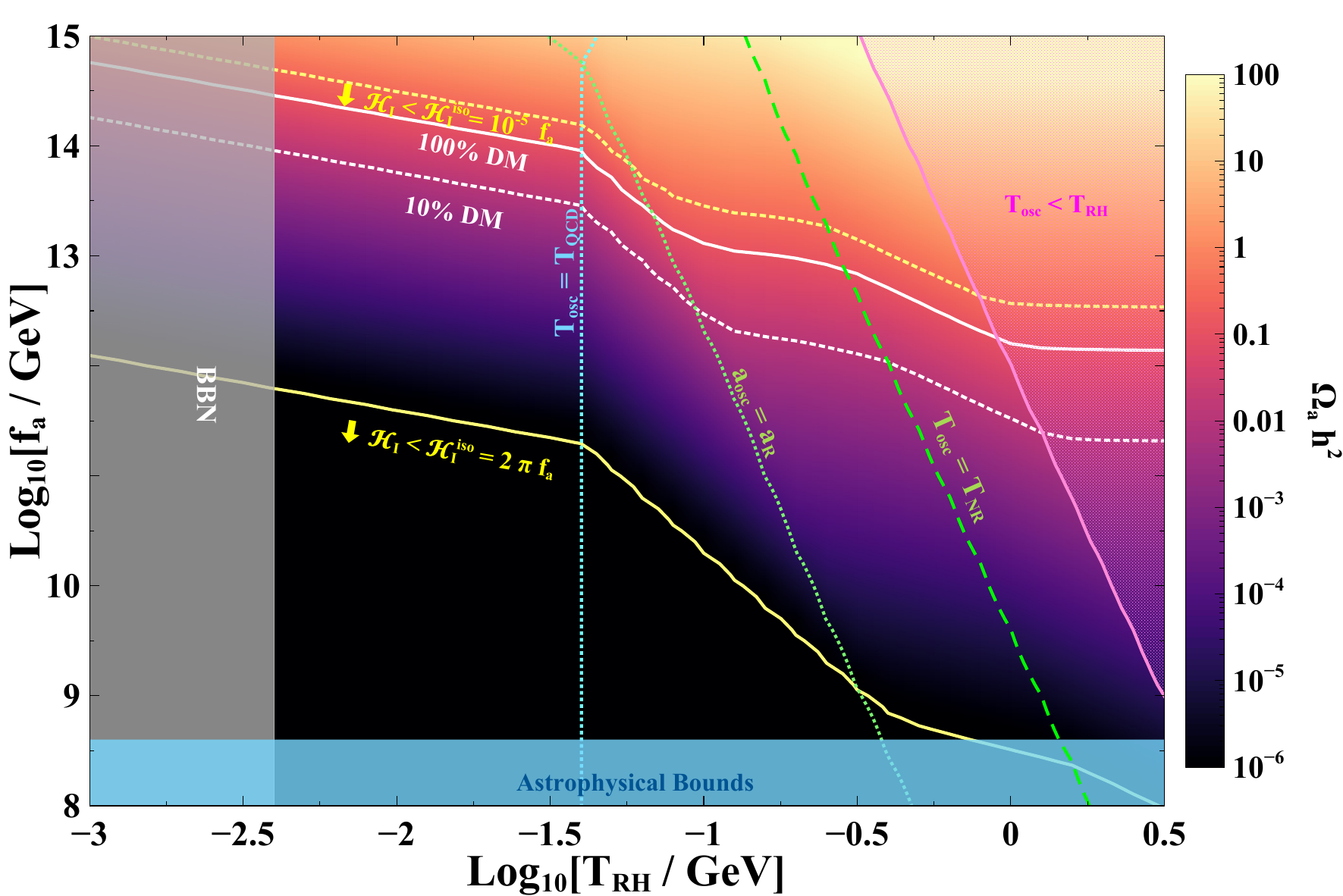}
    \caption{Seesaw Reheating (Vacuum Misalignment): Parameter space in vacuum misalignment case for seesaw reheating with $n=2$ and $T_{\rm NR}=10^3\, T_{\rm RH}$, $T_{\rm max}=10^4 \, T_{\rm RH}$ (left), $T_{\rm NR}=2\,T_{\rm RH}$, $T_{\rm max}=4 \,T_{\rm RH}$ (right).}
    \label{fig:RHN_vac}
\end{figure}

  Following the same procedure, we obtain the axion relic abundance by solving Eqs.~\eqref{energyden}--\eqref{Axion_DM} numerically, using the Hubble and temperature evolutions given by Eqs.~\eqref{eq:Hubble_RHN} and \eqref{eq:Temperature_RHN} respectively,  and accounting for entropy production during
reheating through Eq.~\eqref{eq:rho_a_today_general}. The numerical results are in close agreement with the analytical
expressions for the relic abundance presented in
Appendix~\ref{sec:seesaw-vac}. The resulting parameter space for axion production via vacuum misalignment is shown in the $T_{\rm RH}$--$f_a$ plane in Fig.~\ref{fig:RHN_vac} for two choices of the RHN evolution:  $T_{\rm NR}=10^3\, T_{\rm RH}$ and $T_{\rm max}=10^4 \, T_{\rm RH}$ in the left panel and $T_{\rm NR}=2\,T_{\rm RH}$ and $T_{\rm max}=4 \,T_{\rm RH}$ in the right panel. As in the previous case, the color bar shows the axion DM abundance $\Omega_a h^2$, while the solid and dotted white contours correspond to $100\%$ and $10\%$ of the observed dark matter abundance, respectively, for $\theta_i=1$. The cyan dotted line denotes $T_{\rm osc}=T_{\rm QCD}$, while the pink shaded regions indicate $T_{\rm osc}<T_{\rm RH}$. For the left panel, we take  $T_{\rm NR}=10^3\, T_{\rm RH}$ and $T_{\rm max}=10^4 \, T_{\rm RH}$ for which the RHN-dominated phase extends over a sufficiently long period, and the axion starts oscillating during the non-relativistic RHN-dominated regime. The resulting evolution is therefore similar to the $n=2$ case discussed for the bosonic and fermionic reheating scenarios in Fig.~\ref{fig:vacuum_misalignment}. For the right panel, we take $T_{\rm NR}=2\,T_{\rm RH}$ and $T_{\rm max}=4 \,T_{\rm RH}$, such that the RHN-dominated phase is relatively shorter. The green dotted and dashed contours indicate the boundaries $a_{\rm osc}=a_{\rm R}$ and $T_{\rm osc}=T_{\rm NR}$, respectively. To the left of the dotted line, the axion starts oscillating while the inflaton field still dominates the energy density, considering $n=2$ in the monomial inflaton potential. Between the dotted and dashed green contours, axion oscillations begin during the relativistic RHN-dominated phase, while to the right of the dashed green contour they begin during the non-relativistic RHN-dominated phase. This allows us to see the different scaling of the axion abundance along the $100\%$ and $10\%$ dark matter contours depending on the regime in which the axion starts oscillating. The yellow contours correspond to the isocurvature bounds mentioned earlier.

\subsection{Kinetic Misalignment During Reheating}
As we consider the scenario in which the PQ symmetry is broken before or during inflation and the axion field can acquire an initial angular velocity at the end of inflation, taking us to the regime of the kinetic misalignment mechanism \cite{Co:2019jts, Chang:2019tvx}. Since the entropy production during reheating changes $Y_\theta$, the comoving density corresponding to the Noether charge in Eq. \eqref{eq:Ytheta}, it is more convenient to work with the conserved comoving PQ charge,
\begin{equation}
N_\theta=n_\theta a^3=a^3f_a^2\dot{\theta}=\mathrm{const},
\label{Ntheta}
\end{equation}
where the initial angular velocity is treated as a free parameter without specifying its underlying generation mechanism \footnote{See \cite{Co:2019jts, Chang:2019tvx, Co:2023mhe, Eroncel:2024rpe, Lee:2024bij, Lee:2023dtw} for discussions related to the origin of the initial angular velocity of axions.}.
%In the original kinetic misalignment scenario, the rotation of the PQ field is generated by higher-dimensional PQ-violating operators ~\cite{Co:2019jts} involving quartic and supersymmetric PQ-field potentials. Alternative realizations include an inflaton-induced potential that initiates the axion rotation immediately after inflation~\cite{Co:2023mhe}, explicit PQ breaking in the early Universe~\cite{Chang:2019tvx}, and PQ-violating potentials during inflation that generate the initial velocity through the dynamics of the PQ field~\cite{Lee:2024bij,Lee:2023dtw}. Related axion-rotation scenarios have also been explored in the context of axiogenesis with a heavy QCD axion~\cite{Co:2022aav}.}\textcolor{red}{mention some underlying mechanisms as a footnote?}. 
The axion starts oscillating when its angular velocity becomes comparable to the axion mass, which we take to be
\begin{equation}
\dot{\theta}(T')=2m_a(T'),
\label{eq:Tprime}
\end{equation}
where $T'$ denotes the delayed onset temperature. Using the conservation of $N_\theta$, the angular velocity at any scale factor can be written as
\begin{equation}
\dot{\theta}(a) = \dot{\theta}_i
\left(
\frac{a_I}{a}
\right)^3.
\end{equation}
From this relation, we can determine when the oscillation condition in Eq.~\eqref{eq:Tprime} is satisfied. While the temperature-dependent axion mass also enters the oscillation condition for vacuum misalignment, the axion field remains essentially frozen before the onset of oscillations, so the earlier cosmological evolution does not affect its initial condition at oscillation. In the kinetic misalignment case, however, the angular velocity continuously evolves as $\dot{\theta}\propto a^{-3}$ from the end of inflation. Therefore, the evolution of $\dot{\theta}$ depends on the entire cosmological history, making the background evolution relevant for determining the onset of oscillations and, consequently, the final axion relic abundance, as we will see below.

Since the axion evolution in kinetic misalignment depends on the cosmological history from the end of inflation, we fix the inflationary Hubble scale to $\mathcal{H}_I=0.1\,f_a$, which is consistent with the pre-inflationary Peccei--Quinn symmetry-breaking scenario considered here. In the absence of a specified mechanism for generating the initial axion charge, we impose a conservative requirement that the energy stored in the axion field at the end of inflation does not exceed the total energy density of the Universe,
\begin{equation}
\frac12 f_a^2\dot{\theta}_i^2
<
3M_{\rm P}^2 \mathcal{H}_I^2,
\label{const_con}
\end{equation}
where $\mathcal{H}_I$ denotes the Hubble parameter at the end of inflation. This ensures that the post-inflationary Universe remains dominated by the inflaton condensate until reheating and that the assumed cosmological background is self-consistent. For the benchmark choice $\mathcal{H}_I=0.1 f_a$ adopted throughout this work for kinetic misalignment mechanism, the above consistency condition translates into
\begin{equation}
\dot{\theta}_i
<
\sqrt{6}\,\frac{M_{\rm P}}{f_a} \mathcal{H}_I
\simeq 6\times10^{17}~{\rm GeV}.
\end{equation}
We then proceed to define the critical initial angular velocity as
\begin{equation}
\dot{\theta}_{\rm crit}=2m_a(T_{\rm osc}),
\label{eq:thetadot_crit}
\end{equation}
where $T_{\rm osc}$ is the oscillation temperature obtained for vacuum misalignment. This corresponds to the minimum initial angular velocity required for delayed oscillations and hence for kinetic misalignment to be operational. For $\dot{\theta}_i<\dot{\theta}_{\rm crit}$, the axion does not undergo delayed oscillations and instead follows the usual evolution via vacuum misalignment. Therefore, to have a consistent kinetic misalignment scenario, the initial angular velocity must satisfy
\begin{equation}
\dot{\theta}_{\rm crit}<\dot{\theta}_i<6\times10^{17}~{\rm GeV},
\label{crit}
\end{equation}
where the upper bound follows from the energy-consistency condition in Eq.~\eqref{const_con}. In more general scenarios, like \cite{Lee:2024bij,Lee:2023dtw,Co:2023mhe,Bhandari:2025pzg} for example, where the PQ charge is generated dynamically at a later epoch through interactions with another sector or by an explicit source term, the above bound need not apply directly.
%\textcolor{blue}{It is worth noting that this consistency condition is specific to our assumption of a fixed $f_a$ and an initial axion velocity present at the end of inflation. In scenarios with a dynamical PQ field, the axion velocity can instead be determined by the underlying inflationary dynamics~\cite{Lee:2024bij,Lee:2023dtw,Co:2023mhe}. Alternatively, axion motion can be generated at a much later cosmological epoch, in which case the above end-of-inflation bound need not apply directly~\cite{Bhandari:2025pzg}. Since such model-dependent realizations lie beyond the scope of the present work, we treat the initial angular velocity as an effective phenomenological parameter satisfying the consistency condition in Eq.~\eqref{const_con}}. 
Also, in the kinetic misalignment mechanism, the details related to the isocurvature bound is different than that of vacuum misalignment given in Eq. \eqref{bound}. Unlike the vacuum misalignment mechanism, the axion relic abundance in this case is independent of the initial field displacement $\theta_i$ and is instead determined by the initial velocity $\dot{\theta}_i$. Consequently, the axion field fluctuations generated during inflation do not induce an isocurvature contribution to the axion abundance through $\theta_i$. In principle, however, if the initial axion velocity $\dot{\theta}_i$ is generated dynamically through the dynamics of additional light fields, fluctuations in such fields could induce fluctuations in $\dot{\theta}_i$ and hence contribute to isocurvature perturbations. Since we treat $\dot{\theta}_i$ as a model-independent initial condition and do not specify the underlying mechanism responsible for its generation, we do not consider such contributions. Thus, no isocurvature constraint is imposed on $\mathcal{H}_I$ for the analysis related to the kinetic misalignment mechanism.

We now investigate how the kinetic misalignment dynamics and the resulting axion relic abundance are affected by different reheating scenarios considered in this work, including direct inflaton decays into bosonic and fermionic final states, as well as indirect reheating through right-handed neutrinos.

%\textcolor{Red}{Thet dot for standard cosmo estimate as per our assumptions}

\subsubsection{Direct Reheating}

\begin{figure}[htbp]
    \centering
    \captionsetup[subfigure]{labelformat=simple}
    \begin{subfigure}[b]{0.49\linewidth}
        \centering
        \includegraphics[width=\linewidth]{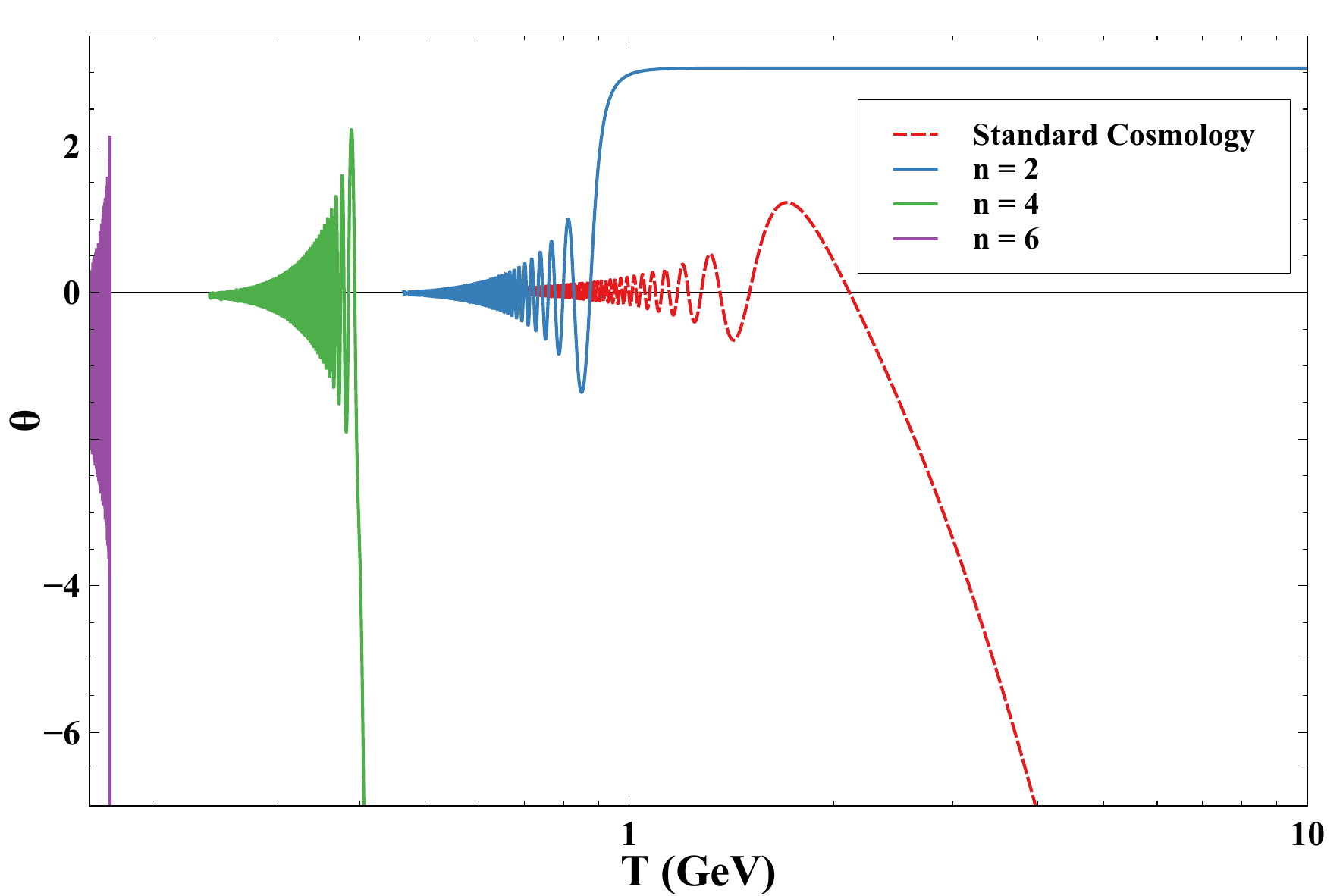}
        \caption{Bosonic reheating (kinetic)}
        \label{fig:axion_evo_bd_kin}
    \end{subfigure}\hfill
    \begin{subfigure}[b]{0.49\linewidth}
        \centering
        \includegraphics[width=\linewidth]{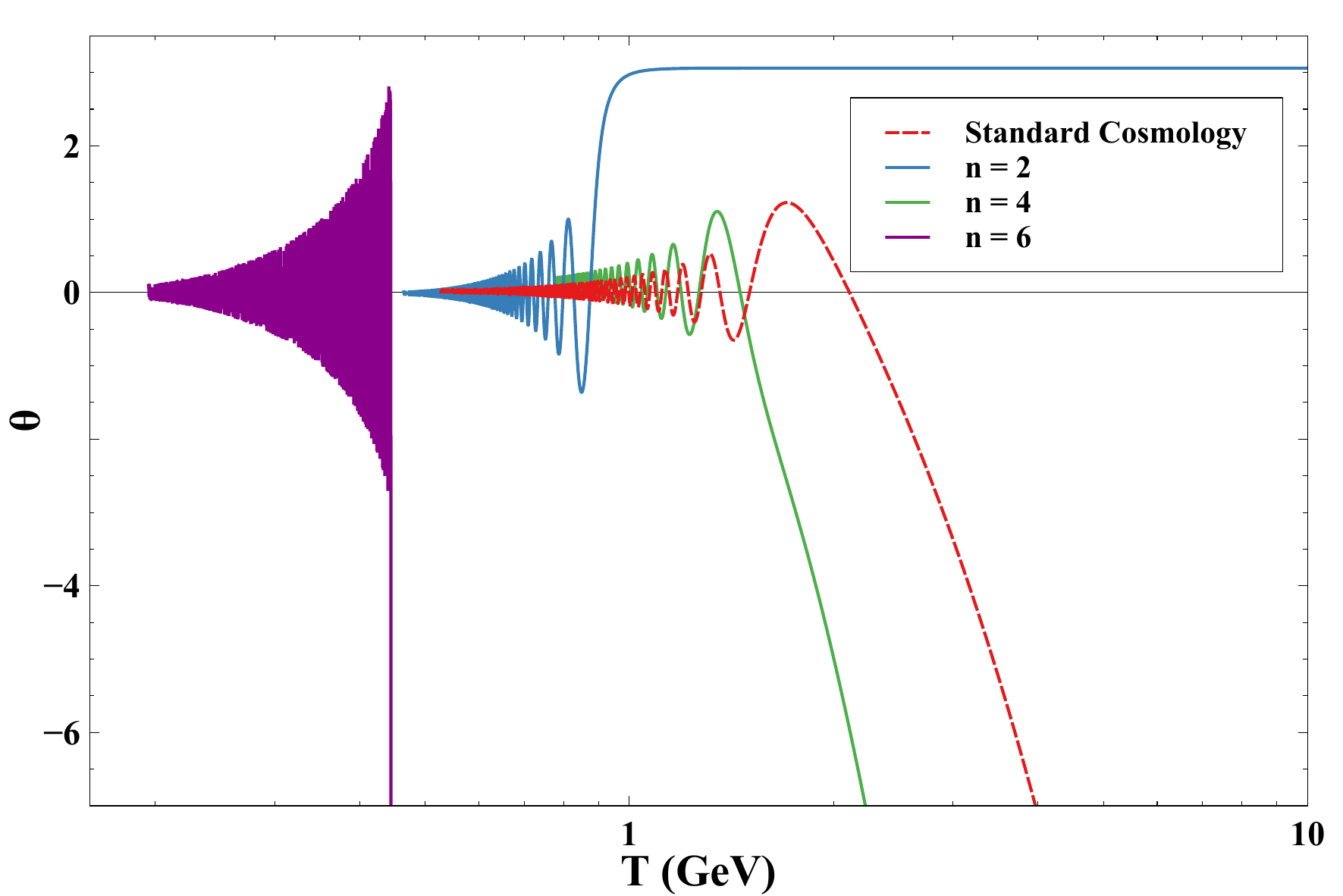}
        \caption{Fermionic reheating (kinetic)}
        \label{fig:axion_evo_fd_kin}
    \end{subfigure}
\caption{Direct Reheating (Kinetic Misalignment): Axion evolution for $f_a = 10^{10}\text{ GeV}$, $T_{\text{RH}} = 0.1\text{ GeV}$, $\mathcal{H}_I = 0.1\,f_a$ and $\dot{\theta_i} = 10^{23}\text{ GeV}$.}
    \label{fig:axion_evolution_grid_2}
\end{figure}

\begin{figure}
    \centering
    \captionsetup[subfigure]{labelformat=simple}
    \begin{subfigure}[b]{0.49\linewidth}
        \centering
        \includegraphics[width=\linewidth]{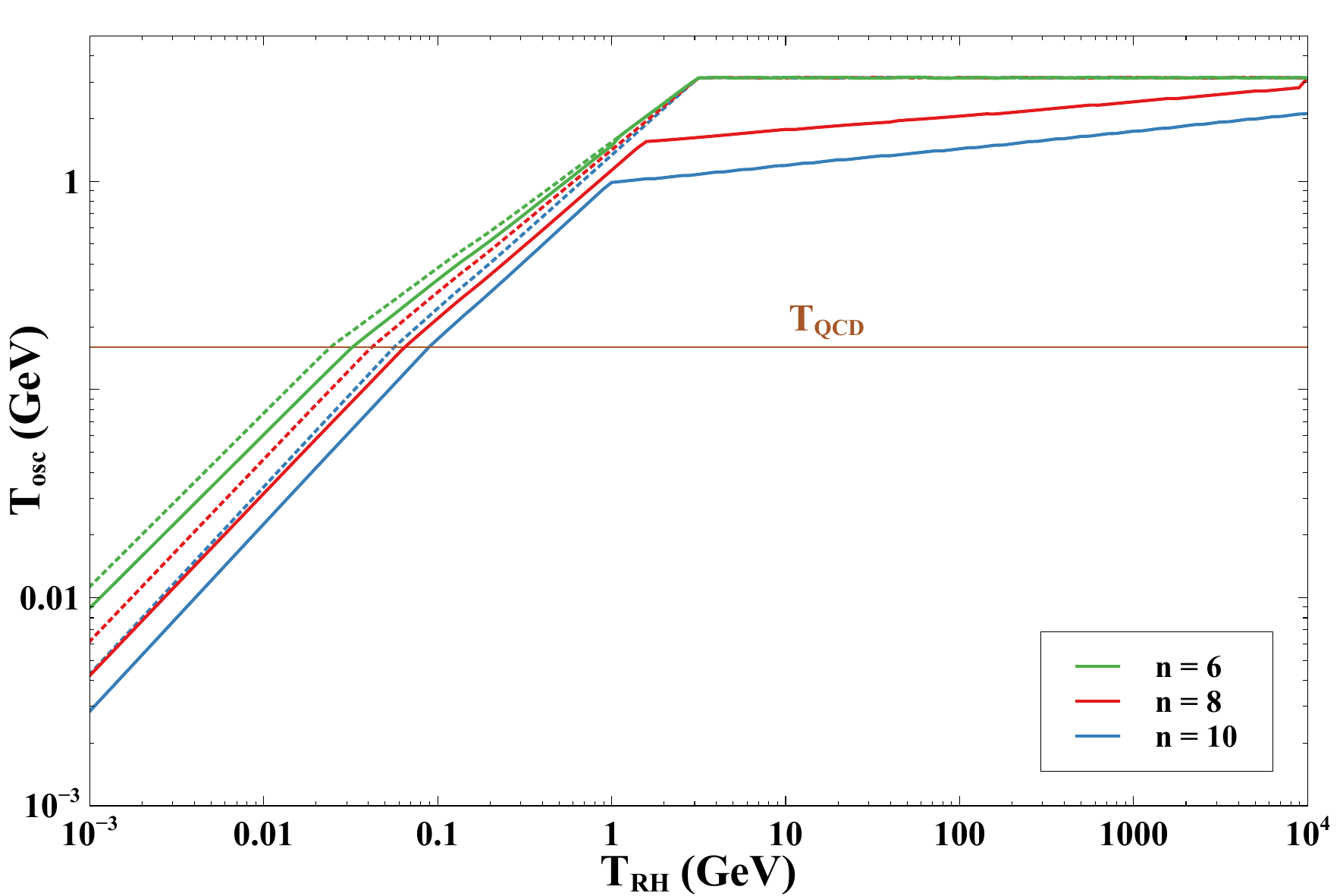}
        \caption{Bosonic reheating (kinetic)}
        \label{fig:axion_evo_bd_kin}
    \end{subfigure}\hfill
    \begin{subfigure}[b]{0.49\linewidth}
        \centering
        \includegraphics[width=\linewidth]{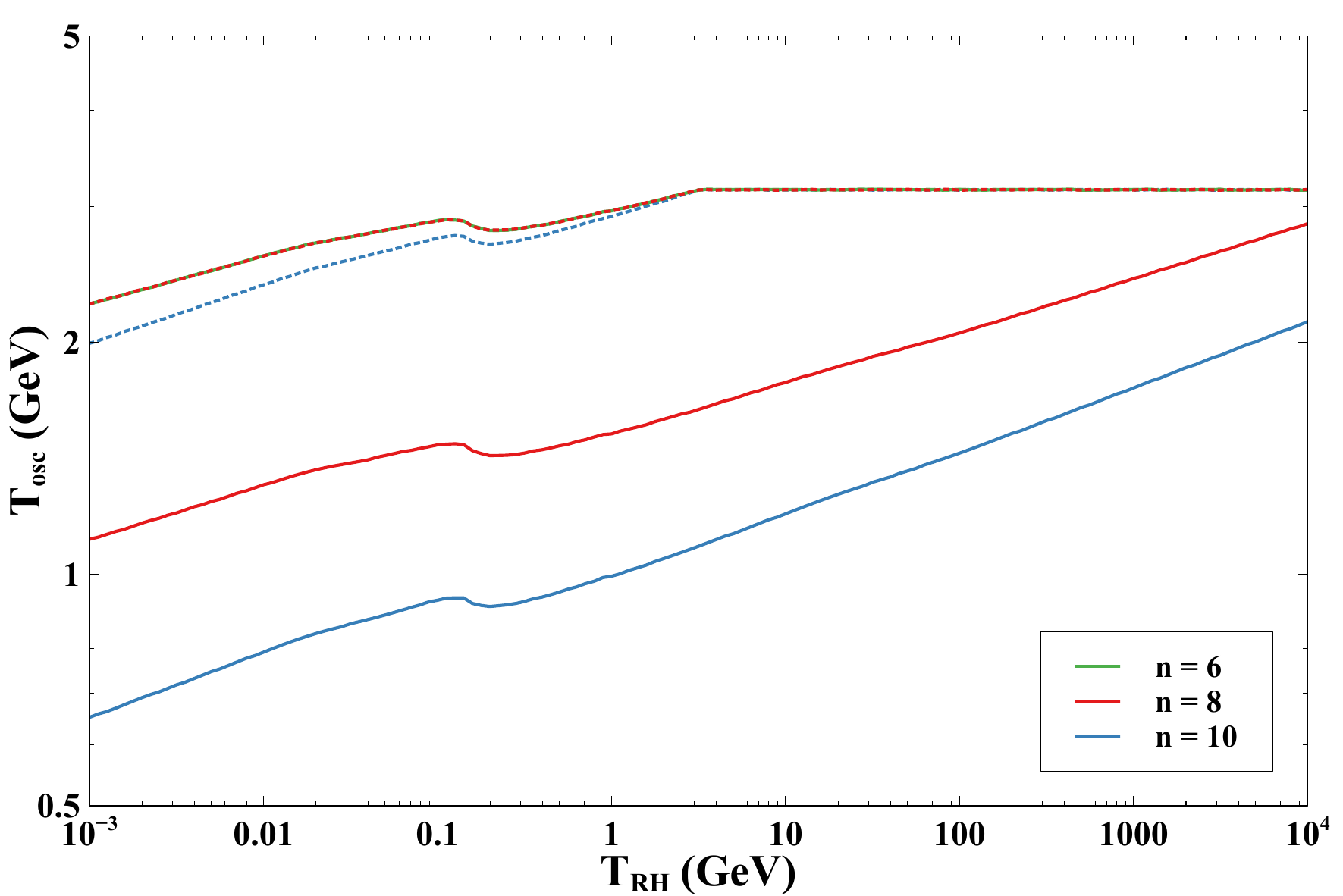}
        \caption{Fermionic reheating (kinetic)}
        \label{fig:axion_evo_fd_kin}
    \end{subfigure}
\caption{Direct Reheating (Kinetic Misalignment): Variaton of $T_{\rm osc}$ with $T_{\rm RH}$ for bosonic (left) and fermionic (right) reheating with different $n$ and $\mathcal{H}_{I} = 0.1  f_{a}$ and $f_a = 10^9$ GeV. The solid lines denote $\dot{\theta_i}= 5 \times 10^{16}$ GeV corresponding to kinetic misalignment compared to dotted lines denoting the vacuum misalignment. }
    \label{fig:Tosc_Kin}
\end{figure}

%\begin{figure}
%    \centering
%    \includegraphics[width=0.49\linewidth]{Figure/Tosc_kin.pdf}
%    \includegraphics[width=0.49\linewidth]{Figure/Tosc_kin_2.pdf}
%    \caption{Direct Reheating (Kinetic Misalignment): Variaton of $T_{\rm osc}$ with $T_{\rm RH}$ for bosonic (left) and fermionic (right) reheating with different $n$ and $\mathcal{H}_{I} = 0.1  f_{a}$ and $f_a = 10^9$ GeV. The solid lines denote $\dot{\theta_i}= 5 \times 10^{16}$ GeV corresponding to kinetic misalignment compared to dotted lines denoting the vacuum misalignment.}
%    \label{fig:Tosc_Kin}
%\end{figure}

Similar to the vacuum misalignment scenario, we solve the axion evolution equation in Eq.~\eqref{thetaEvo} numerically for different values of $n$. Considering both bosonic and fermionic reheating, we use the corresponding Hubble evolution given in Eq.~\eqref{hub_case1} and temperature evolution in Eq.~\eqref{temp_case1}. For the kinetic misalignment case, the resulting axion evolution is shown in Fig.~\ref{fig:axion_evolution_grid_2} for $f_a=10^{10}~{\rm GeV}$, $T_{\rm RH}=0.1~{\rm GeV}$, $\mathcal{H}_I=0.1 f_a$ and $\dot{\theta}_i=10^{23}~{\rm GeV}$.

To further illustrate the difference in the onset of oscillations, we show the corresponding $T_{\rm osc}$--$T_{\rm RH}$ relation in Fig. \ref{fig:Tosc_Kin} for $f_a=10^9~{\rm GeV}$ and $\mathcal{H}_I=0.1 f_a$. The solid lines correspond to the kinetic misalignment case with $\dot{\theta}_i=5\times10^{16}~{\rm GeV}$, while the dotted lines show the corresponding vacuum misalignment result. We find that the $T_{\rm osc}$ dependence on $T_{\rm RH}$ is different in bosonic and fermionic reheating, reflecting the different cosmological evolution of the Hubble parameter and temperature in the two reheating scenarios. In the kinetic misalignment case, this difference in the background evolution directly affects the evolution of the axion angular velocity and hence the temperature at which the delayed oscillation condition is satisfied. The analytical expressions for the oscillation temperature above and below the QCD scale are given in Eqs.~\eqref{eq:Tosc_KM_direct_above}
and \eqref{eq:Tosc_KM_direct_below}, respectively,
in Appendix~\ref{sec:direct-kin}.

\begin{figure}
    \centering
    \captionsetup[subfigure]{labelformat=simple}
    \begin{subfigure}[b]{0.49\linewidth}
        \centering
        \includegraphics[width=\linewidth]{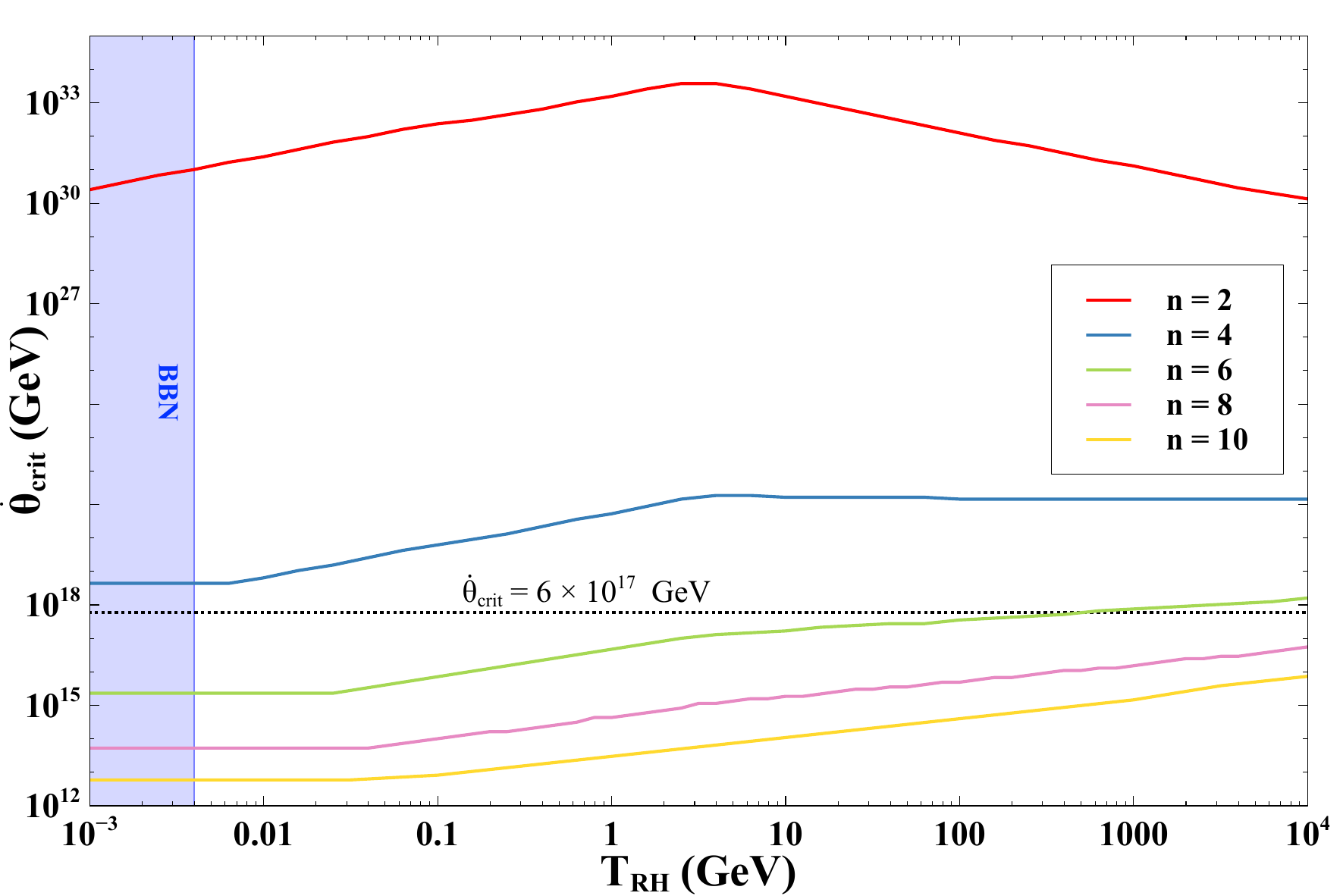}
        \caption{Bosonic reheating (kinetic)}
        \label{fig:axion_evo_bd_kin}
    \end{subfigure}\hfill
    \begin{subfigure}[b]{0.49\linewidth}
        \centering
        \includegraphics[width=\linewidth]{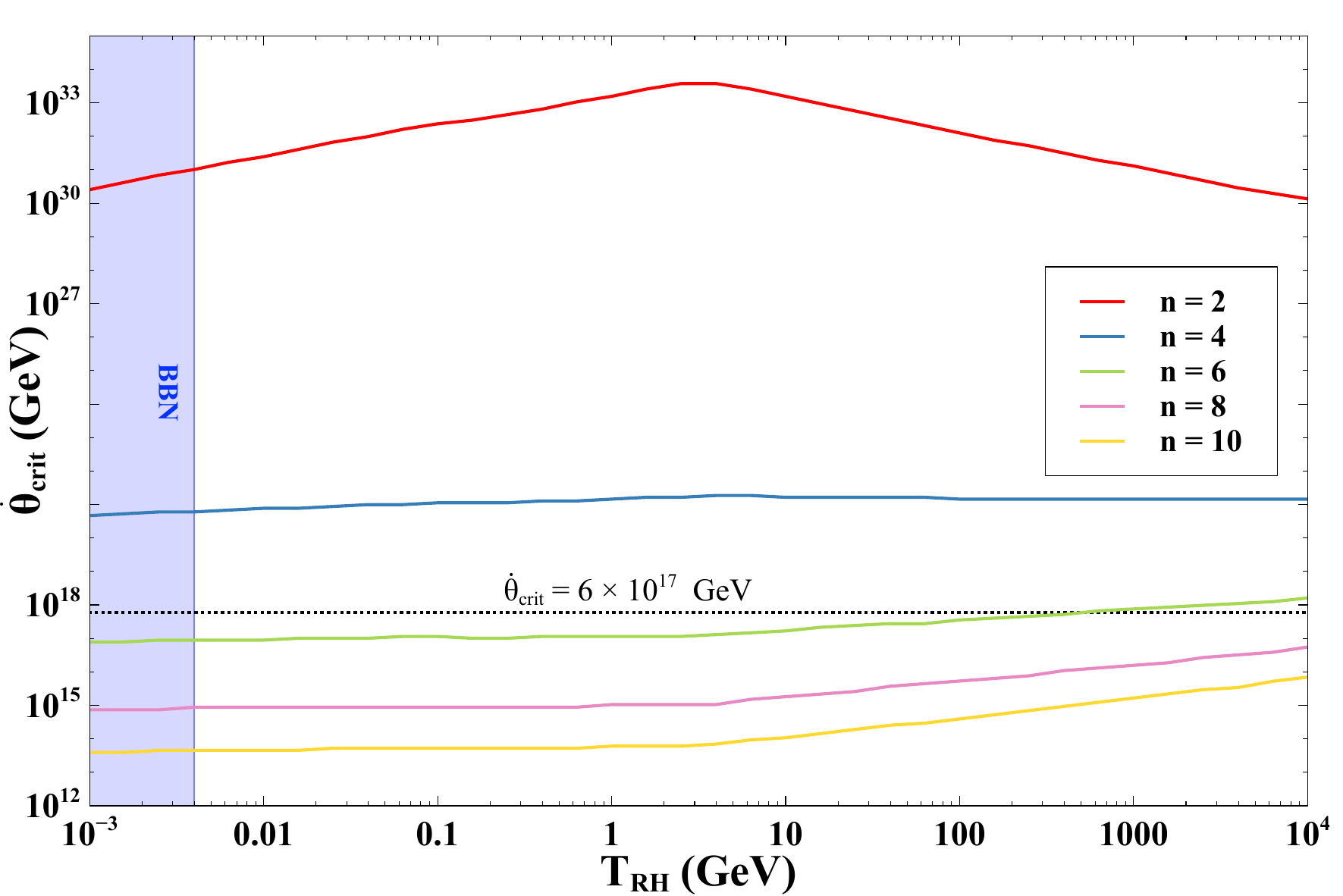}
        \caption{Fermionic reheating (kinetic)}
        \label{fig:axion_evo_fd_kin}
    \end{subfigure}
\caption{Direct Reheating (Kinetic Misalignment): Variation of $\dot{\theta}_{\rm crit}$ with $T_{\rm RH}$ for $f_a = 10^{9}$ GeV considering bosonic (left) and fermionic (right) reheating.}
    \label{fig:crit_decay}
\end{figure}

%\begin{figure}
%    \centering
%    \includegraphics[width=0.49\linewidth]{Figure/theta_crit_B.pdf}
%    \includegraphics[width=0.49\linewidth]{Figure/theta_crit_F.pdf}
    % \includegraphics[width=0.48\linewidth]{Figure/theta_crit2.pdf}
%    \caption{Direct Reheating (Kinetic Misalignment): Variation of $\dot{\theta}_{\rm crit}$ with $T_{\rm RH}$ for $f_a = 10^{9}$ GeV considering bosonic (left) and fermionic (right) reheating.}
%    \label{fig:crit_decay}
%\end{figure}

Now we proceed to examine whether the condition in Eq.~\eqref{eq:thetadot_crit} can be satisfied consistently with the energy bound in Eq.~\eqref{const_con}. We show the critical angular velocity $\dot{\theta}_{\rm crit}$ as a function of $T_{\rm RH}$ in Fig.~\ref{fig:crit_decay}, for $f_a=10^{9}~{\rm GeV}$. The left and right panels correspond to bosonic and fermionic reheating, respectively. We impose the lower bound on $T_{\rm RH}$ from BBN and the lower bound on $f_a$ from astrophysical observations. From the figure, we find that for $n=2$ and $n=4$, $\dot{\theta}_{\rm crit}$ exceeds the upper bound on the initial angular velocity obtained from Eq.~\eqref{const_con} throughout the entire parameter space. Therefore, these cases cannot realize kinetic misalignment consistently within our assumptions. For $n\geq6$, the critical angular velocity can satisfy the consistency bound, and we therefore proceed to determine the corresponding axion relic abundance numerically for benchmark value $\dot{\theta}_i = 5\times10^{16}~{\rm GeV},$ which satisfies the above consistency requirement.

\begin{figure}[t]
    \centering
    % n = 8
    \includegraphics[width=0.49\linewidth]{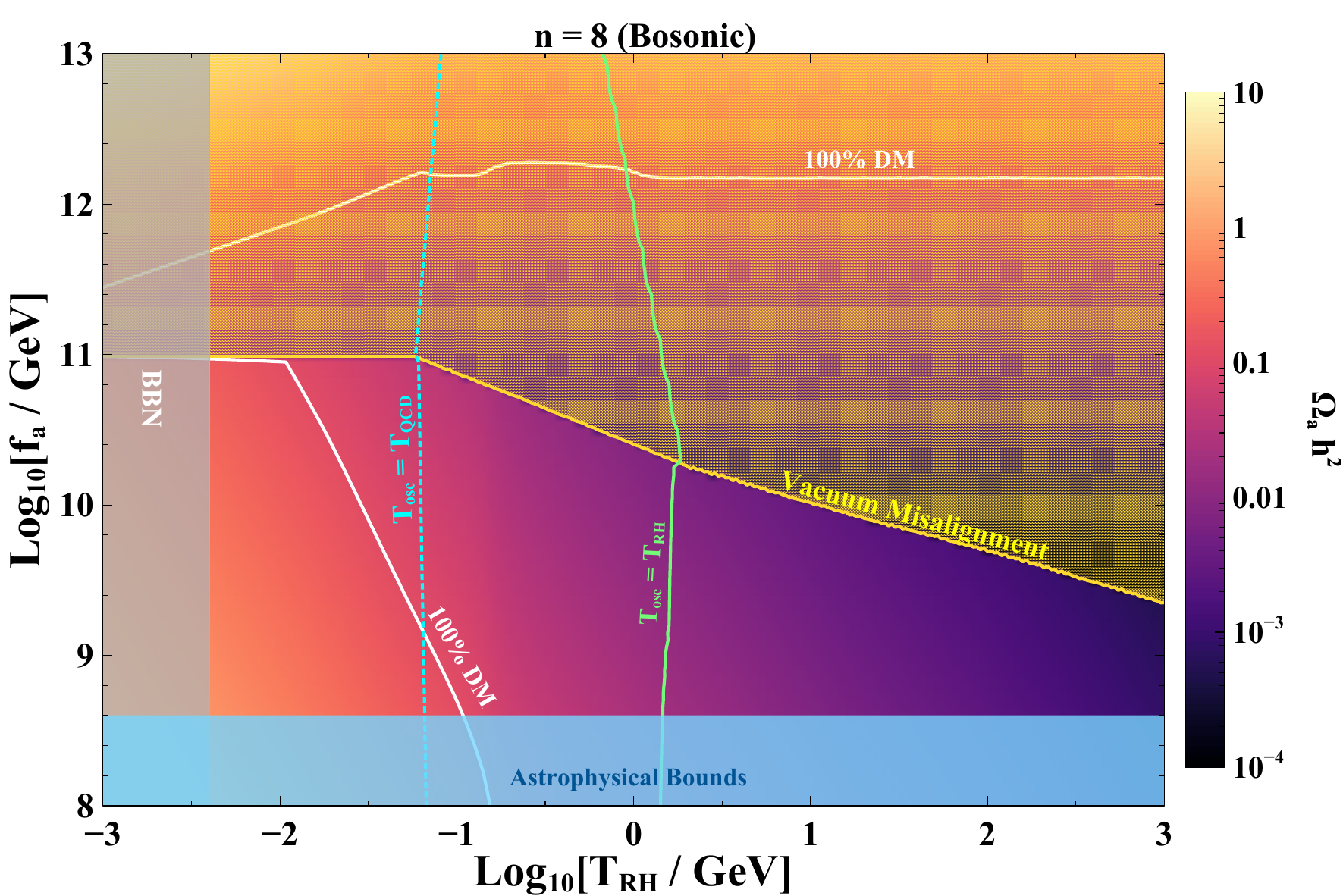}
    \includegraphics[width=0.49\linewidth]{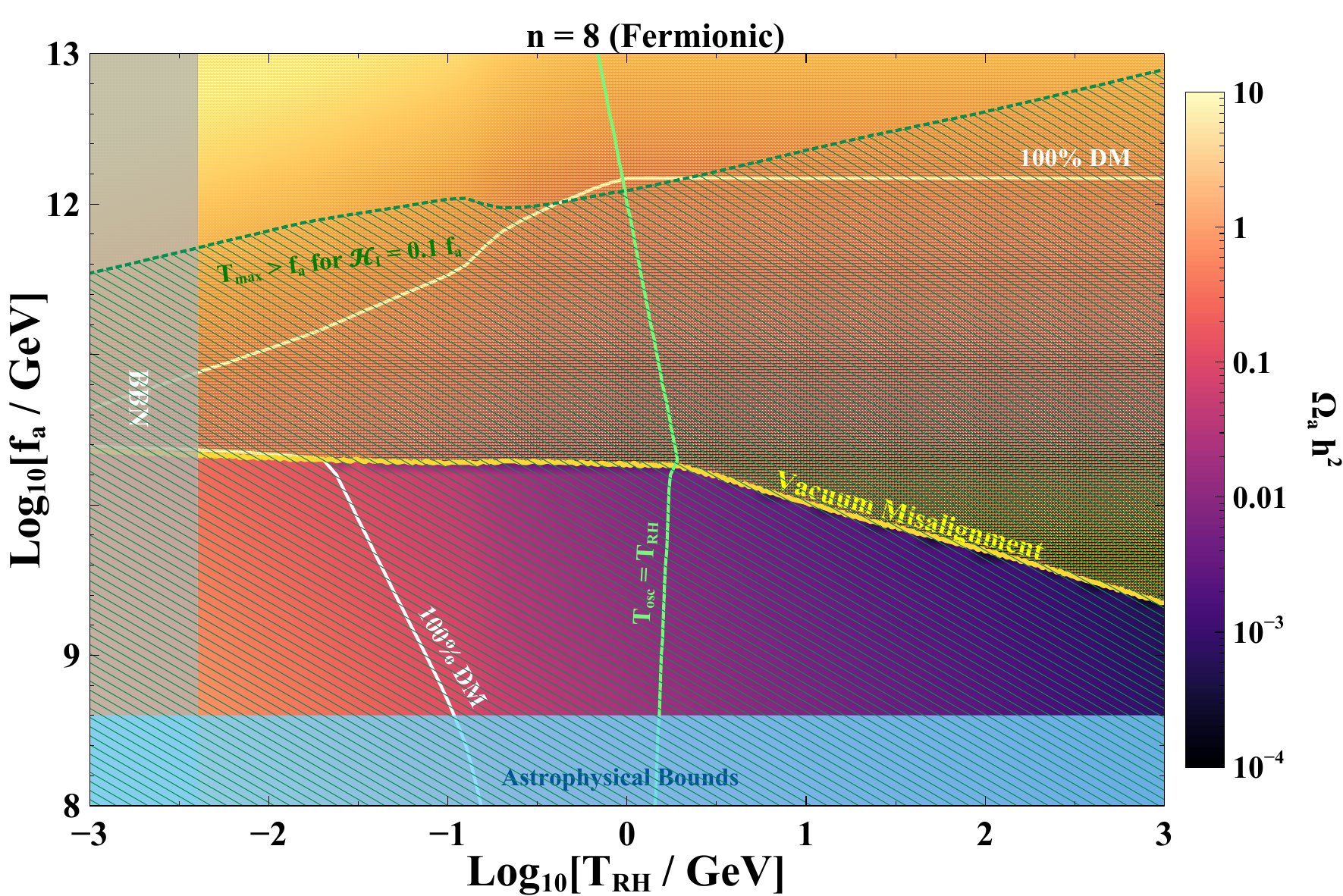}\\
    % n = 10
    \includegraphics[width=0.49\linewidth]{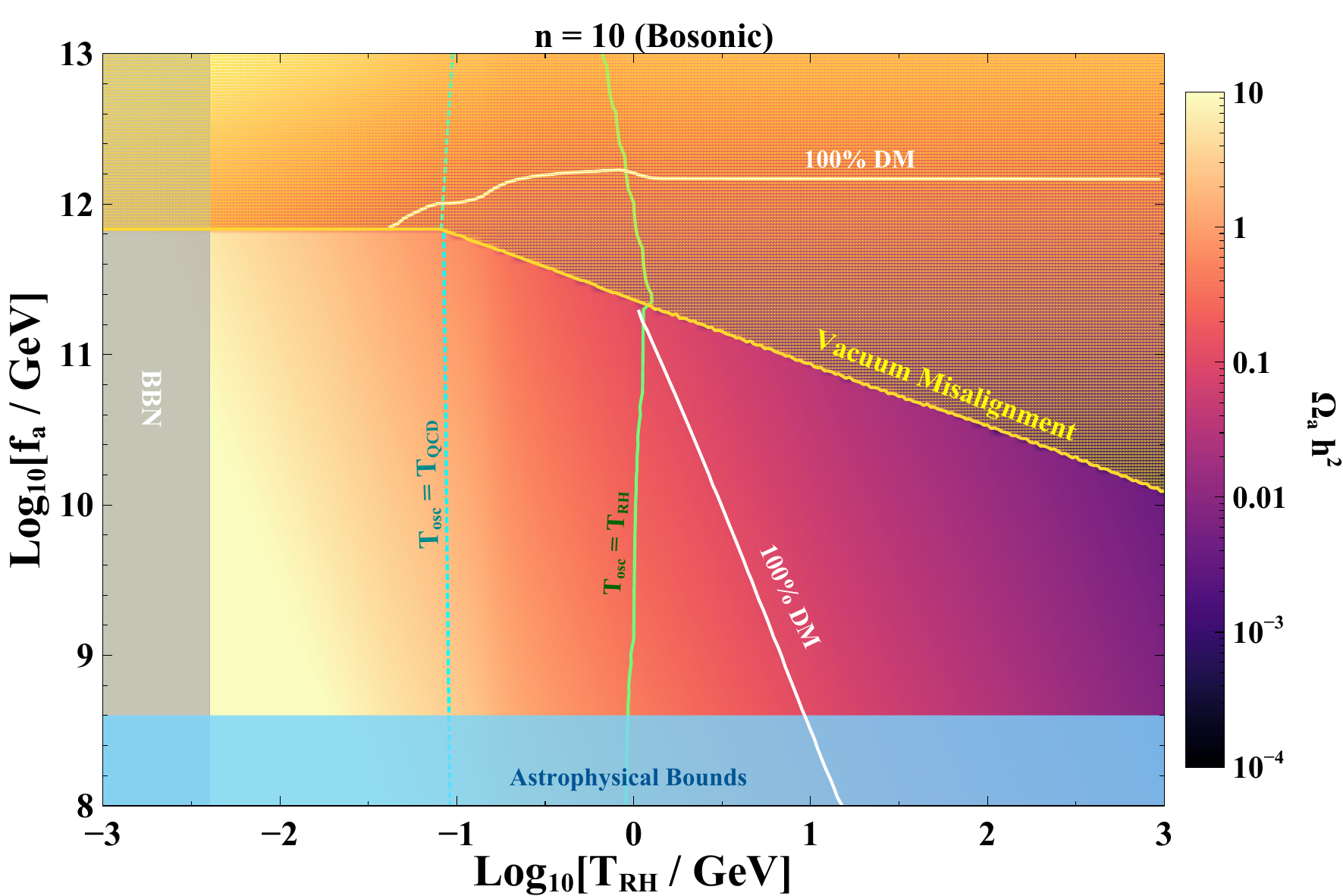}
    \includegraphics[width=0.49\linewidth]{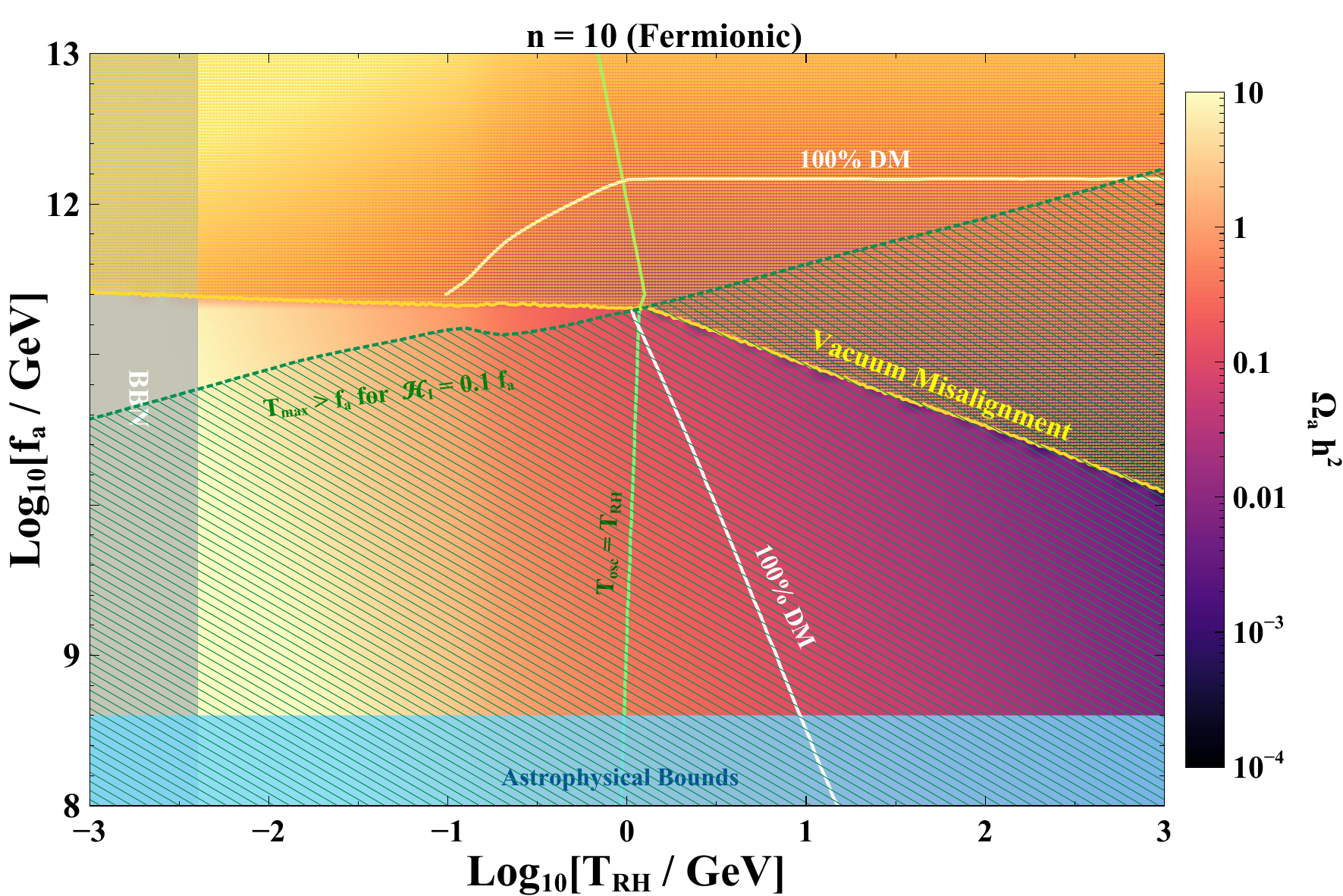}
    
    \caption{Direct Reheating (Kinetic Misalignment): Parameter space in the kinetic misalignment scenario in $T_{\rm RH}$--$f_a$ plane for bosonic (left column) and fermionic (right column) reheating. The top and bottom rows correspond to $n=8$ and $n=10$, respectively. We fix $\dot{\theta}_i=5\times10^{16}~{\rm GeV}$ and $\mathcal{H}_I=0.1 f_a$. The solid white contour denotes the parameter space yielding the observed axion dark matter abundance. The yellow solid contour corresponds to the critical value of $f_a$ for which $\dot{\theta}_i=\dot{\theta}_{\rm crit}$ at a given $T_{\rm RH}$ such that kinetic misalignment is realized below this contour and vacuum misalignment above it. See text for details related to the other bounds and contours. }
    \label{fig:KMM_parameter_space}
\end{figure}

Using the oscillation temperature obtained from Eq.~\eqref{eq:Tprime}, we solve Eqs.~\eqref{energyden}--\eqref{Axion_DM} numerically to obtain the axion relic abundance for the allowed $n=6$ case, considering both bosonic and fermionic reheating. We find that the axion abundance is below the observed dark matter abundance for the initial benchmark value of angular velocities. Reproducing the observed abundance would require a larger initial angular velocity, $\dot{\theta}_i$, which would exceed the upper bound from Eq.~\eqref{const_con}. Thus, although $n=6$ can satisfy the conditions for kinetic misalignment, it cannot produce the observed axion abundance without violating the energy-consistency bound for the benchmark choice $\mathcal{H}_I=0.1 f_a$.

The resulting axion relic abundance for $n=8$ and $n=10$ is shown in Fig.~\ref{fig:KMM_parameter_space}. The color bar denotes the axion DM abundance $\Omega_a h^2$, while the solid white contours correspond to $100\%$ of the observed dark matter abundance for the benchmark choice of initial velocity $\dot{\theta}_i=5\times10^{16}~{\rm GeV}$. In the kinetic-misalignment scenario, the numerically obtained axion relic abundance is in close agreement with the analytical expression in Eq.~\eqref{eq:Omega_KM_direct} for $\mathcal C_{\rm KM}=2$, derived in Appendix~\ref{sec:direct-kin}. Although the oscillation temperatures differ for bosonic and fermionic reheating, the analytical abundance has the same form at fixed $n$ and initial conditions. The cyan dotted contour denotes $T_{\rm osc}=T_{\rm QCD}$, while the yellow solid contour indicates the boundary $\dot{\theta}_i=\dot{\theta}_{\rm crit}$ such that the region below this contour corresponds to kinetic misalignment and the yellow-shaded region above it corresponding to vacuum misalignment. The constraint from symmetry restoration $T_{\rm max}<f_a$, discussed earlier, is applicable for fermionic reheating. For the benchmark values of $\mathcal{H}_I$ and $\dot{\theta}_i$ adopted here, the parameter space satisfying the observed dark matter abundance is excluded by the $T_{\rm max}<f_a$ constraint for both $n=8$ and $n=10$ in the fermionic reheating case. In both the reheating scenarios, we also find that the value of $T_{\rm RH}$ along the $100\%$ dark matter contour increases with increasing $n$ in the regime where kinetic misalignment dominates. For the benchmark choice $\mathcal{H}_I=0.1f_a$, the inflationary upper bound in Eq.~\eqref{Hmax} gives $f_a^{\max}=4.7\times10^{14}\,{\rm GeV}$, which exceeds the range of $f_a$ displayed in Fig.~\ref{fig:KMM_parameter_space}; hence, this constraint does not restrict the parameter space shown in the figure.

\subsubsection{Seesaw Reheating}

\begin{figure}
    \centering
    \includegraphics[width=0.49\linewidth]{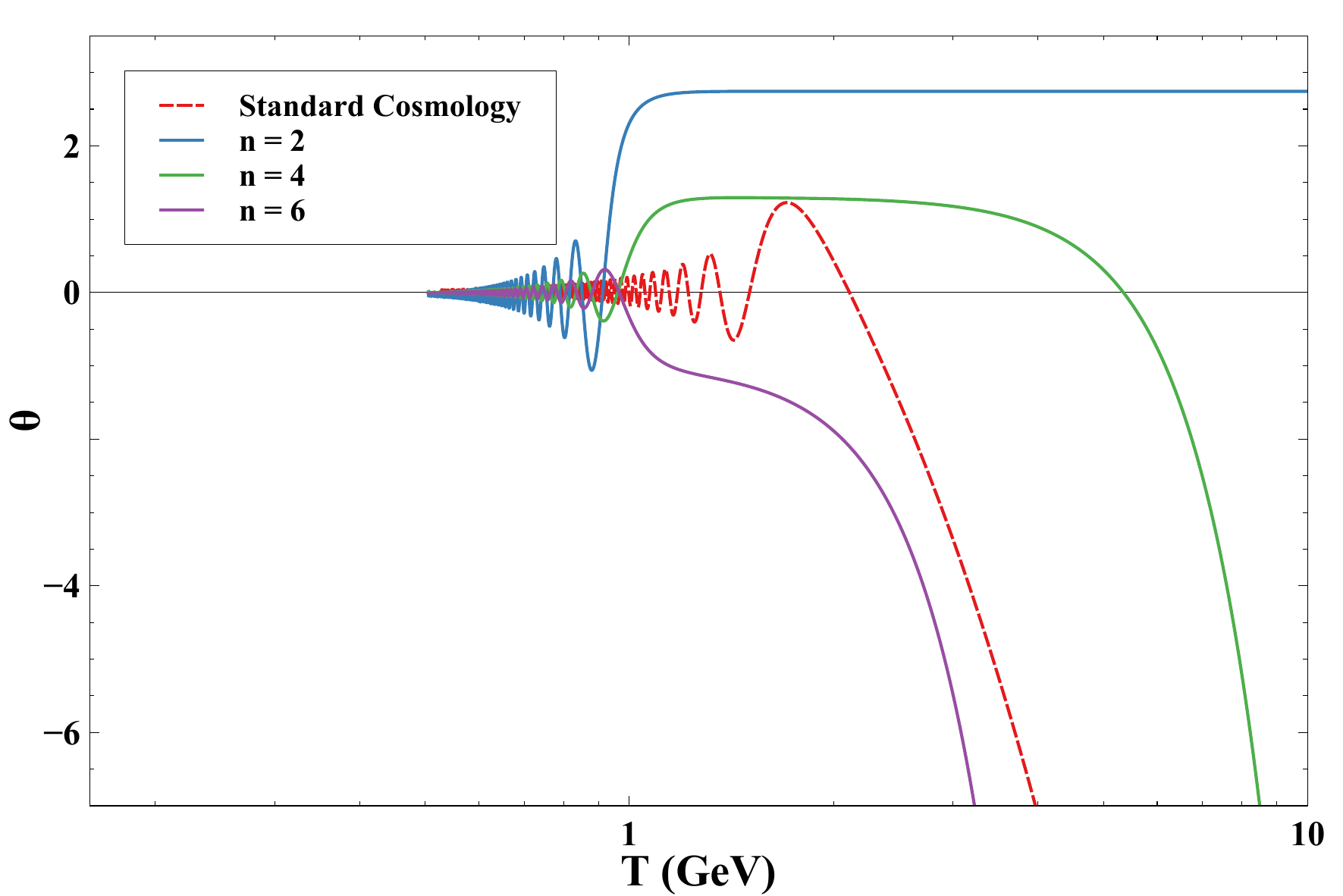}
      \includegraphics[width=0.49\linewidth]{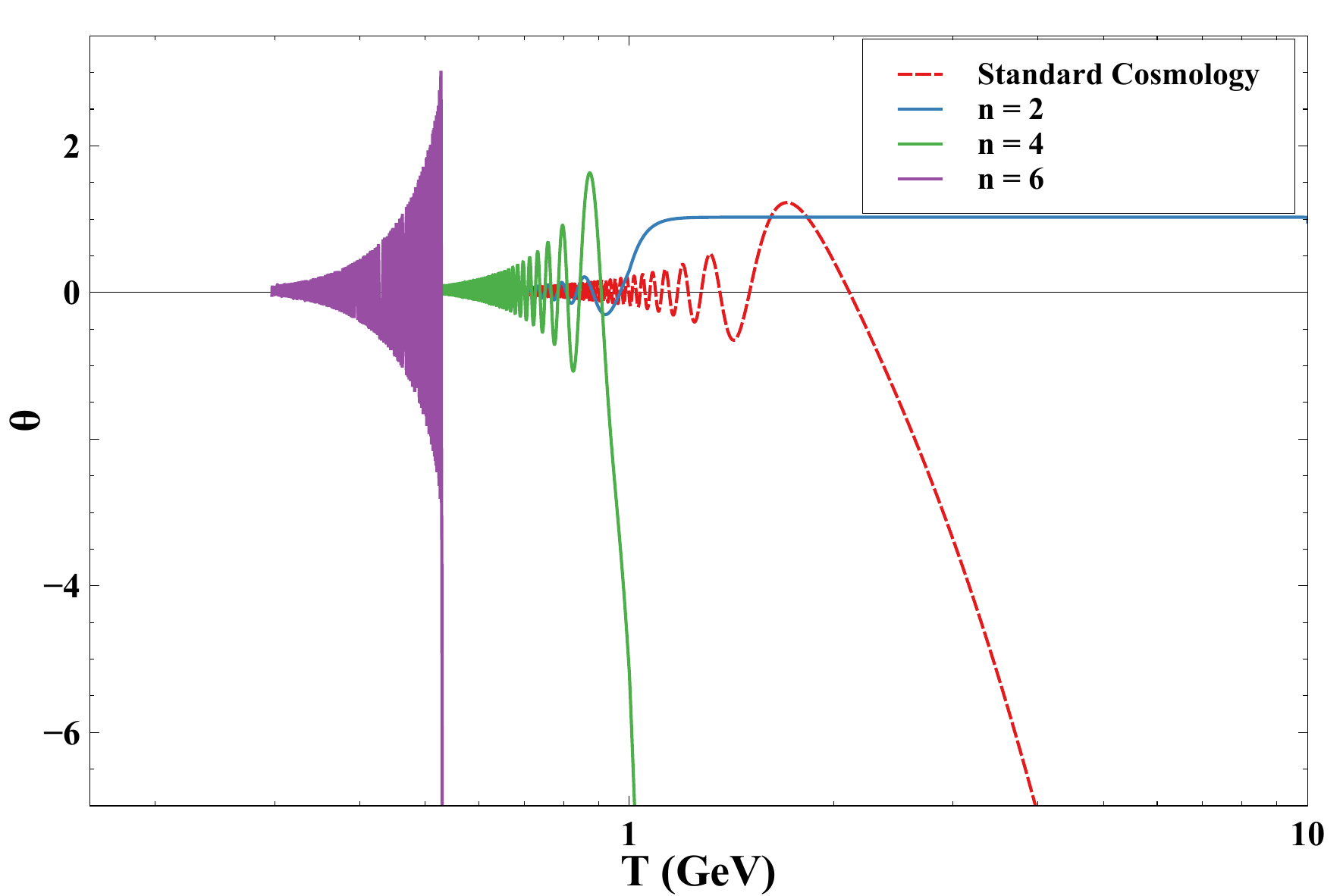}
    
    \caption{Seesaw Reheating (Kinetic Misalignment): Axion evolution for different $n$ and $T_{\rm RH} = 0.1 $ GeV, $T_{\rm NR}=10^3\, T_{\rm RH}$, $T_{\rm max}=10^4 \, T_{\rm RH}$  (left) and $T_{\rm NR}=10 T_{\rm RH}$, $T_{\rm max}=10^2 T_{\rm RH}$ (right). Both the panels correspond to $\mathcal{H}_I=0.1f_a$, $f_a = 10^{10}$ GeV and $\dot{\theta}_i=10^{23}~{\rm GeV}$.}
    \label{fig:RHN_evo_kin}
\end{figure}

We again solve the axion evolution equation given in Eq.~\eqref{thetaEvo} numerically for different values of $n$ considering indirect reheating via RHN decay. However, in the present case, the Hubble and temperature evolution are different from the previous scenario and are given by Eqs.~\eqref{eq:Hubble_RHN} and \eqref{eq:Temperature_RHN}, respectively. In Fig.~\ref{fig:RHN_evo_kin}, we show the axion evolution for different values of $n$ with $T_{\rm RH}=0.1~{\rm GeV}$, $\mathcal{H}_I=0.1 f_a$, $f_a = 10^{10}$ GeV  and $\dot{\theta}_i=10^{23}~{\rm GeV}$. We consider two different choices of the maximum temperature and the temperature corresponding to the relativistic-to-non-relativistic transition of the RHNs, namely  $T_{\rm NR}=10^3\, T_{\rm RH}$ and $T_{\rm max}=10^4 \, T_{\rm RH}$ (left panel) and $T_{\rm max}=10\,T_{\rm NR}=10\,T_{\rm RH}$ (right panel).

\begin{figure}
    \centering
    \includegraphics[width=0.49\linewidth]{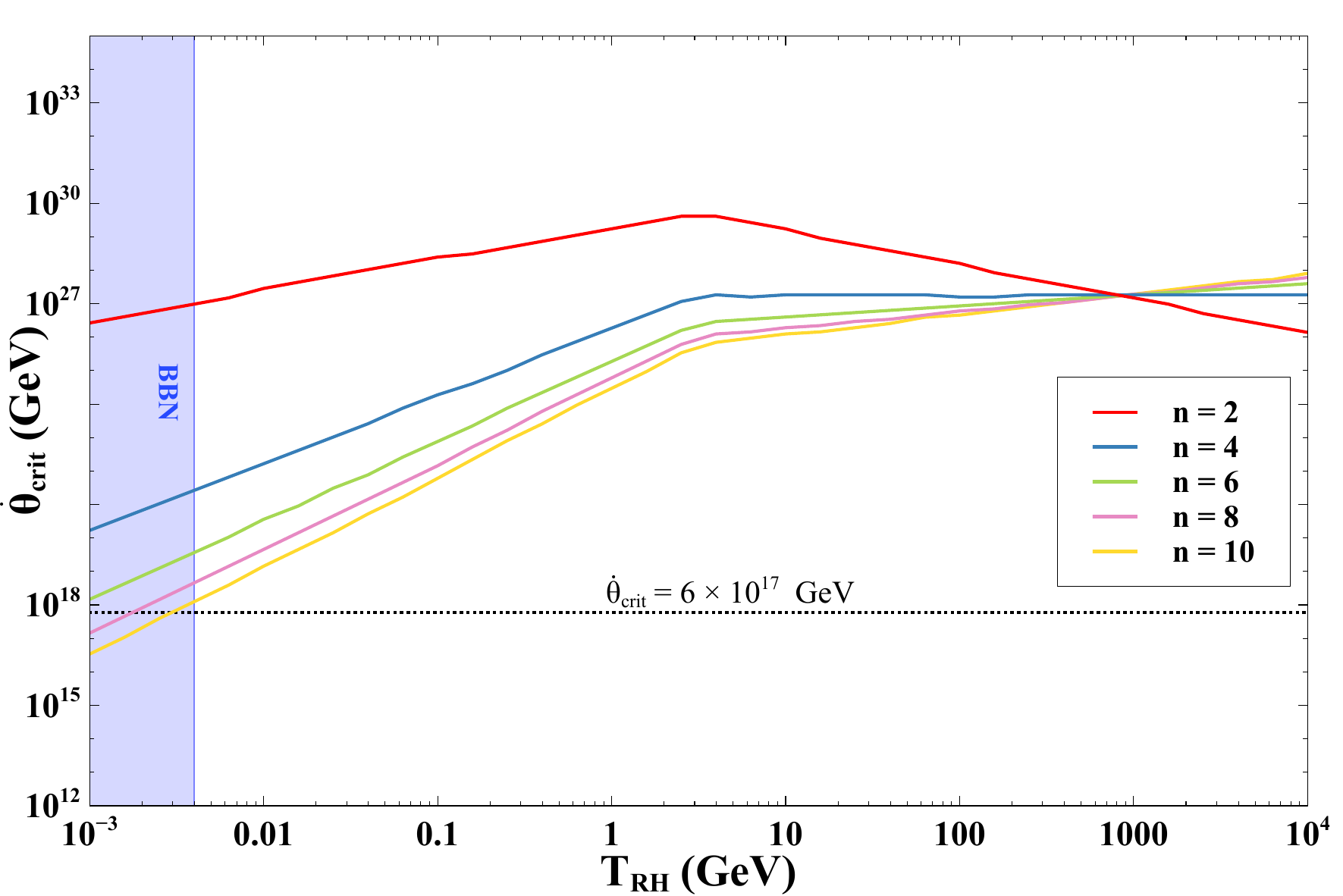}
    \includegraphics[width=0.49\linewidth]{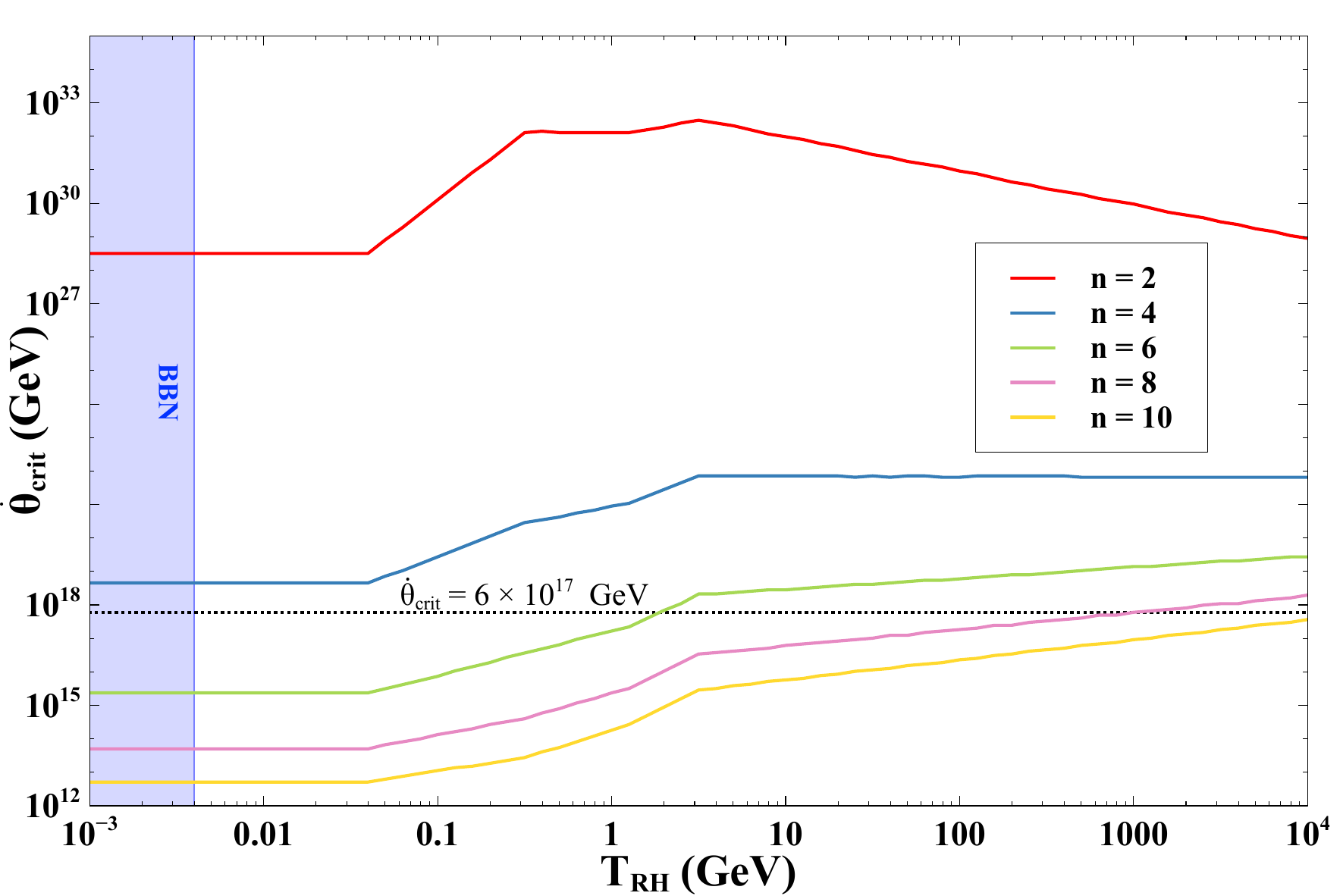}
    \caption{ Seesaw Reheating (Kinetic Misalignment): Variation of $\dot{\theta}_{\rm crit}$ with reheating temperature $T_{\rm RH}$ for different $n$ with $T_{\rm NR}=10^3T_{\rm RH}$, $T_{\rm max}=10^4 T_{\rm RH}$ (left) and $T_{\rm NR}=2\,T_{\rm RH}$, $T_{\rm max}=4 \,T_{\rm RH}$ (right). Both the panels correspond to $\mathcal{H}_I=0.1f_a$ and $f_a = 10^{9}$ GeV. }
    \label{fig:RHN_thetacrit}
\end{figure}

In the case of vacuum misalignment with seesaw reheating discussed in \ref{subc412}, for benchmark values corresponding to the left panel of Fig.~\ref{fig:RHN_evo_kin} we found that the axion evolution does not significantly depend on the value of $n$ in the inflaton potential, as the onset of oscillation is mainly determined by the Hubble evolution at the time when the axion starts oscillating. In contrast, for kinetic misalignment, the axion evolution depends on the entire evolution of $\dot{\theta}$ and hence can retain a dependence on the inflaton potential through $n$, rather than being determined only by the conditions at the onset of oscillations. For the left panel of Fig.~\ref{fig:RHN_evo_kin}, we see that the axion starts oscillating at nearly the same temperature for all values of $n$, despite the different evolution of $\theta$. This can be understood from the corresponding evolution of $\theta_{\rm crit}$ shown in the left panel of Fig.~\ref{fig:RHN_thetacrit}. In this case, the axion does not have sufficient kinetic energy to cross the potential barrier and enter the kinetic-misalignment regime. Therefore, the evolution remains effectively similar to that of vacuum misalignment, resulting in nearly the same oscillation temperature for different values of $n$. This is mainly due to the longer period of RHN domination, which leads to stronger damping of $\dot{\theta}$. The longer duration of RHN domination can also be understood from the corresponding values of $a_{\rm R}$ and $a_{NR}$ obtained from Eq.~\eqref{scale_fac}. As a result, a significant amount of the initial kinetic energy is lost before the axion can cross the potential barrier, and the dependence on the inflaton potential through $n$ becomes less relevant for the onset of oscillations.

On the other hand, in the right panel of Fig.~\ref{fig:RHN_evo_kin}, the RHN-dominated phase is shorter, while the $\phi^n$ component dominates the evolution for a larger part of the axion evolution. Consequently, the damping of $\dot{\theta}$ is weaker, allowing the axion to retain sufficient kinetic energy to cross the potential barrier and enter the kinetic-misalignment regime. In this case, the axion evolution becomes sensitive to the Hubble evolution throughout its history, and hence to the value of $n$. This explains the different evolution of $\theta$ for different $n$ in the right panel of Fig.~\ref{fig:RHN_evo_kin}. This dependence on the preceding expansion history is explicit in the analytical oscillation temperatures in
Eqs.~\eqref{eq:Tosc_KM_seesaw_above} and \eqref{eq:Tosc_KM_seesaw_below}, derived in Appendix~\ref{sec:seesaw-kin} for the different stages of seesaw reheating. These expressions apply when the initial angular velocity is sufficient to realize kinetic misalignment.

\begin{figure}
    \centering
    \includegraphics[width=0.49\linewidth]{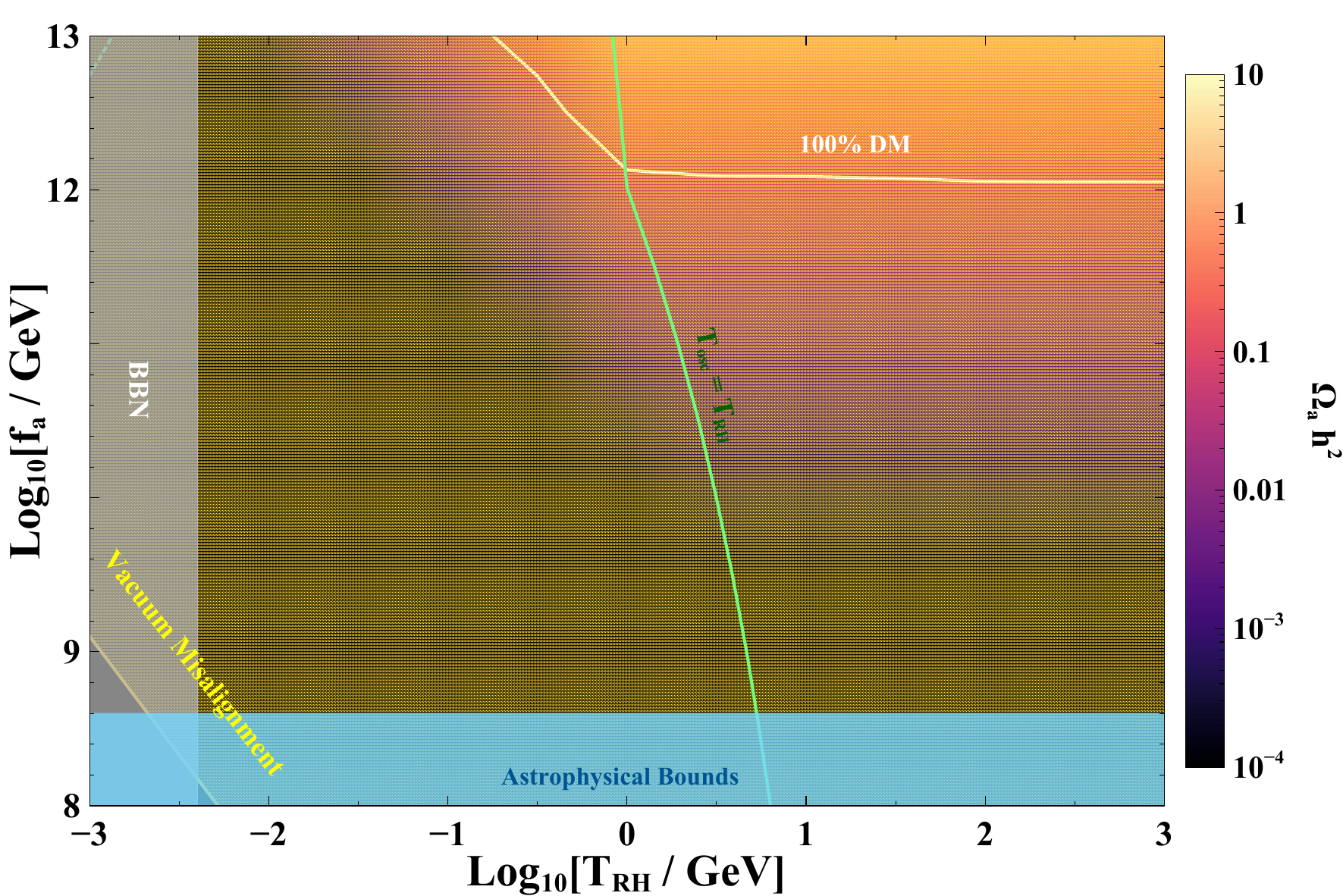}
    \includegraphics[width=0.49\linewidth]{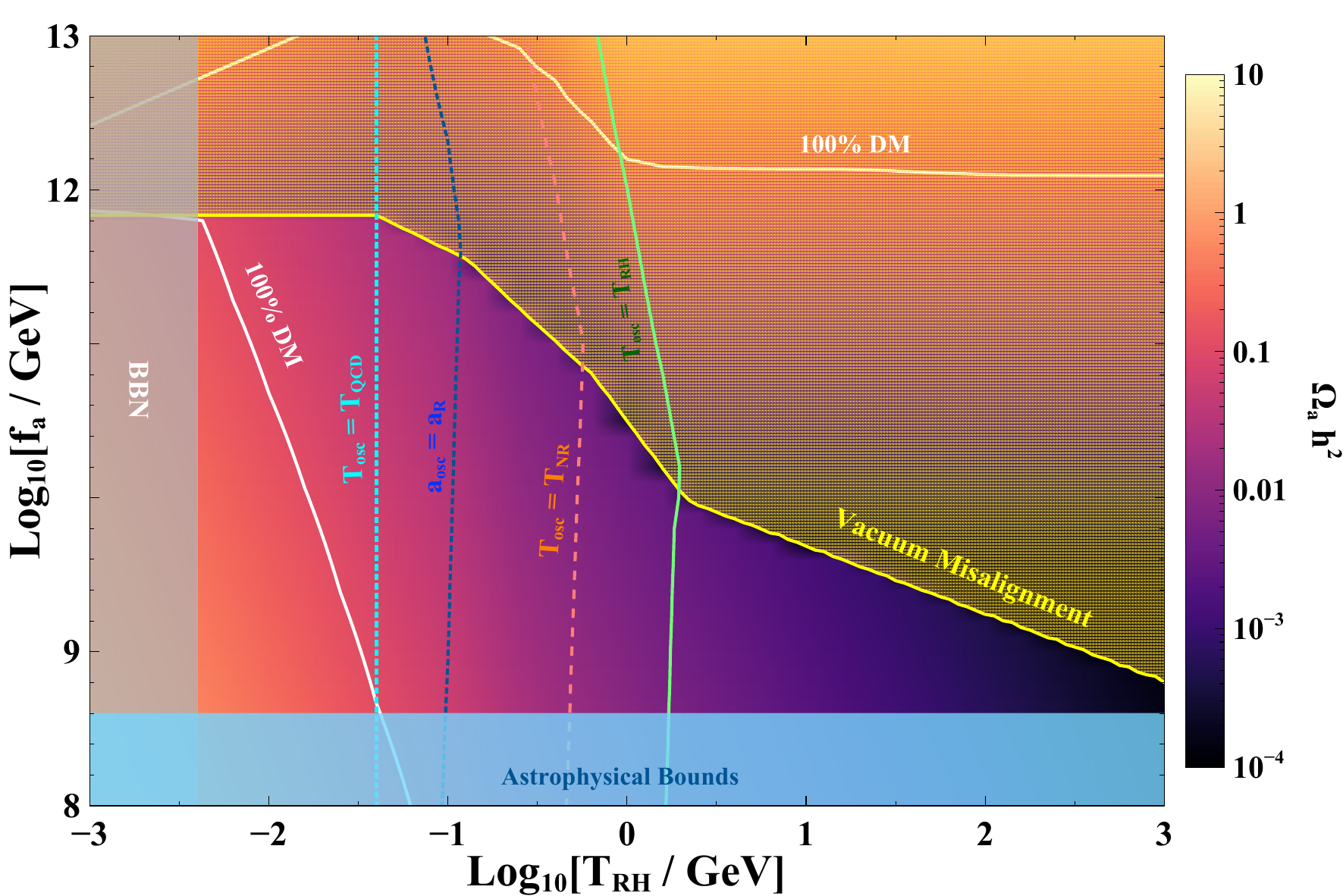}
   \caption{ Seesaw Reheating (Kinetic Misalignment): Axion relic abundance for $n=10$, with $\dot{\theta}_i=5\times10^{16}~{\rm GeV}$ and $\mathcal{H}_I=0.1\,f_a$. The left and right panels correspond to BP1: $T_{\rm NR}=10^3T_{\rm RH}$, $T_{\rm max}=10^4T_{\rm RH}$, and BP2: $T_{\rm NR}=2\,T_{\rm RH}$, $T_{\rm max}=4\,T_{\rm RH}$, respectively. The solid white contour denotes the parameter space yielding the observed axion dark matter abundance. The yellow solid contour corresponds to the critical value of $f_a$ for which $\dot{\theta}_i=\dot{\theta}_{\rm crit}$ at a given $T_{\rm RH}$, with kinetic misalignment realized below this contour and vacuum misalignment above it. See text for details. }
    \label{fig:RHN_kin}
\end{figure}

We next examine whether the condition in Eq.~\eqref{eq:thetadot_crit} can be satisfied consistently with the energy bound in Eq.~\eqref{const_con}. In Fig.~\ref{fig:RHN_thetacrit}, we show the critical angular velocity $\dot{\theta}_{\rm crit}$ as a function of the reheating temperature $T_{\rm RH}$ for $f_a=10^{9}~{\rm GeV}$. The left and right panels correspond to $T_{\rm NR}=10^3\,T_{\rm RH}$, $T_{\rm max}=10^4\,T_{\rm RH}$ and $T_{\rm NR}=2\,T_{\rm RH}$, $T_{\rm max}=4\,T_{\rm RH}$, respectively. In the left panel, the longer period of RHN domination leads to significant damping of the axion motion. As a result, $\dot{\theta}_{\rm crit}$ remains too large to satisfy the kinetic-misalignment condition within the parameter space consistent with the BBN and energy-density bounds, even for $n\,= \,10$. Thus, there is effectively no viable parameter space for kinetic misalignment in up to $n=10$. In contrast, the RHN-dominated phase is short-lived for the benchmark choices in the right panel, resulting in a weaker damping and allowing kinetic misalignment for $n\geq6$ within the allowed parameter space. This can be clearly seen by considering two benchmark choices for the RHN thermal history. We define benchmark BP1 by $T_{\rm NR}=10^3T_{\rm RH}$, $T_{\rm max}=10^4T_{\rm RH}$, corresponding to a longer period of RHN domination, while benchmark BP2 is defined by $T_{\rm NR}=2\,T_{\rm RH}$, $T_{\rm max}=4\,T_{\rm RH}$, where the RHN-dominated phase is much shorter. Using the oscillation temperature obtained from Eq.~\eqref{eq:Tprime}, we solve Eqs.~\eqref{energyden}--\eqref{Axion_DM} numerically to obtain the axion relic abundance for these two benchmark points. For BP1, the axion remains under-abundant for all the considered values $n=6,8,10$. In contrast, for BP2, the observed relic abundance can be obtained for $n=10$ with $\dot{\theta}_i=5\times10^{16}~{\rm GeV}$ and $\mathcal{H}_I=0.1f_a$. This can be seen from Fig.~\ref{fig:RHN_kin}. For $n=10$, BP1 does not yield any viable parameter space for the observed dark matter abundance, while BP2 allows the axion to account for the observed relic abundance for relatively low reheating temperatures, from $\mathcal{O}(0.1)~{\rm GeV}$ down to the BBN bound, as indicated by the solid white contour. In the region where kinetic misalignment is realized, the numerical
abundance is in close agreement with Eq.~\eqref{eq:Omega_KM_seesaw}, with its normalized $n=10$ form given in Eq.~\eqref{eq:Omega_KM_seesaw_n10} for $\mathcal C_{\rm KM}=2$. The derivation in Appendix~\ref{sec:seesaw-kin} shows explicitly how the abundance retains a dependence on the preceding reheating history. As in the previous case, the color bar shows the axion abundance $\Omega_a h^2$. The cyan dotted contour denotes $T_{\rm osc}=T_{\rm QCD}$, while the blue dotted and orange dashed contours indicate the boundaries $a_{\rm osc}=a_{\rm R}$ and $T_{\rm osc}=T_{\rm NR}$, respectively. To the left of the green dotted contour, the axion starts oscillating while the inflaton field still dominates the energy density. Between the green dotted and dashed contours, axion oscillations begin during the relativistic RHN-dominated phase, while to the right of the green dashed contour the oscillations begin during the non-relativistic RHN-dominated phase.

\section{Detection aspects}
\label{sec:detection}
In this section we discuss the detection prospects of our scenario via observations of primordial gravitational wave (PGW) of inflationary origin or axion signatures in typical axion detection experiments.

\subsection{Inflationary primordial gravitational waves}

One of the promising avenues for probing the extended reheating period affecting axion oscillations in a non-trivial way is through the primordial gravitational-wave background generated during inflation. Quantum tensor fluctuations produced during inflation are stretched beyond the Hubble horizon by the rapid exponential expansion, resulting in a nearly scale-invariant tensor power spectrum on super-horizon scales. Following the end of inflation, the comoving Hubble horizon grows, allowing these tensor modes to re-enter the horizon at different epochs depending on their wavelengths. Modes that re-enter during the radiation-dominated era give rise to a nearly scale-invariant present-day GW energy density spectrum. In contrast, modes re-entering during the reheating epoch experience a different expansion history, leading to a characteristic tilt in the GW spectrum over the corresponding frequency range. The slope of this tilt is determined by the effective equation-of-state parameter during reheating. Consequently, the present-day inflationary GW spectrum carries direct information about the reheating history and depends on both the reheating temperature, $T_{\rm RH}$, and the reheating equation-of-state parameter, $\omega_{\phi}$.

The PGW spectrum with an inflationary origin have been extensively studied in several earlier works including~\cite{Giovannini:1998bp,Giovannini:1999bh,Riazuelo:2000fc,Seto:2003kc,Boyle:2007zx,Stewart:2007fu,Li:2021htg,Caprini:2018mtu,Figueroa:2019paj,Bernal:2020ywq,Caldwell:2022qsj,Gouttenoire:2021jhk,Haque:2021dha,Maity:2024cpq}. The stochastic PGW background is conveniently characterized by the GW energy density per logarithmic interval of comoving wavenumber, normalized to the critical energy density of the Universe, $\rho_{\rm crit}=3M_P^2\mathcal{H}^2$. It is defined as
\begin{eqnarray}\label{eq:gw6}
\Omega_{\rm GW}(t,k)\equiv\frac{1}{\rho_{\rm crit}}\frac{d\rho_{\rm GW}(t,k)}{d\ln k}.
\end{eqnarray}
For primordial tensor perturbations, the GW energy density evolves according to
\begin{eqnarray}\label{eq:gw7}
\Omega_{\rm GW}(a,k)=\frac{k^2}{12\,a^2\mathcal{H}^2}\,
\mathcal{P}_{T,\rm prim}\,\mathcal{T}(a,k),
\end{eqnarray}
where $\mathcal{T}(a,k)$ denotes the tensor transfer function accounting for the evolution of tensor modes after horizon re-entry. The primordial tensor power spectrum is parameterized as
\begin{eqnarray}\label{eq:gw8}
\mathcal{P}_{T,\rm prim}
=r\,\mathcal{P}_{\zeta}(k_*)
\left(\frac{k}{k_*}\right)^{n_T},
\end{eqnarray}
with $k_*=0.05\,{\rm Mpc}^{-1}$ being the Planck pivot scale and $\mathcal{P}_{\zeta}(k_*)\simeq2.1\times10^{-9}$ is the scalar amplitude. Here, $r$ and $n_T$ represent the tensor-to-scalar ratio and tensor spectral index, respectively. Current PLANCK observations constrain $r<0.036$ \cite{Planck:2018jri}, while single-field slow-roll inflation predicts the consistency relation $n_T\simeq-r/8$. Since the resulting tilt is negligibly small within the allowed range of $r$, we set $n_T=0$ throughout our analysis.

The transfer function relates the primordial tensor amplitude of the modes to its value at a later epoch after the modes become sub-horizon and is given by
\begin{eqnarray}\label{eq:gw9}
\mathcal{T}(a,k)=\frac12\left(\frac{a_{\rm H}}{a}\right)^2,
\end{eqnarray}
where $a_{\rm H}$ denotes the scale factor at horizon crossing, defined through
$k=a_{\rm H}\mathcal{H}(a_{\rm H})$. Since different Fourier modes re-enter the horizon at different epochs, the present-day transfer function $\mathcal{T}(a_0,k)$ encodes valuable information about the expansion history of the early Universe. Consequently, the present-day GW energy density spectrum is
\begin{eqnarray}\label{eq:gw10}
\Omega_{\rm GW}(k)\equiv\Omega_{\rm GW}(a_0,k)
=\frac{k^2}{24\,a_0^2\mathcal{H}_0^2}
\mathcal{P}_{T,\rm prim}
\left(\frac{a_{\rm H}}{a_0}\right)^2.
\end{eqnarray}
The subsequent evolution of the GW spectrum depends on whether a given mode re-enters the horizon during reheating or the radiation-dominated era. In the following, we discuss the GW spectrum for both direct reheating and seesaw reheating.

\subsection*{Direct reheating}

Assuming an inflationary potential of the form $V(\phi)\propto\phi^n$, the present-day frequency corresponding to a mode that re-enters the horizon at $a_{\rm H}$ can be obtained as
\begin{eqnarray}\label{eq:gw11}
f\equiv\frac{k}{2\pi a_0}
=\sqrt{\frac{g_*(T_{\rm RH})}{360}}
\left(\frac{g_{*,s}(T_0)}{g_{*,s}(T_{\rm RH})}\right)^{1/3}
\frac{T_0T_{\rm RH}}{M_P}
\begin{cases}
\left(\dfrac{a_{\rm RH}}{a_{\rm H}}\right)^{\frac{2(n-1)}{n+2}},
& a_I<a_{\rm H}\le a_{\rm RH},\\[0.4cm]
\dfrac{a_{\rm RH}}{a_{\rm H}},
& a_{\rm RH}<a_{\rm H}\le a_{\rm EQ},
\end{cases}
\end{eqnarray}
where $a_{\rm EQ}$ denotes the scale factor at matter-radiation equality. Correspondingly, the present-day GW energy density is
\begin{eqnarray}\label{eq:gw12}
\Omega_{\rm GW}(a_{\rm H})
&\simeq&
\Omega_{\gamma,0}
\frac{\mathcal{P}_{T,\rm prim}}{24}
\frac{g_*(T_{\rm H})}{g_*^\gamma(T_0)}
\left(\frac{g_{*,s}(T_0)}{g_{*,s}(T_{\rm H})}\right)^{4/3}
\nonumber\\
&&\times
\begin{cases}
\dfrac{g_*(T_{\rm RH})}{g_*(T_{\rm H})}
\left(\dfrac{g_{*,s}(T_{\rm H})}{g_{*,s}(T_{\rm RH})}\right)^{4/3}
\left(\dfrac{a_{\rm RH}}{a_{\rm H}}\right)^{\frac{2(n-4)}{n+2}},
&
a_I<a_{\rm H}\le a_{\rm RH},
\\[0.4cm]
1,
&
a_{\rm RH}<a_{\rm H}\le a_{\rm EQ}.
\end{cases}
\end{eqnarray}
Combining Eqs.~\eqref{eq:gw11} and \eqref{eq:gw12}, the present-day GW spectrum as a function of frequency becomes
\begin{eqnarray}\label{eq:gw13}
\Omega_{\rm GW}(f)
&\simeq&
\Omega_{\gamma,0}
\frac{\mathcal{P}_{T,\rm prim}}{24}
\frac{g_*(T_{\rm H})}{g_*^\gamma(T_0)}
\left(\frac{g_{*,s}(T_0)}{g_{*,s}(T_{\rm H})}\right)^{4/3}
\nonumber\\
&&\times
\begin{cases}
\dfrac{g_*(T_{\rm RH})}{g_*(T_{\rm H})}
\left(\dfrac{g_{*,s}(T_{\rm H})}{g_{*,s}(T_{\rm RH})}\right)^{4/3}
\left(\dfrac{f}{f_{\rm RH}}\right)^{\frac{n-4}{n-1}},
&
f_{\rm RH}<f\le f_{\rm max},
\\[0.4cm]
1,
&
f_{\rm EQ}<f\le f_{\rm RH}.
\end{cases}
\end{eqnarray}
Here, $\Omega_{\gamma,0}$ and $T_0$ denote the present-day photon energy density fraction and CMB temperature, respectively. The characteristic frequencies $f_{\rm RH}$ and $f_{\rm max}$ correspond to tensor modes re-entering the horizon at the end of reheating and at the end of inflation, respectively,
\begin{eqnarray}
f_{\rm RH}
&=&
\sqrt{\frac{g_*(T_{\rm RH})}{360}}
\left(\frac{g_{*,s}(T_0)}{g_{*,s}(T_{\rm RH})}\right)^{1/3}
\frac{T_0T_{\rm RH}}{M_P}, \label{eq:frh_direct}
\\
f_{\rm max}
&=&
\frac{\mathcal{H}_I}{2\pi}
\left(\frac{g_{*,s}(T_0)}{g_{*,s}(T_{\rm RH})}\right)^{1/3}
\left(
\frac{\pi T_{\rm RH}^2}{M_P \mathcal{H}_I}
\sqrt{\frac{g_{*s}(T_{\rm RH})}{90}}
\right)^{\frac{n+2}{3n}}
\frac{T_0}{T_{\rm RH}}.
\end{eqnarray}

\subsection*{Seesaw reheating}

Unlike the case of direct reheating, where the Universe evolves with a single effective equation of state throughout the reheating epoch, the scenario considered here involves a sequence of distinct cosmological eras characterized by different equations of state. Initially, the Universe evolves with $w = \frac{n-2}{n+2}$,
which is subsequently followed by a radiation-dominated era with
$w = \frac{1}{3}$, corresponding to relativistic RHNs. This is followed by a matter-dominated era characterized by $w = 0$, when the RHNs become non-relativistic. Consequently, the GW spectrum receives contributions from the successive cosmological eras, and the resulting GW amplitude can be expressed as
\begin{eqnarray}\label{eq:gw13}
\Omega_{\rm GW}(a_{\rm H})
&\simeq&
\Omega_{\gamma,0}
\frac{\mathcal{P}_{T,\rm prim}}{24}
\frac{g_*(T_{\rm H})}{g_*^\gamma(T_0)}
\left(\frac{g_{*,s}(T_0)}{g_{*,s}(T_{\rm H})}\right)^{4/3}
\nonumber\\
&&\times
\begin{cases}
\dfrac{g_*(T_{\rm RH})}{g_*(T_{\rm H})}
\left(\dfrac{g_{*,s}(T_{\rm H})}{g_{*,s}(T_{\rm RH})}\right)^{4/3}
\left(\dfrac{a_{\rm RH}}{a_{\rm NR}}\right)^{\frac{2(n-4)}{n+2}}\,\left(\dfrac{a_{\rm R}}{a_{\rm H}}\right)^{-1},
&
a_I<a_{\rm H}\le a_{\rm R},
\\[0.4 cm]
\dfrac{g_*(T_{\rm RH})}{g_*(T_{\rm H})}
\left(\dfrac{g_{*,s}(T_{\rm H})}{g_{*,s}(T_{\rm RH})}\right)^{4/3}
\left(\dfrac{a_{\rm RH}}{a_{\rm NR}}\right)^{\frac{2(n-4)}{n+2}},
&
a_{\rm R}<a_{\rm H}\le a_{\rm NR},
\\[0.4 cm]
\dfrac{g_*(T_{\rm RH})}{g_*(T_{\rm H})}
\left(\dfrac{g_{*,s}(T_{\rm H})}{g_{*,s}(T_{\rm RH})}\right)^{4/3}
\left(\dfrac{a_{\rm RH}}{a_{\rm H}}\right)^{\frac{2(n-4)}{n+2}},
&
a_{\rm NR}<a_{\rm H}\le a_{\rm RH},
\\[0.4cm]
1,
&
a_{\rm RH}<a_{\rm H}\le a_{\rm EQ}.
\end{cases}
\end{eqnarray}
In terms of frequency,
\begin{eqnarray}\label{eq:gw14}
\Omega_{\rm GW}(f)
&\simeq&
\Omega_{\gamma,0}
\frac{\mathcal{P}_{T,\rm prim}}{24}
\frac{g_*(T_{\rm H})}{g_*^\gamma(T_0)}
\left(\frac{g_{*,s}(T_0)}{g_{*,s}(T_{\rm H})}\right)^{4/3}
\nonumber\\
&&\times
\begin{cases}
\dfrac{g_*(T_{\rm RH})}{g_*(T_{\rm H})}
\left(\dfrac{g_{*,s}(T_{\rm H})}{g_{*,s}(T_{\rm RH})}\right)^{4/3}
\left(\dfrac{f_{\rm NR}}{f_{\rm RH}}\right)^{-2}\, \left(\frac{f}{f_{\rm R}}\right)^{\frac{n-4}{n-1}},
&
f_{\rm R}< f \le f_{\rm max},
\\[0.4 cm]
\dfrac{g_*(T_{\rm RH})}{g_*(T_{\rm H})}
\left(\dfrac{g_{*,s}(T_{\rm H})}{g_{*,s}(T_{\rm RH})}\right)^{4/3}
\left(\dfrac{f_{\rm NR}}{f_{\rm RH}}\right)^{-2},
&
f_{\rm NR}< f \le f_{\rm R},
\\[0.4 cm]
\dfrac{g_*(T_{\rm RH})}{g_*(T_{\rm H})}
\left(\dfrac{g_{*,s}(T_{\rm H})}{g_{*,s}(T_{\rm RH})}\right)^{4/3}
\left(\dfrac{f}{f_{\rm RH}}\right)^{-2},
&
f_{\rm RH}< f \le f_{\rm NR},
\\[0.4cm]
1,
&
f_{\rm EQ}< f \le f_{\rm RH}.
\end{cases}
\end{eqnarray}
Here, $f_{\rm R}$ and $f_{\rm NR}$ denote the frequencies corresponding to modes that re-enter the horizon at the end of inflaton domination and at the end of relativistic RHN domination, respectively. While the expressions for $f_{\rm RH}$ remain the same as in Eq. \eqref{eq:frh_direct}, the expression for $f_{\rm NR}$, $f_{R}$ and $f_{\rm max}$ becomes
\begin{eqnarray}
    f_{\rm NR} = f_{\rm RH} \left(\frac{T_{\rm NR}}{T_{\rm RH}}\right)^{4/3}, \,\,\,\,\,\,\,\,\,
    f_{\rm R} = f_{\rm NR} \left(\frac{T_{\rm R}}{T_{\rm NR}}\right)^{4}, \,\,\,\,\,\,\,\,\,
    f_{\rm max}  =  f_{R} \left(\frac{\mathcal{H}_{I}}{\mathcal{H}(a_{R})}\right)^{\frac{2(n-1)}{n+2}},
\end{eqnarray}
respectively, where $\mathcal{H}(a_{R}) = \sqrt{\frac{\pi^2}{30}g_{*}(T_{\rm RH})} \frac{T^2_{\rm RH}}{M_{\rm P}} \left(\frac{T_{\rm NR}}{T_{\rm RH}}\right)^{4} \left(\frac{T_{\rm RH}}{T_{\rm NR}}\right)^{8}$.

\begin{figure}
    \centering
    \captionsetup[subfigure]{labelformat=simple}
    \begin{subfigure}[b]{0.5\linewidth}
        \centering
        \includegraphics[width=\linewidth]{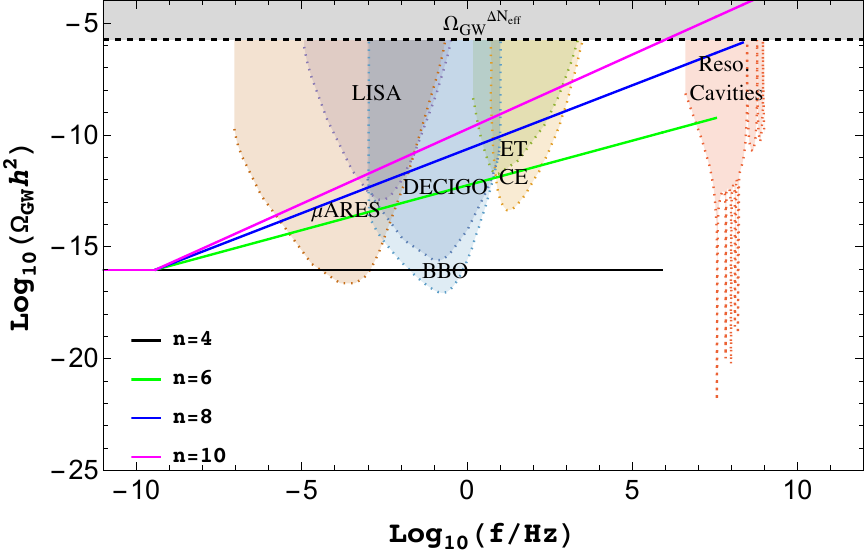}
        \caption{Direct Reheating}
        \label{sfig:13a}
    \end{subfigure}\hfill
    \begin{subfigure}[b]{0.5\linewidth}
        \centering
        \includegraphics[width=\linewidth]{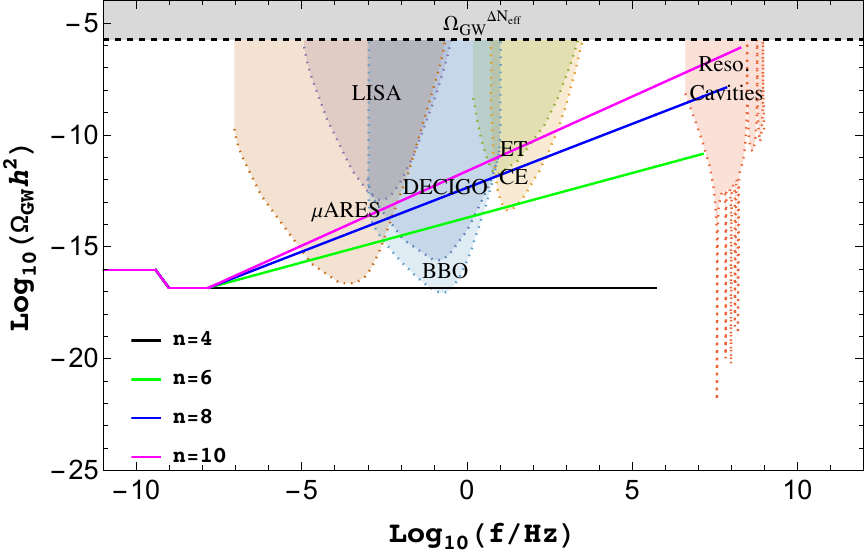}
        \caption{Seesaw Reheating}
        \label{sfig:13b}
    \end{subfigure}
    \caption{PGW spectrum for different values of $n$ for direct reheating (left panel) and seesaw reheating (right panel). In both panels, we consider $T_{\rm RH} = 0.1$ GeV and $\mathcal{H}_{I} = 10^{9}$ GeV. For seesaw reheating in the right panel, we take $T_{\rm max} = 2 T_{\rm NR}= 4 T_{\rm RH}$.}
    \label{fig:GW_spectrum}
\end{figure}

%\begin{figure}
%    \centering
%    \includegraphics[width=0.49\linewidth]{Figure/GW_direct_reheating.pdf}
%     \includegraphics[width=0.49\linewidth]{Figure/GW_RHN_reheating.pdf}
%    \caption{PGW spectrum for different values of $n$ for direct reheating (left panel) and seesaw reheating (right panel). In both panels, we consider $T_{\rm RH} = 0.1$ GeV and $\mathcal{H}_{I} = 10^{9}$ GeV. For seesaw reheating in the right panel, we take $T_{\rm max} = 2 T_{\rm NR}= 4 T_{\rm RH}$.}
%    \label{fig:GW_spectrum}
%\end{figure}

For our analysis, we estimate the detectability of the PGW signal using the signal-to-noise ratio (SNR), $\mathcal{S}$, computed over the observation time $t_{\rm obs}$ and the experimental frequency range $[f_{\rm min}^{\rm exp},\,f_{\rm max}^{\rm exp}]$~\cite{Maggiore:1999vm,Allen:1996vm,Allen:1997ad}:
\begin{align}
\mathcal{S} =
\left[
n_{\rm det}\, t_{\rm obs}
\int_{f_{\rm min}^{\rm exp}}^{f_{\rm max}^{\rm exp}}
df\,
\left(
\frac{\Omega_{\rm GW}(f)\,h^2}
{\Omega_{\rm expt}(f)\,h^2}
\right)^2
\right]^{1/2}.
\end{align}
Here, $n_{\rm det}$ depends on the search method: $n_{\rm det}=1$ for auto-correlation analyses, whereas $n_{\rm det}=2$ for cross-correlation searches. We consider a predicted GW signal detectable when its SNR exceeds the threshold $\mathcal{S}>10$. For LISA~\cite{LISACosmologyWorkingGroup:2022kbp}, and DECIGO~\cite{Yagi:2011wg}, we assume observation times of $4$ years whereas for BBO~\cite{Seto:2001qf}, CE~\cite{LIGOScientific:2016wof}, ET~\cite{Punturo_2010}, and $\mu$ARES~\cite{Sesana:2019vho}, we assume it to be $10$ years.

Fig. \ref{fig:GW_spectrum} shows the PGW spectra for direct reheating (left panel) and seesaw reheating (right panel). For direct reheating, the flat spectrum corresponding to $n=4$ can be probed by future experiments such as BBO and $\mu$ARES. Owing to the increasingly blue-tilted nature of the GW spectra for $n=6$, $8$, and $10$, these spectra can potentially fall within the sensitivity ranges of several future experiments, including LISA, DECIGO, ET, CE, and resonant-cavity~\cite{Herman:2022fau} experiments. The maximum frequency of the PGW spectrum depends on the Hubble scale at the end of inflation, $\mathcal{H}_I$. Consequently, for sufficiently large values of $\mathcal{H}_I$, the spectra for $n=8$ and $n=10$ can already be excluded by the bound on the contribution of GWs to the effective number of relativistic degrees of freedom, $N_{\rm eff}$ \cite{Planck:2018vyg}.

For seesaw reheating, shown in the right panel, we consider $T_{\rm max}=2 T_{\rm NR}=4 T_{\rm RH}$. In this case, the PGW spectrum exhibits a red-tilted behavior between $f_{\rm RH}$ and $f_{\rm NR}$, followed by a flat spectrum between $f_{\rm NR}$ and $f_{\rm R}$. As a result, the presence of both relativistic and non-relativistic RHN-dominated eras modifies the PGW spectrum at intermediate frequencies. In particular, the spectra for $n=8$ and $n=10$ remain consistent with the $N_{\rm eff}$ bound for the same values of $\mathcal{H}_I$ considered in the left panel. It is worth mentioning that most of the previous papers including \cite{Barman:2026yhc} treated RHNs to be non-relativistic from the epoch of their production, and hence this feature of multiple kinks in the PGW spectrum was absent in their analysis.

\begin{figure}
    \centering
      \captionsetup[subfigure]{labelformat=simple}
    \begin{subfigure}{0.49\linewidth}
        \centering
        \includegraphics[width=\linewidth]{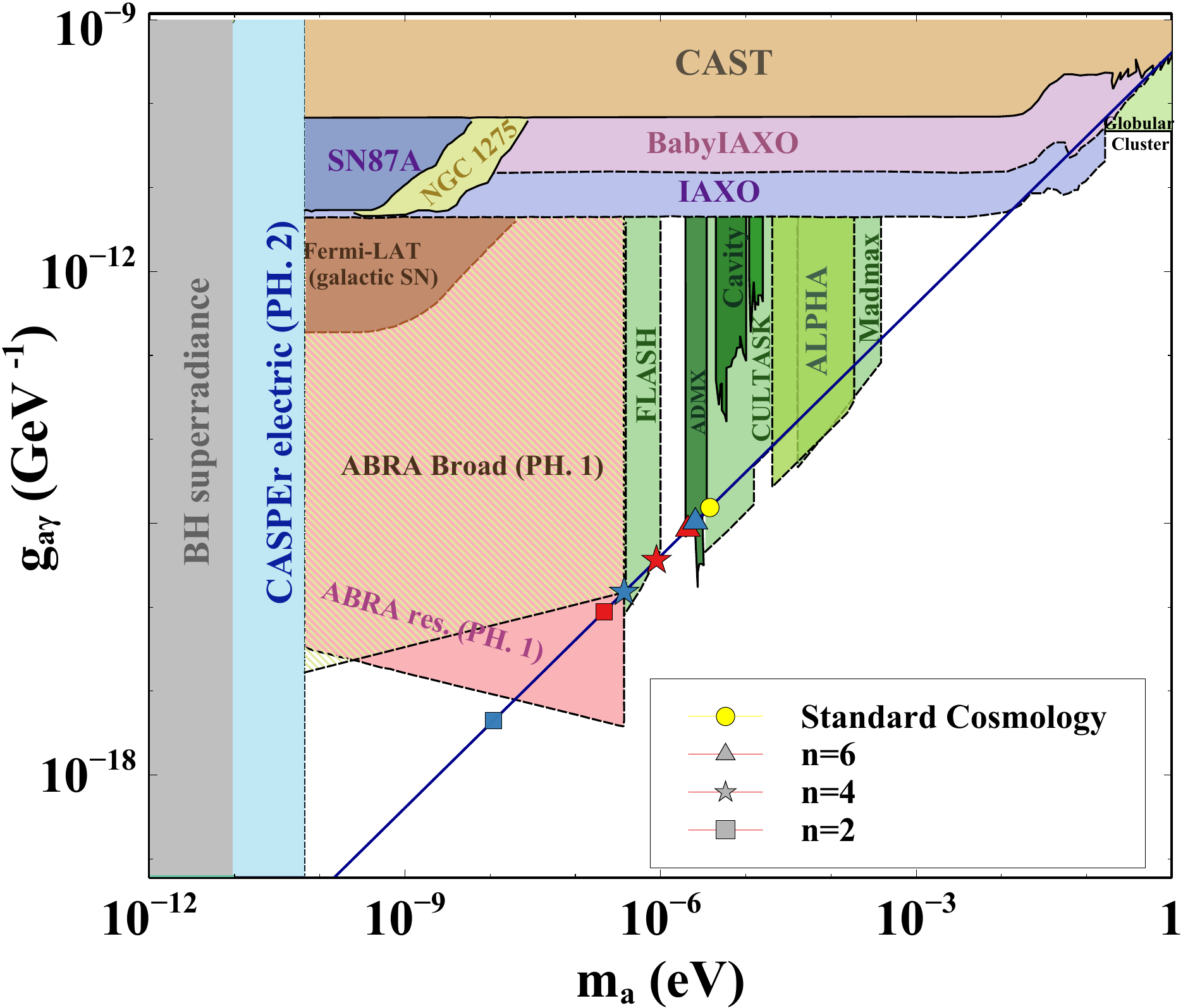}
        \caption{Bosonic (vacuum)}
    \end{subfigure}
    \hfill
    \begin{subfigure}{0.49\linewidth}
        \centering
        \includegraphics[width=\linewidth]{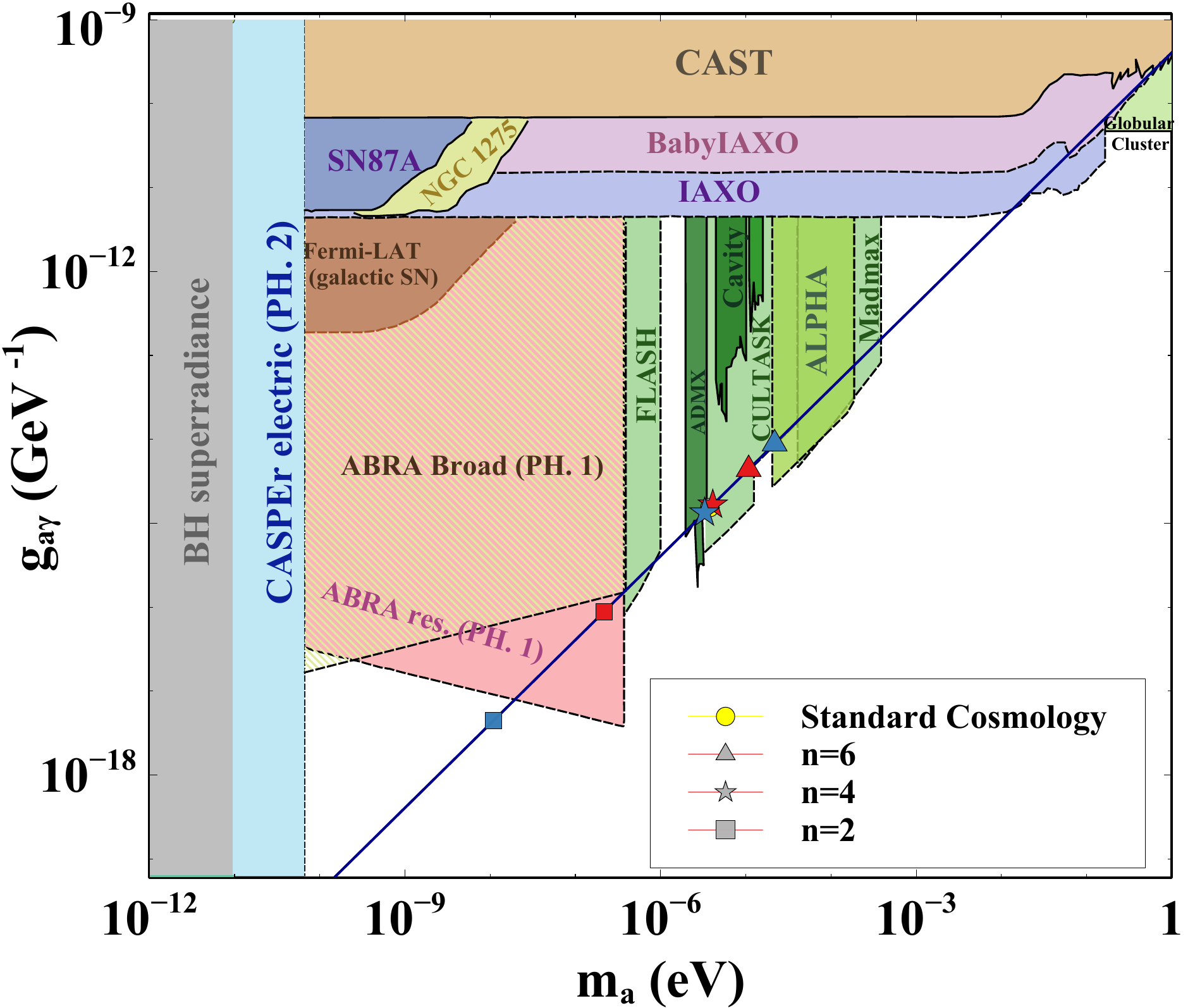}
        \caption{Fermionic (vacuum)}
    \end{subfigure}

    \vspace{0.3cm}

    \begin{subfigure}{0.49\linewidth}
        \centering
        \includegraphics[width=\linewidth]{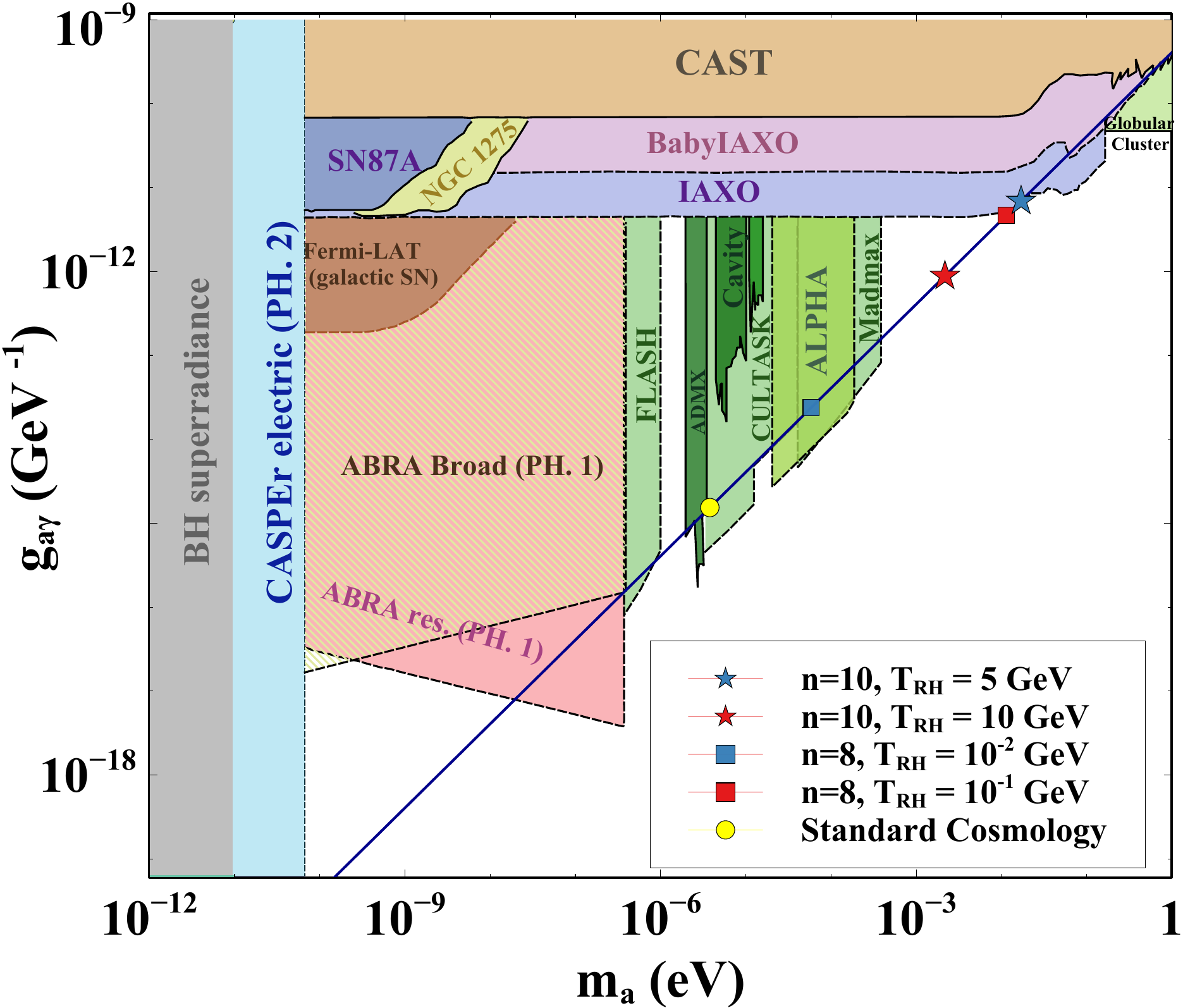}
        \caption{Bosonic (kinetic)}
    \end{subfigure}
    \hfill
    \begin{subfigure}{0.49\linewidth}
        \centering
        \includegraphics[width=\linewidth]{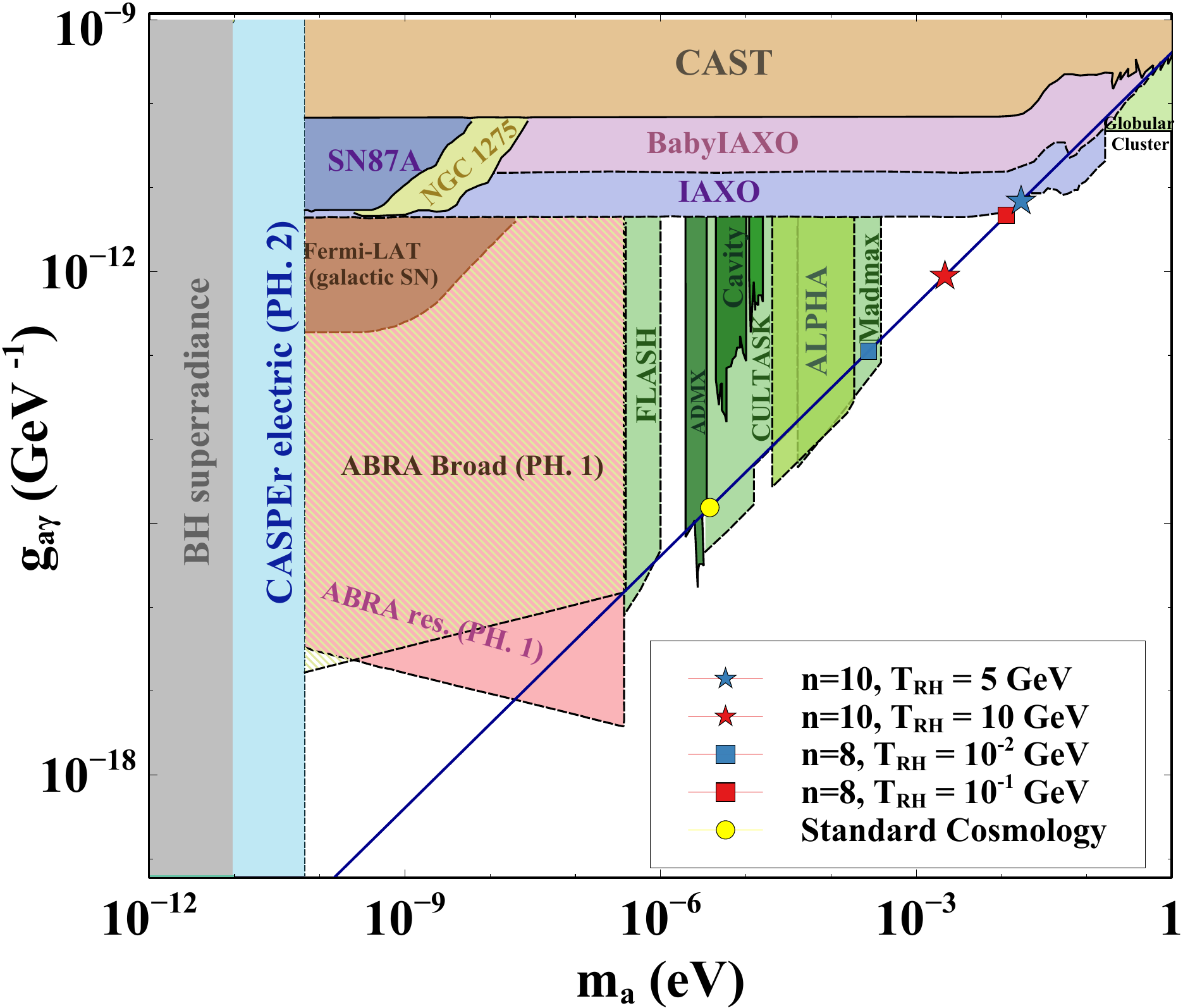}
        \caption{Fermionic (kinetic)}
    \end{subfigure}

    \caption{
    Axion-photon coupling $g_{a\gamma}$ as a function of the axion mass $m_a$ for the vacuum misalignment (top panels) and kinetic misalignment (bottom panels) mechanisms. The left and right panels correspond to reheating through bosonic and fermionic decays, respectively. In the top panels, the different marker shapes indicate the choice of the inflaton potential $V(\phi)\propto\phi^n$ and the red and blue markers correspond to $T_{\rm RH}=0.1$ and $0.01\,{\rm GeV}$, respectively. In the bottom panels, the values of $n$ and $T_{\rm RH}$ are indicated explicitly in the legend. The yellow $\circ$ denote the axion-photon coupling and mass corresponding to the observed dark matter abundance from the conventional vacuum misalignment mechanism in the standard cosmological history.
    }
    \label{fig:det1}
\end{figure}

\subsection{Axion-Photon coupling}

\begin{figure}
    \centering
      \captionsetup[subfigure]{labelformat=simple}
    \begin{subfigure}{0.49\linewidth}
        \centering
        \includegraphics[width=\linewidth]{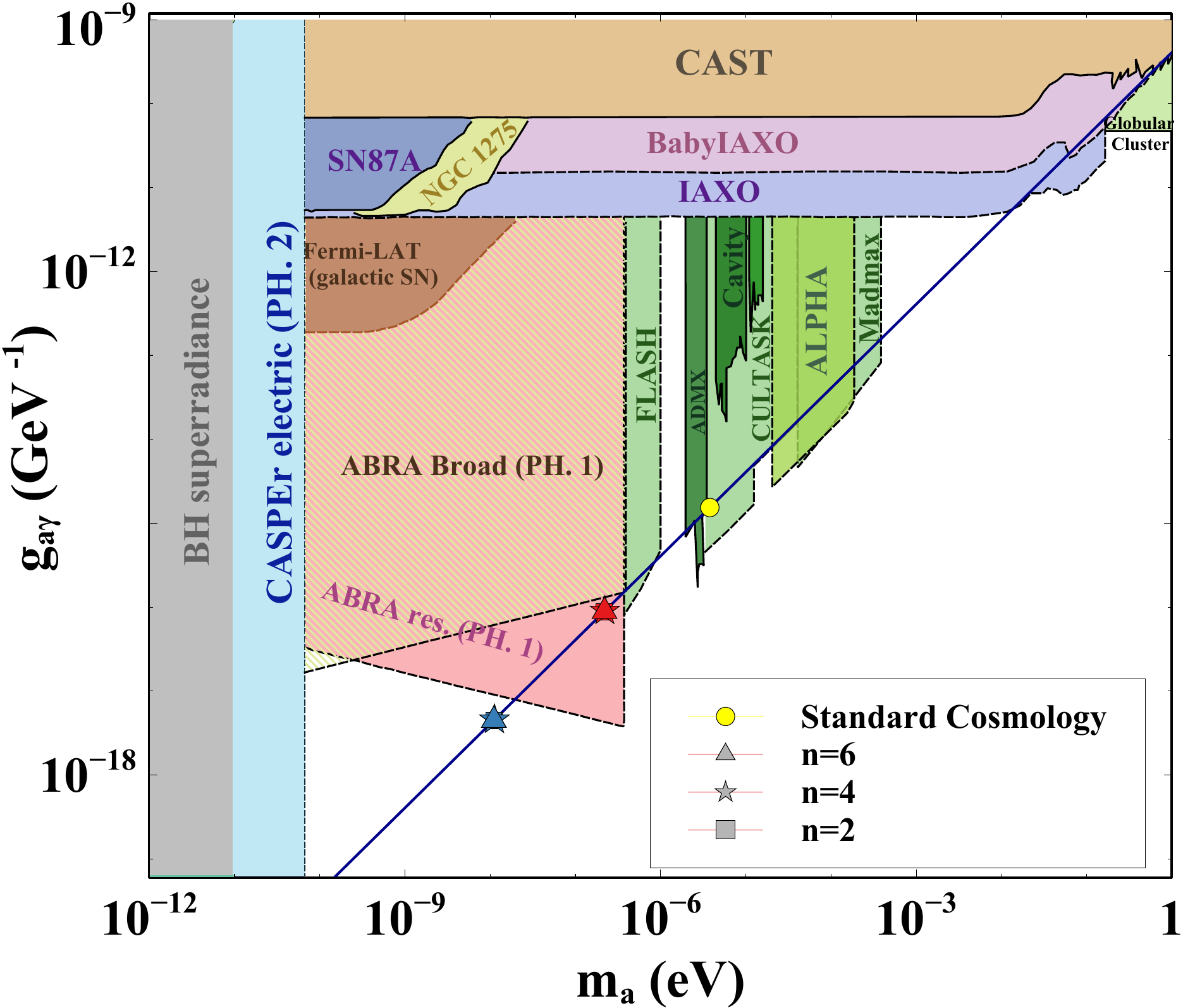}
        \caption{Seesaw reheating BP1 (vacuum)}
    \end{subfigure}
    \hfill
    \begin{subfigure}{0.49\linewidth}
        \centering
        \includegraphics[width=\linewidth]{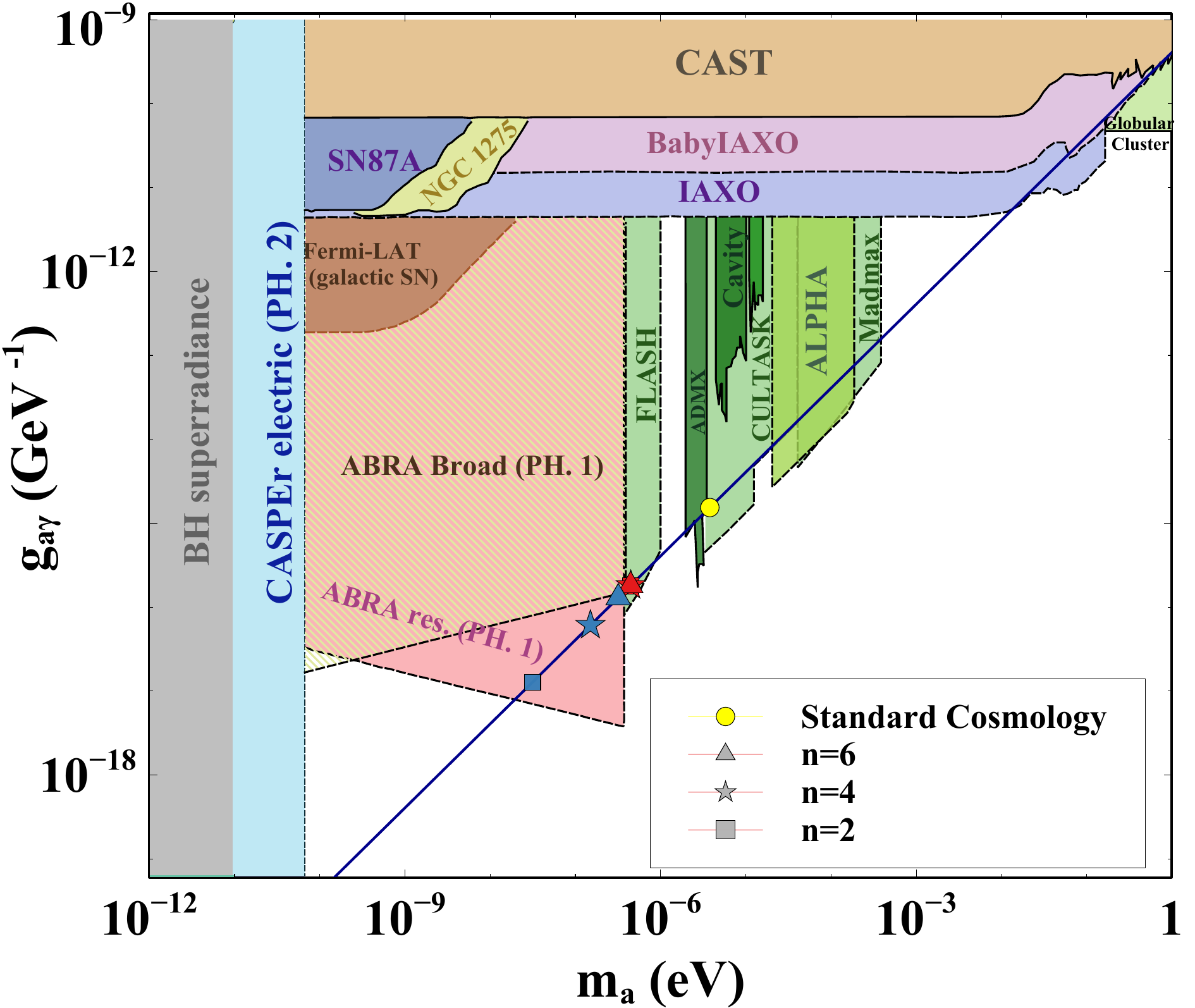}
        \caption{Seesaw reheating BP2 (vacuum)}
    \end{subfigure}

    \vspace{0.3cm}

    \begin{subfigure}{0.49\linewidth}
        \centering
        \includegraphics[width=\linewidth]{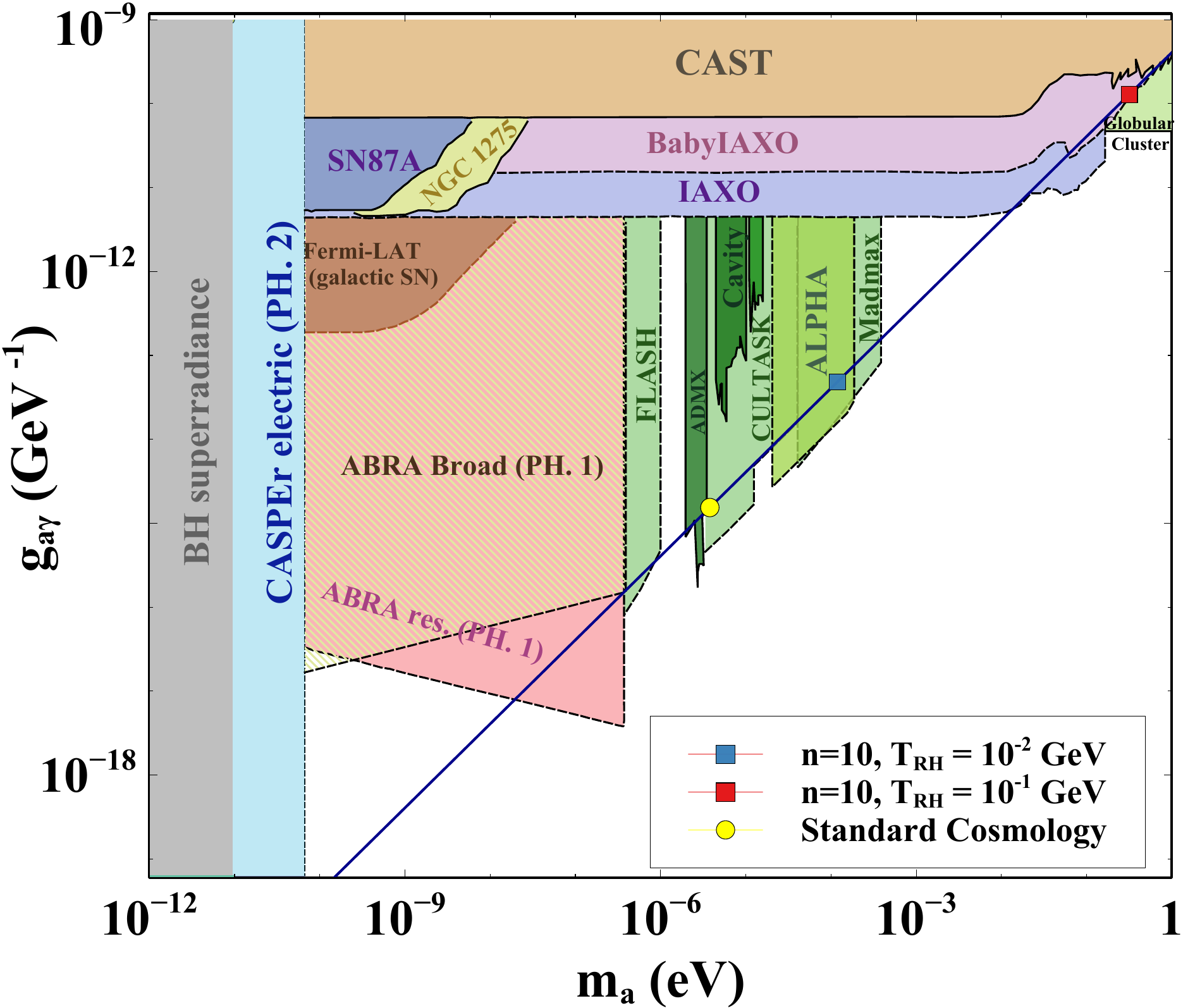}
        \caption{Seesaw reheating (kinetic)}
    \end{subfigure}
%    \hfill
%    \begin{subfigure}{0.49\linewidth}
%        \centering
%        \includegraphics[width=\linewidth]{Figure/det4}
%        \caption{Fermionic (kinetic)}
 %   \end{subfigure}

    \caption{
    Axion-photon coupling $g_{a\gamma}$ as a function of the axion mass $m_a$ for the vacuum misalignment (top panels) and kinetic misalignment (bottom panel) mechanisms in the seesaw reheating scenario. The left and right top panels correspond to
    BP1: $T_{\rm NR}=10^3T_{\rm RH}$ and $T_{\rm max}=10^4T_{\rm RH}$, and
    BP2: $T_{\rm NR}=2T_{\rm RH}$ and $T_{\rm max}=4T_{\rm RH}$, respectively.
    In the top panels, the different marker shapes indicate the choice of the inflaton potential $V(\phi)\propto\phi^n$ and the red and blue markers correspond to $T_{\rm RH}=0.1$ and $0.01\,{\rm GeV}$, respectively.
    The bottom panel shows the result for $T_{\rm NR}=10^3T_{\rm RH}$ and $T_{\rm max}=10^4T_{\rm RH}$, with the values of $n$ and $T_{\rm RH}$ indicated explicitly in the legend. The yellow $\circ$ denotes the axion-photon coupling and mass corresponding to the observed dark matter abundance from the conventional vacuum misalignment mechanism in the standard cosmological history.
    }
    \label{fig:RHN_det}
\end{figure}

The axion detection prospects in most experiments primarily rely on axion-photon coupling. The axion-photon interaction is parametrized as 
\begin{equation}
    \mathcal{L}_{a\gamma} \supset \frac{1}{4} g_{a \gamma } F_{\mu \nu} \tilde{F}^{\mu \nu},
\end{equation}
where, $g_{a \gamma } = (\frac{E}{N} - 1.92)\frac{\alpha}{2 \pi f_a}$ \cite{Srednicki:1985xd} with $E, N$ representing electromagnetic and color anomaly coefficients, respectively. Fig. \ref{fig:det1} and \ref{fig:RHN_det} show the parameter space in the plane of axion-photon coupling $g_{a\gamma}$ and axion mass $m_a$ including the bounds and sensitivities of different experiments or observables. The line labeled as KSVZ refers to the model's predicted axion-photon coupling. The current experimental bounds on the axion-photon coupling from various experiments or observables are shown by the solid colored contours (CAST \cite{CAST:2017uph}, SN87A \cite{Turner:1987by, Burrows:1988ah, Raffelt:1987yt}, NGC 1275 \cite{Fermi-LAT:2016nkz}, ADMX \cite{ADMX:2006kgb, ADMX:2019uok}, Globular clusters \cite{Ayala:2014pea}) whereas future experimental sensitivities or observables are shown by the dashed contours (CASPEr \cite{Budker:2013hfa}, FLASH \cite{Alesini:2023qed}, ABRACADABRA \cite{Ouellet:2018beu}, CULTASK \cite{Lee:2020cfj, Semertzidis:2019gkj}, MADMAX \cite{Caldwell:2016dcw}, IAXO \cite{IAXO:2019mpb},  Fermi-LAT \cite{Meyer:2016wrm}, ALPHA  \cite{PhysRevLett.123.141802, PhysRevD.107.055013}, black hole (BH) superradiance \cite{Cardoso:2018tly}).

%\begin{figure}
%    \centering
%    \includegraphics[width=0.49\linewidth]{Figure/RHN_det1.pdf}
%    \includegraphics[width=0.49\linewidth]{Figure/RHN_det2.pdf}
%    \includegraphics[width=0.49\linewidth]{Figure/RHN_det3.pdf}

%    \caption{
%    Axion-photon coupling $g_{a\gamma}$ as a function of the axion mass $m_a$ for the vacuum misalignment (top panels) and kinetic misalignment (bottom panel) mechanisms in the seesaw reheating scenario. The left and right top panels correspond to    $T_{\rm NR}=10^3T_{\rm RH}$ and $T_{\rm max}=10^4T_{\rm RH}$, and $T_{\rm NR}=2T_{\rm RH}$ and $T_{\rm max}=4T_{\rm RH}$, respectively. In the top panels, the different marker shapes indicate the choice of the inflaton potential $V(\phi)\propto\phi^n$ and the red and blue markers correspond to $T_{\rm RH}=0.1$ and $0.01\,{\rm GeV}$, respectively. The bottom panel shows the result for $T_{\rm NR}=10^3T_{\rm RH}$ and $T_{\rm Max}=10^4T_{\rm RH}$, with the values of $n$ and $T_{\rm RH}$ indicated explicitly in the legend. The yellow $\circ$ denotes the axion-photon coupling and mass corresponding to the observed dark matter abundance from the conventional vacuum misalignment mechanism in the standard cosmological history.}
%    \label{fig:RHN_det}
%\end{figure}

The top and bottom panels of Figs.~\ref{fig:det1} and \ref{fig:RHN_det} correspond to vacuum and kinetic misalignment mechanisms respectively. For Fig.~\ref{fig:det1}, the left and right panels correspond to bosonic and fermionic reheating, respectively, while Fig.~\ref{fig:RHN_det} shows the corresponding results for the seesaw reheating scenario. In the panels for vacuum-misalignment scenarios, the different marker shapes indicate the choice of the inflaton potential $V(\phi)\propto\phi^n$, with the red and blue markers corresponding to $T_{\rm RH}=0.1$ and $0.01\,{\rm GeV}$, respectively. In the panels for kinetic-misalignment scenario, the values of $n$ and $T_{\rm RH}$ are indicated explicitly in the legend. The yellow $\circ$ denotes the axion-photon coupling and mass corresponding to the observed dark matter abundance from the conventional misalignment mechanism (vacuum or kinetic) in the standard cosmological history. For different marker points in Figs.~\ref{fig:det1} and \ref{fig:RHN_det} consistent with the observed DM relic, the axion-photon coupling is considered to be the model-independent one which scales with $f_a$ and hence with axion mass $m_a$. 

\begin{figure}
    \centering
    \captionsetup[subfigure]{labelformat=simple}
    \begin{subfigure}[b]{0.49\linewidth}
        \centering
        \includegraphics[width=\linewidth]{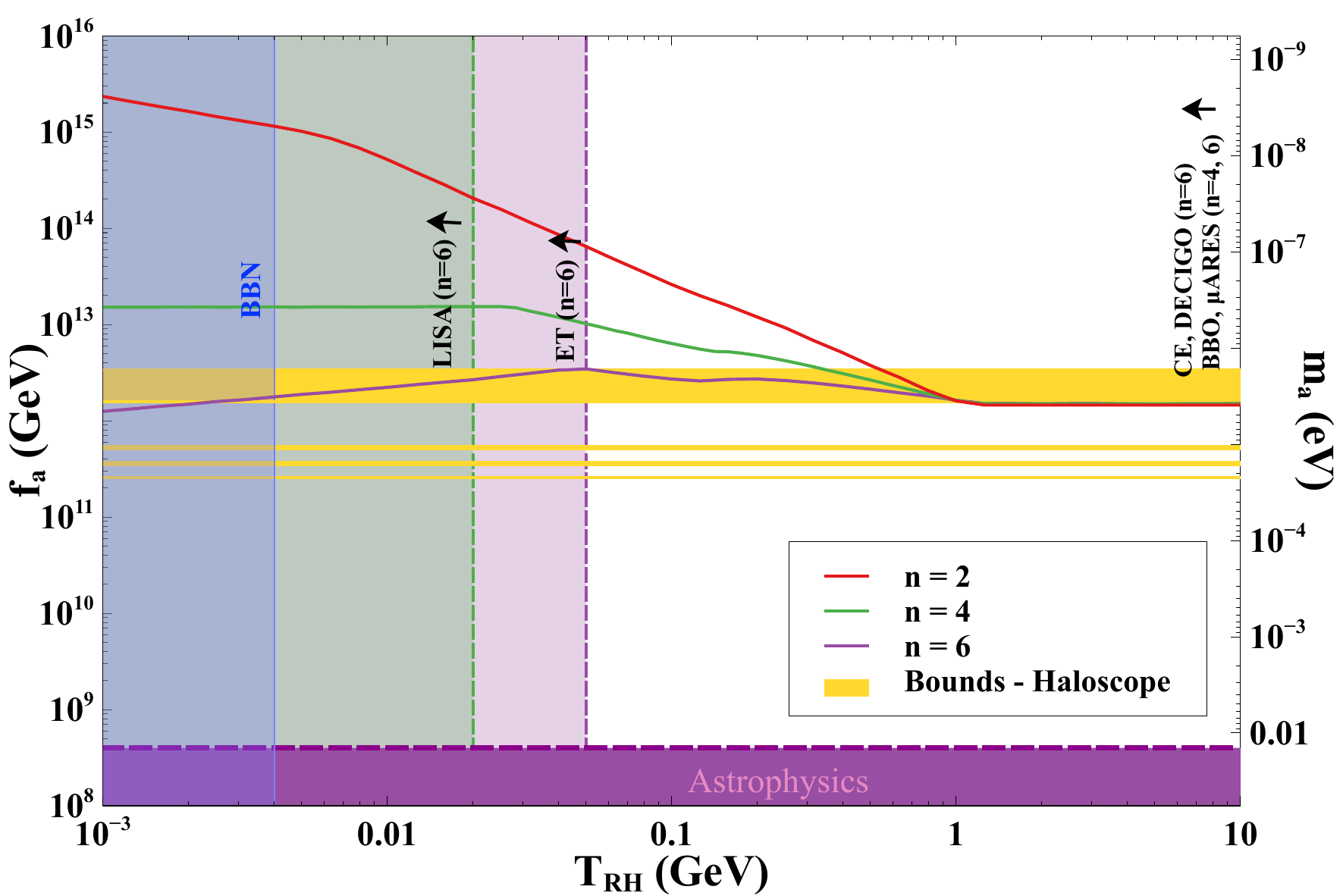}
        \caption{Bosonic reheating (vacuum)}
        \label{fig:axion_evo_bd_kin}
    \end{subfigure}\hfill
    \begin{subfigure}[b]{0.49\linewidth}
        \centering
        \includegraphics[width=\linewidth]{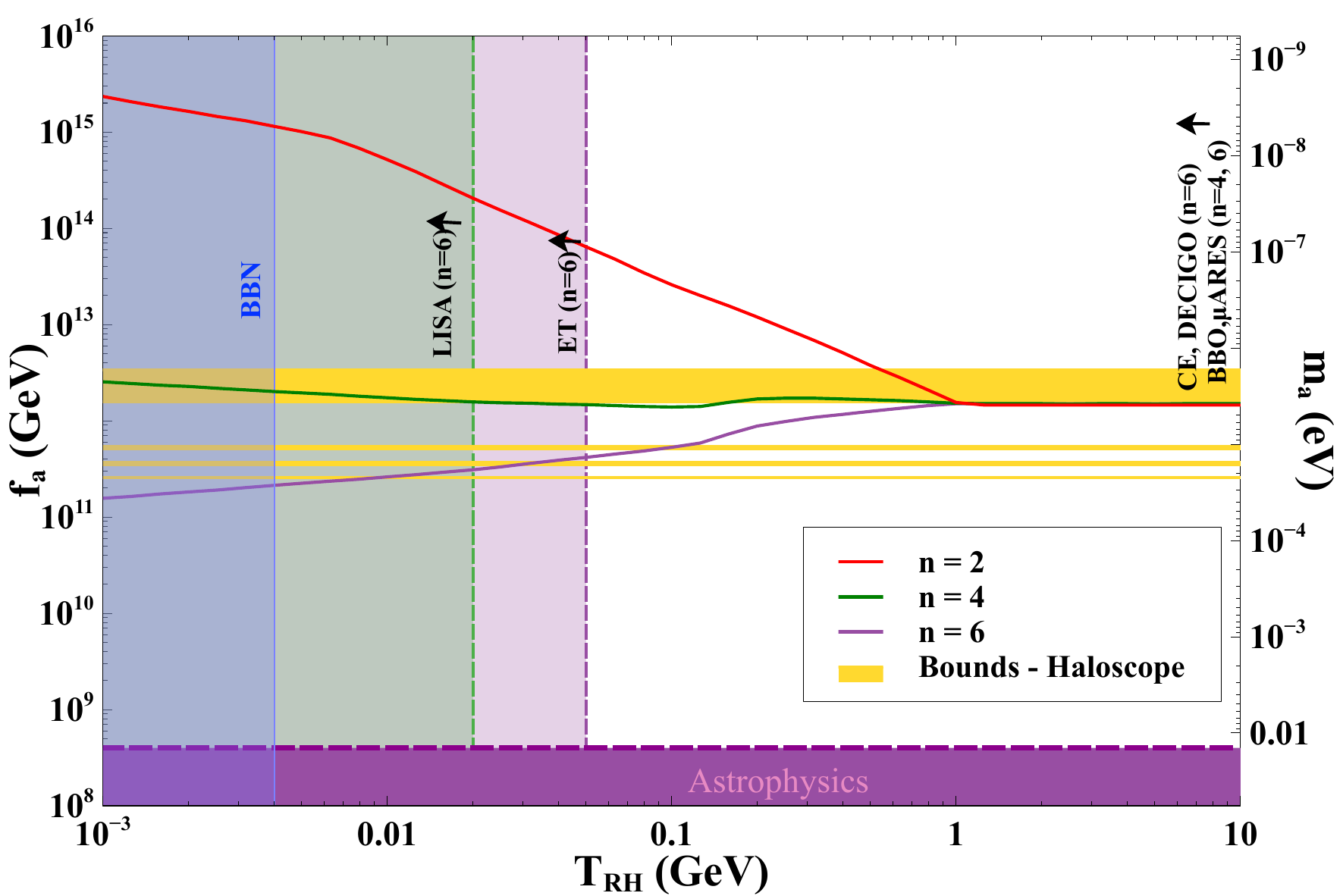}
        \caption{Fermionic reheating (vacuum)}
        \label{fig:axion_evo_fd_kin}
    \end{subfigure}
\caption{Parameter space in the $T_{\text{RH}}$--$f_a$ plane for standard vacuum misalignment with bosonic (left) and fermionic (right) reheating. Solid contours correspond to the observed 100\% axion relic abundance ($\Omega_a h^2 \approx 0.12$) for different $n$. Vertical shaded regions indicate future GW sensitivities, while the horizontal yellow shaded region indicates existing bounds from haloscope searches. See text for details.}
    \label{fig:sum_vac}
\end{figure}

For the direct reheating scenario shown in Fig.~\ref{fig:det1}, vacuum misalignment generally requires a lower axion mass than the standard-cosmology prediction, except for the fermionic decay case with $n=6$, whereas kinetic misalignment requires a higher axion mass. The dependence of the required axion mass on $n$ and $T_{\rm RH}$ follows the corresponding relic-abundance behavior discussed in Figs.~\ref{fig:sum_vac} and \ref{fig:summary_kin}. For the indirect or seesaw reheating scenario in Fig.~\ref{fig:RHN_det}, the left and right top panels correspond to $T_{\rm NR}=10^3 \, T_{\rm RH}$, $T_{\rm max}=10^4 \, T_{\rm RH}$ and $T_{\rm NR}=2 \,T_{\rm RH}$, $T_{\rm max}=4\,T_{\rm RH}$, respectively. For the chosen benchmark in top left panel, the vacuum misalignment result is essentially independent of $n$ and is primarily controlled by $T_{\rm RH}$. For the benchmark in top right panel, the required axion mass is lower than the one in standard cosmology and depends on both $n$ and $T_{\rm RH}$, consistent with the behavior observed in Fig.~\ref{fig:RHN_sum}. For kinetic misalignment shown in the bottom panel, the required axion mass is again higher than the standard-cosmology vacuum-misalignment prediction. In particular, the parameter points corresponding to $T_{\rm RH}=0.1~{\rm GeV}$ are excluded, owing to the higher axion masses required in this case, as can also be understood from Fig.~\ref{fig:RHN_kin_sum}.

%%\begin{figure}
%    \centering
%    \includegraphics[width=0.49\linewidth]{Figure/summary1.pdf}
%%    \includegraphics[width=0.49\linewidth]{Figure/summary2.pdf}
 %   \caption{Parameter space in the $T_{\text{RH}}$--$f_a$ plane for standard vacuum misalignment with bosonic (left) and fermionic (right) reheating. Solid contours correspond to the observed 100\% axion relic abundance ($\Omega_a h^2 \approx 0.12$) for different $n$. Vertical shaded regions indicate future GW sensitivities, while horizontal shaded regions show bounds and projections from axion experiments. See text for details.}
%\label{fig:sum_vac}
%\end{figure}

\section{Summary and Conclusion}
\label{sec:conclude}
In this section, we first summarize our key results for both vacuum and kinetic misalignment mechanisms while considering all three possible reheating scenarios.

% \begin{figure}
%     \centering
%     \includegraphics[width=0.48\linewidth]{Figure/summary_n2.pdf}
%     \includegraphics[width=0.48\linewidth]{Figure/summary_n46.pdf}
%      \includegraphics[width=0.48\linewidth]{Figure/summary_n46FD.pdf}

%     \caption{\cite{adams2023axiondarkmatter}}
%     \label{fig:Tosc}
% \end{figure}

Fig. \ref{fig:sum_vac} shows the summary of vacuum misalignment with direct reheating. We show the parameter space in the $T_{\rm RH}$--$f_a$ plane for bosonic and fermionic reheating in the left and right panels, respectively. The solid curves correspond to the parameter choices yielding the observed axion relic abundance, $\Omega_a h^2\simeq0.12$, for different values of $n$. The QCD axion mass $m_a$, which is related to $f_a$ through Eq. \eqref{axion_mass}, is shown on the right vertical axis. The entire parameter space depicted lies within the projected
sensitivity of searches based on the axion--photon coupling, while the horizontal yellow shaded region indicates existing
bounds from haloscope searches. Projected sensitivities from different GW experiments are shown as dashed vertical contours with the leftward arrow pointing to the region within reach of a particular GW detector for a particular inflaton potential.
% The horizontal shaded regions indicate constraints and projected sensitivities from axion searches, while the vertical shaded regions correspond to constraints and sensitivities from GW experiments which depend on the reheating temperature.
% In particular, the horizontal regions include the QCD-axion parameter ranges targeted by different axion searches with the yellow region corresponding to bounds from haloscope searches, sensitivities of proposed searches using the axion--photon coupling, and sensitivities of proposed searches exploiting alternative axion couplings, respectively.
The purple region at the bottom denotes astrophysical constraints, while the vertical shaded region towards left corresponds to the lower bound on reheating temperature from BBN. Since the axion experimental and astrophysical constraints primarily depend on $f_a$ (or equivalently $m_a$), whereas the GW sensitivities depend predominantly on $T_{\rm RH}$, these experiments probe complementary directions in the parameter space. For $n=6$, the entire parameter space can be probed by BBO, $\mu$ARES, CE, and DECIGO, while the regions with $T_{\rm RH}<0.05$ GeV and $T_{\rm RH}<0.02$ GeV can be probed by ET and LISA, respectively. For $n=4$, owing to the nearly flat nature of the GW spectrum, BBO and $\mu$ARES are sensitive to the entire parameter space. In contrast, for $n=2$, the red-tilted GW spectrum significantly suppresses the signal at higher frequencies, rendering the entire parameter space inaccessible to future GW experiments. We find that a significant portion of the parameter space yielding the observed axion relic abundance is either excluded by existing constraints or lies within projected axion sensitivities. For bosonic reheating, a large fraction of the $n=6$ parameter space is excluded, together with smaller portions for $n=2$ and $n=4$. For fermionic reheating, the entire $n=4$ parameter space is excluded, while parts of the $n=2$ and $n=6$ regions are also constrained. Thus, requiring the axion to constitute the full dark matter abundance in an extended reheating scenario already allows us to constrain several choices of the reheating history and inflaton potential.

\begin{figure}
    \centering
    \captionsetup[subfigure]{labelformat=simple}
    \begin{subfigure}[b]{0.49\linewidth}
        \centering
        \includegraphics[width=\linewidth]{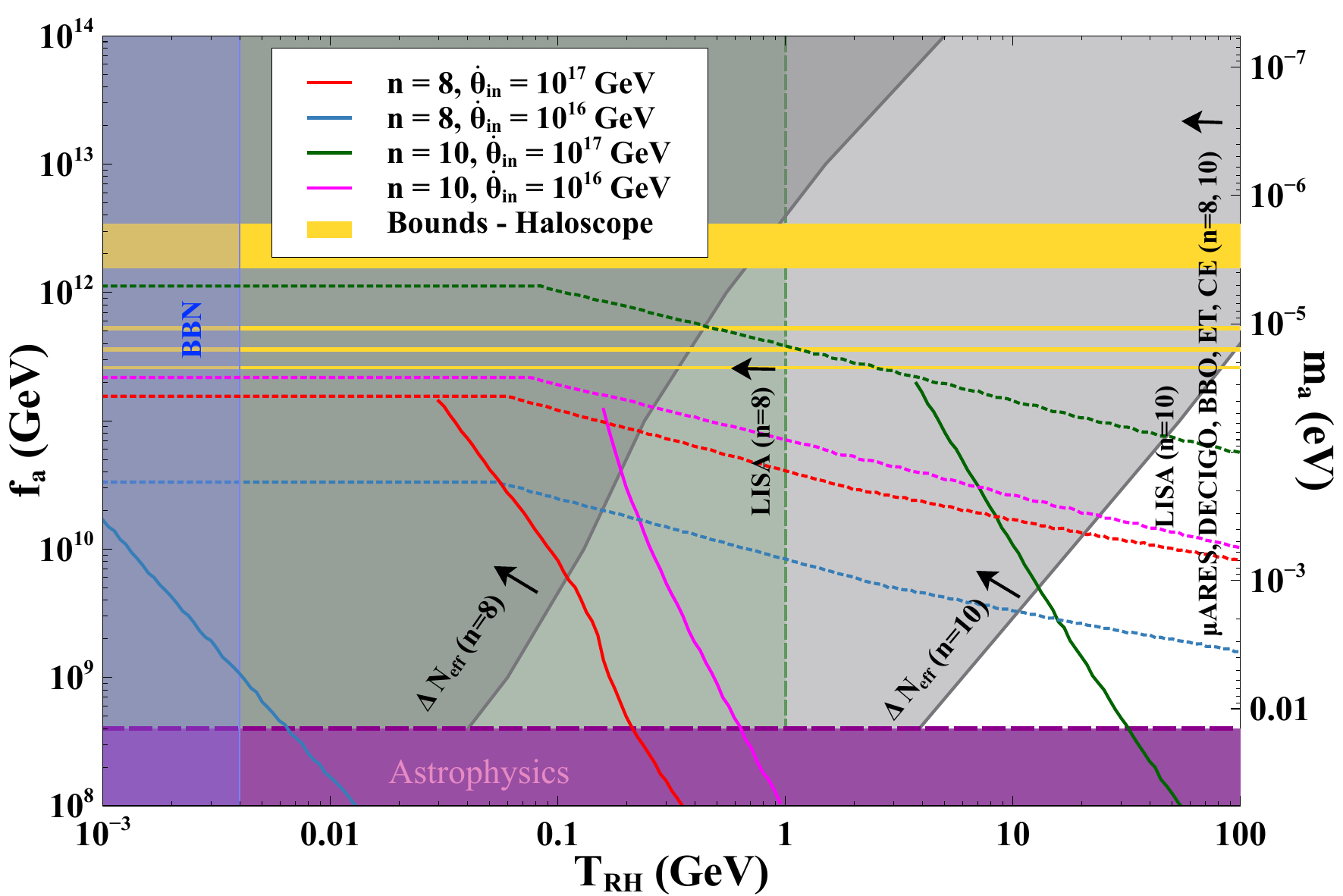}
        \caption{Bosonic reheating (kinetic)}
        \label{sfig:17a}
    \end{subfigure}\hfill
    \begin{subfigure}[b]{0.49\linewidth}
        \centering
        \includegraphics[width=\linewidth]{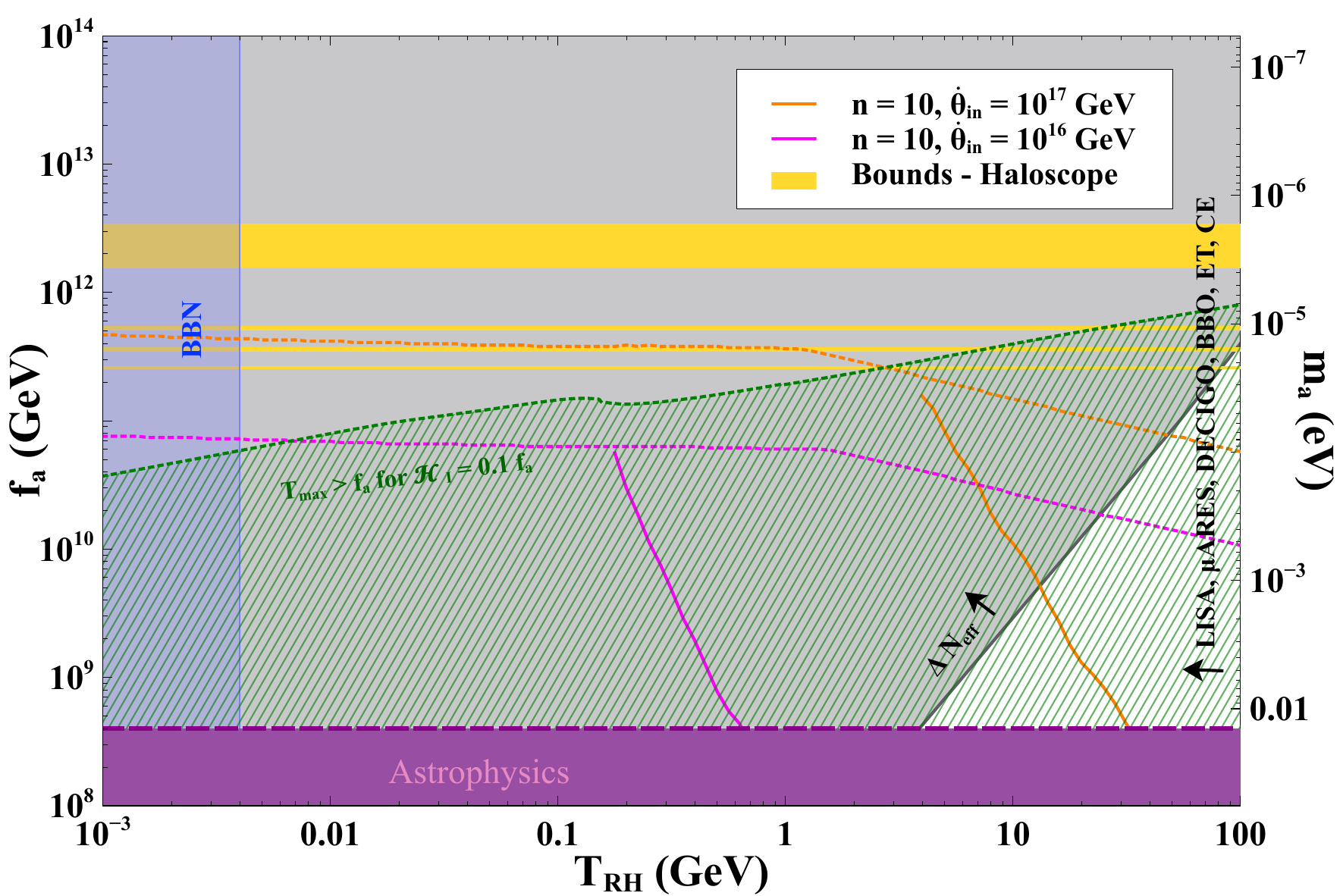}
        \caption{Fermionic reheating (kinetic)}
        \label{sfig:17b}
    \end{subfigure}
    \caption{Parameter space in the $T_{\text{RH}}$--$f_a$ plane for kinetic misalignment with bosonic (left) and fermionic (right) reheating. Solid contours correspond to the observed 100\% axion relic abundance ($\Omega_a h^2 \approx 0.12$)  for benchmark values of $\dot{\theta}$ and $n$. Vertical shaded regions indicate future GW and sensitivities, while the horizontal yellow shaded region indicates existing
bounds from haloscope searches. See text for details related to other contours and bands. }
    \label{fig:summary_kin}
\end{figure}

%\begin{figure}
%    \centering
%    \includegraphics[width=0.49\linewidth]{Figure/summary3.pdf}
%       \includegraphics[width=0.49\linewidth]{Figure/sum_Kin_FD.pdf}
%    \caption{Parameter space in the $T_{\text{RH}}$--$f_a$ plane for kinetic misalignment with bosonic (left) and fermionic (right) reheating. Solid contours correspond to the observed 100\% axion relic abundance ($\Omega_a h^2 \approx 0.12$)  for benchmark values of $\dot{\theta}$ and $n$. Vertical shaded regions indicate future GW and sensitivities, while horizontal shaded regions show bounds and projected reach from axion experiments. See text for details related to other contours and bands.}
%    \label{fig:summary_kin}
%\end{figure}

\begin{figure}
    \centering
    \includegraphics[width=0.49\linewidth]{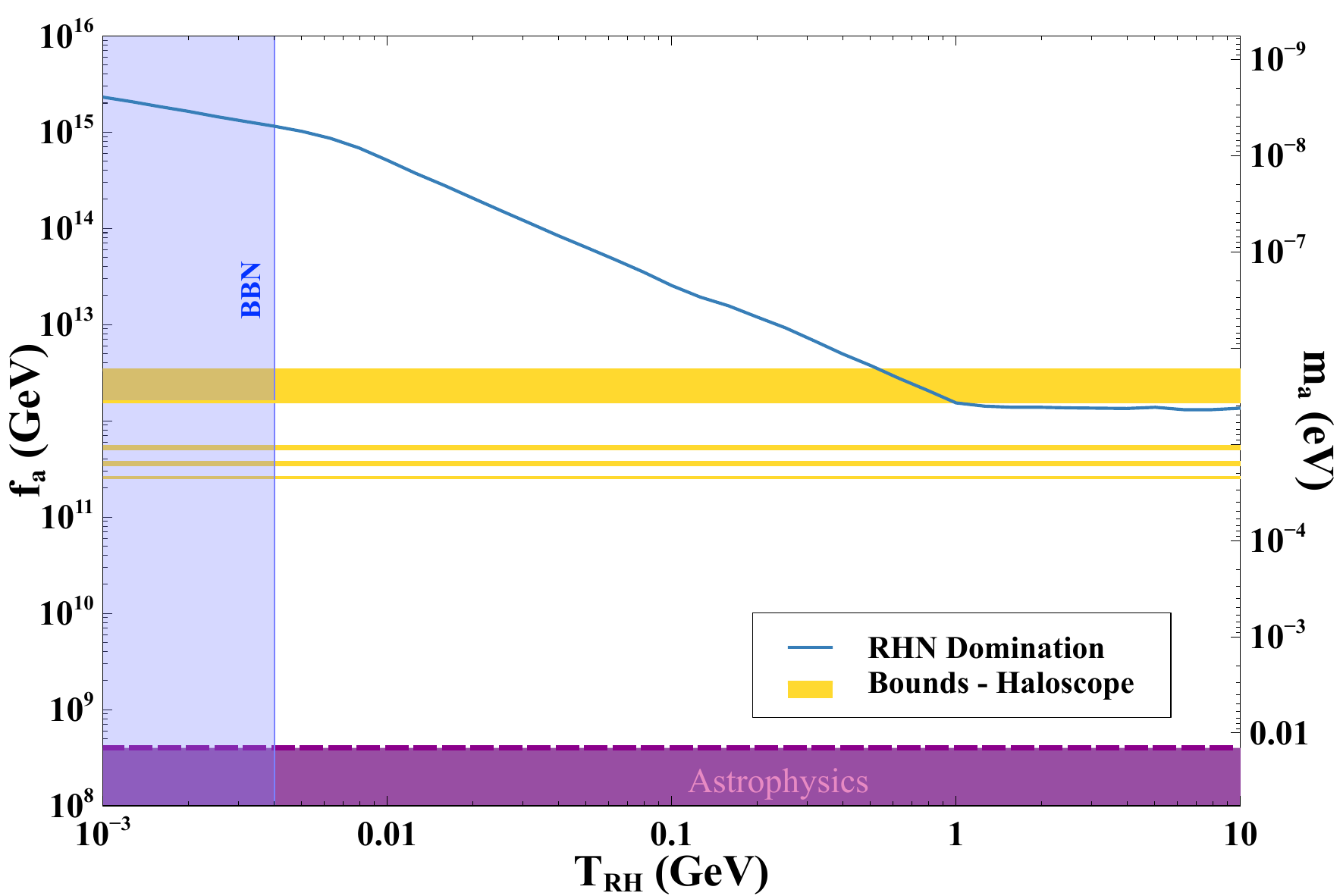}
    \includegraphics[width=0.49\linewidth]{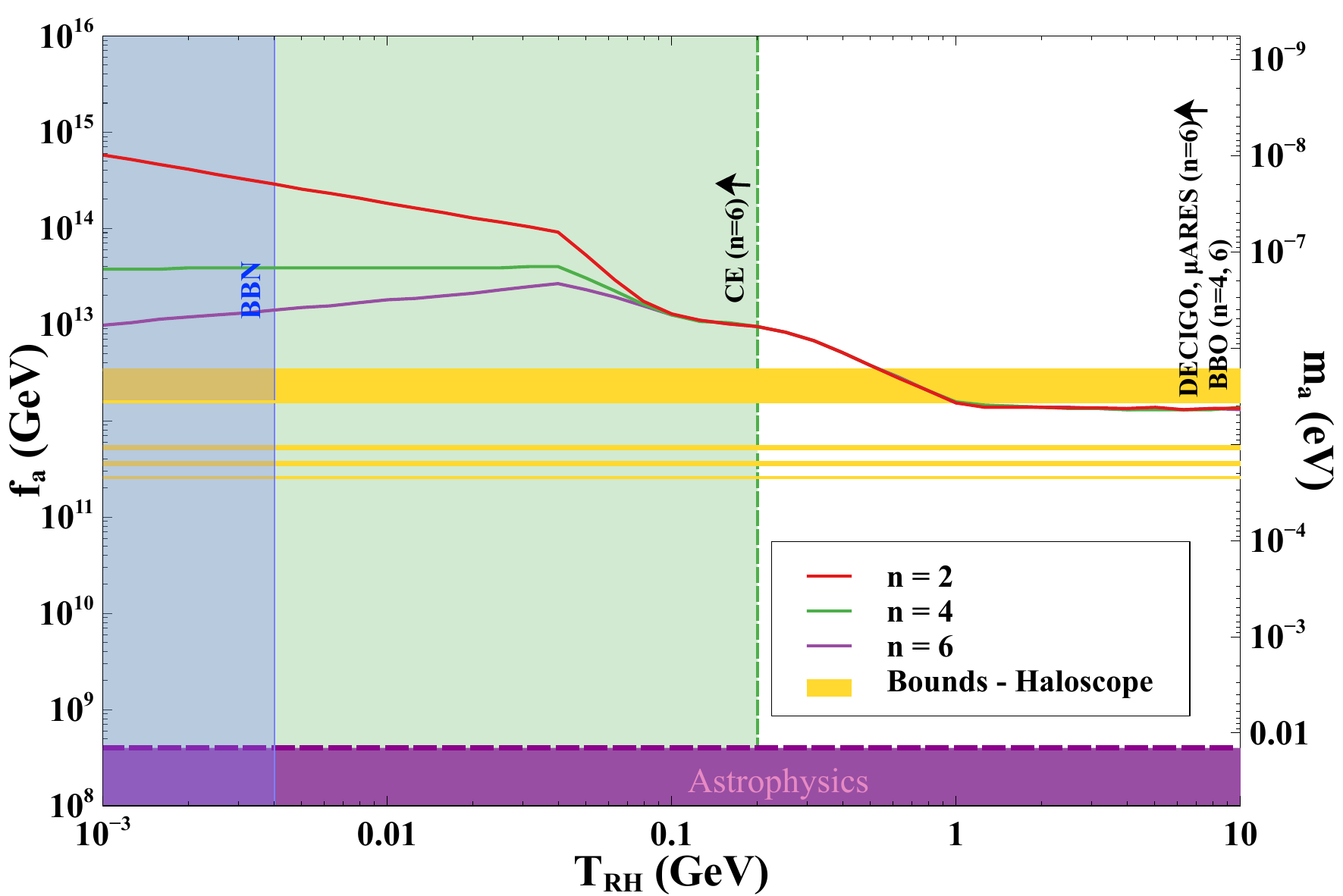}
    \caption{Parameter space in the $T_{\text{RH}}$--$f_a$ plane for standard vacuum misalignment with seesaw reheating for $T_{\rm NR}=10^3\,T_{\rm RH}$, $T_{\rm max}=10^4 \,T_{\rm RH}$ (left) and $T_{\rm NR}=2\,T_{\rm RH}$, $T_{\rm max}=4 \,T_{\rm RH}$ (right). Solid contours correspond to the observed 100\% axion relic abundance ($\Omega_a h^2 \approx 0.12$) for different $n$. Vertical shaded regions indicate future GW sensitivities, while the horizontal yellow shaded region indicates existing
bounds from haloscope searches.}
    \label{fig:RHN_sum}
\end{figure}

In Fig.~\ref{fig:summary_kin}, we show the parameter space for kinetic misalignment for the two choices of the initial axion velocity, $\dot{\theta}_i$, and for $n=8$ and $10$, as indicated in the legend. The axion-related experimental constraints and sensitivities are the same as those discussed for Fig.~\ref{fig:sum_vac}. We further show the exclusion regions arising from the contribution of GWs to $N_{\rm eff}$. These regions are represented by gray-shaded areas bounded by solid gray lines. To distinguish the projected sensitivities of future GW experiments from the current bounds arising from the GW contribution to $N_{\rm eff}$, we use dashed lines in Fig. \ref{fig:sum_vac} and solid lines in Fig. \ref{fig:summary_kin}, respectively. Since the GW spectrum is more strongly blue-tilted for $n=10$ than for $n=8$, the resulting constraint is correspondingly stronger for $n=10$. Throughout this analysis, we fix the inflationary Hubble scale to $\mathcal{H}_I=0.1f_a$. The dotted curves indicate the boundary above which, for a given $\dot{\theta}_i$, the axion abundance is dominated by the vacuum misalignment contribution rather than kinetic misalignment; consequently, the kinetic-misalignment preferred parameter-space contours terminate at these boundaries. The requirement of satisfying the GW contribution to $N_{\rm eff}$ strongly constrains both $T_{\rm RH}$ and $f_a$ in the bosonic reheating case, with the surviving parameter space favoring a relatively heavier axion for the larger $n$. On the other hand, for fermionic reheating, with our benchmark choice of $\mathcal{H}_I=0.1f_a$, the $n=10$ case remains completely excluded for both the values of $\dot{\theta}_i$, primarily due to the symmetry non-restoration requirement $T_{\rm max}<f_a$. Higher values of $n$ may open up some parameter space; however, the corresponding contribution of GWs to $N_{\rm eff}$ is expected to provide a stronger constraint, while the behaviour of the $T_{\rm max}<f_a$ bound will become weaker. However, we restrict our study to $n\leq10$.

%For bosonic reheating, we find that the parameter space remains viable against the $N_{\rm eff}$ constraint only for the larger initial velocity, $\dot{\theta}_i=10^{17}~{\rm GeV}$, and even then only a small region survives. For $n=8$, the allowed region corresponds approximately to $T_{\rm RH}\sim0.1$--$0.2~{\rm GeV}$ and $f_a\sim4\times10^8$--$6\times10^9~{\rm GeV}$, while for $n=10$ it occurs at much higher reheating temperatures, $T_{\rm RH}\sim 13$--$30~{\rm GeV}$, with $f_a\sim4\times10^8$--$5\times10^9~{\rm GeV}$. Thus, the requirement of satisfying the GW contribution to $N_{\rm eff}$ strongly constrains both $T_{\rm RH}$ and $f_a$, with the surviving parameter space favouring a relatively heavier axion for the larger $n$. For fermionic reheating, for our benchmark choice of $\mathcal{H}_I=0.1f_a$, the $n=10$ case remains completely excluded for both considered values of $\dot{\theta}_i$, primarily due to the symmetry non-restoration requirement $T_{\rm max}<f_a$. Higher values of $n$ may open up some parameter space; however, the corresponding contribution of GWs to $N_{\rm eff}$ is expected to provide a stronger constraint, while the behaviour of the $T_{\rm max}<f_a$ bound wll become weaker. However effect of increasing n further has not been investigated in the present analysis. We therefore restrict our study to $n\leq10$.

\begin{figure}
    \centering
    \includegraphics[width=0.5\linewidth]{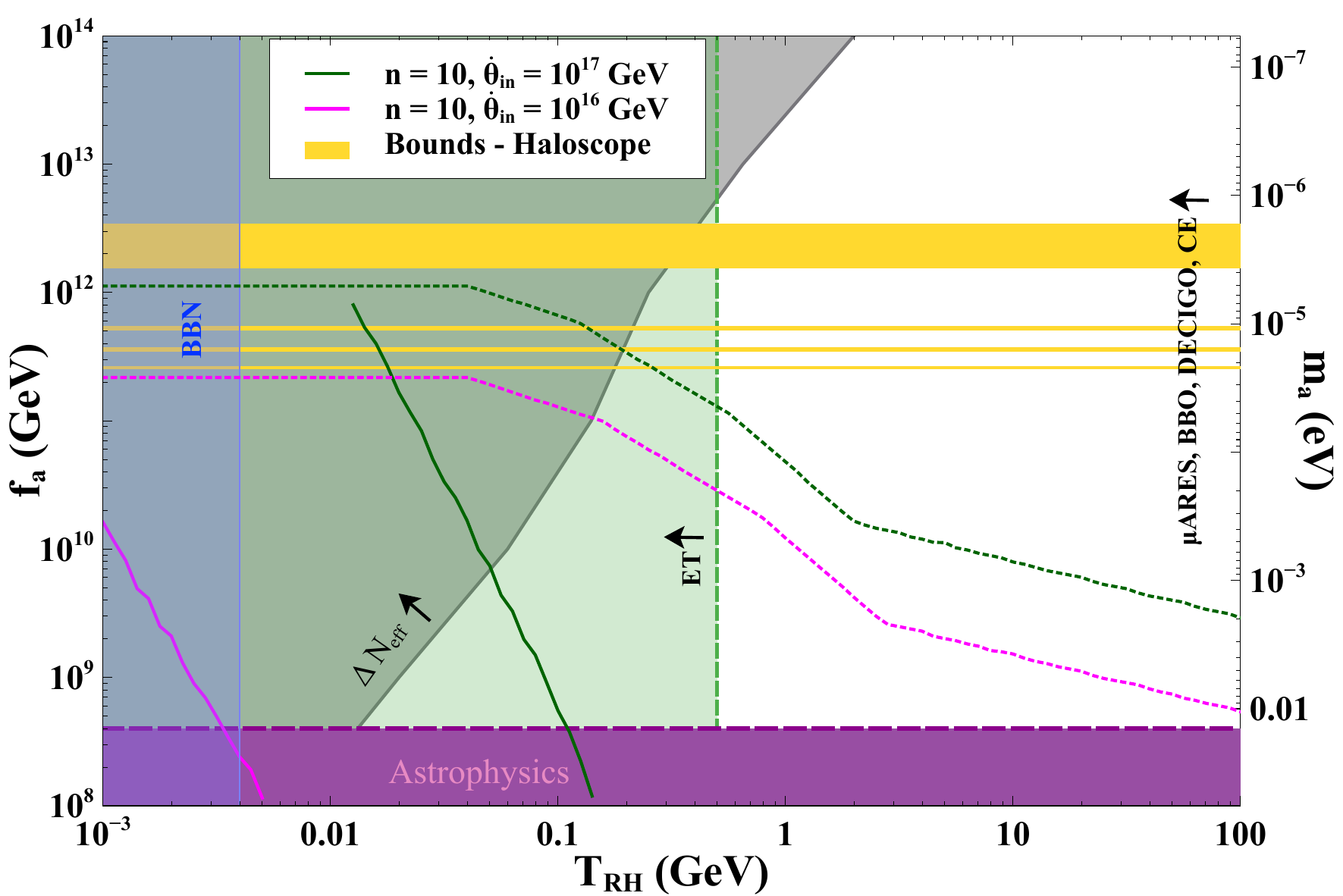}
    \caption{Parameter space in the $T_{\text{RH}}$--$f_a$ plane for standard kinetic misalignment with seesaw reheating for $T_{\rm NR}=2\,T_{\rm RH}$, $T_{\rm max}=4 \,T_{\rm RH}$. Solid curves correspond to the observed 100\% axion relic abundance ($\Omega_a h^2 \approx 0.12$) for benchmark values of $\dot{\theta}$ and $n$. Vertical shaded regions indicate future GW sensitivities, while the horizontal yellow shaded region indicates existing bounds from haloscope searches. See text for details related to other contours and bands.}
    \label{fig:RHN_kin_sum}
\end{figure}

For the seesaw reheating scenario, the corresponding parameter space for vacuum misalignment is shown in Fig.~\ref{fig:RHN_sum} for two different choices of the RHN evolution parametrized in terms of $T_{\rm NR}, T_{\rm max}$. The parameter space in the left panel lies beyond the reach of future GW experiments. This is primarily due to the prolonged period of non-relativistic RHN domination, which significantly red-tilts the PGW spectrum and suppresses the signal in the relevant frequency range. In this case, the axion starts oscillating during the non-relativistic RHN-dominated era, making the evolution qualitatively similar to that obtained for direct reheating with $n=2$. In the right panel, the RHN-dominated phase is relatively short. For larger $T_{\rm RH}$, axion oscillations occur during RHN domination, resulting in an approximately common relic-abundance curve for different values of $n$. At lower $T_{\rm RH}$, however, the onset of oscillations occurs during the $\phi^n$-dominated phase, leading to a dependence on $n$. 
%In contrast to the direct-reheating case shown in Fig.~\ref{fig:sum_vac}, the axion-experiment constraints are less restrictive here, with the parameter space for $n=4$ and $n=6$ largely avoiding the corresponding exclusions. 
The shorter RHN-dominated period also results in a less suppressed GW spectrum, allowing future experiments such as CE, DECIGO, BBO, and $\mu$ARES to probe a portion of the parameter space, as shown in the figure.

Finally, Fig.~\ref{fig:RHN_kin_sum} shows the parameter space for kinetic misalignment in the seesaw reheating scenario. The exclusion region arising from the contribution of GWs to $N_{\rm eff}$ lies towards the left of the grey colored contour. The remaining allowed parameter space can be probed by future GW observatories, including ET, CE, DECIGO, BBO, and $\mu$ARES. As discussed above, the dotted curves mark the boundary between the kinetic- and vacuum-misalignment regimes, with kinetic misalignment applicable below the curves; hence, the solid lines denoting the relic abundance terminate at these boundaries. As in Fig.~\ref{fig:summary_kin}, a viable region is obtained only for the larger initial velocity $\dot{\theta_i}$. Here, focusing on $n=10$, the combined requirement of satisfying the GW contribution to $N_{\rm eff}$ and obtaining the correct relic abundance through kinetic misalignment allows approximately $T_{\rm RH}\sim0.1$--$0.6~{\rm GeV}$ and $f_a\sim4\times10^8$--$7\times10^9~{\rm GeV}$. Compared with Fig.~\ref{fig:summary_kin}, the allowed range extends to lower reheating temperatures.

To conclude, we have studied the details of QCD axion misalignment via two of the popular mechanisms namely, the vacuum misalignment and kinematic misalignment during an extended period of reheating after the slow-roll inflation. The inflaton field with a monomial potential dominates the energy density of the Universe followed by reheating the Universe by dissipating the energy of the inflaton condensate into radiation directly or indirectly via RHN decay. After finding the allowed parameter space in terms of axion mass or equivalently the axion decay constant, the monomial inflaton potential and reheating temperature from DM, cosmology and other laboratory constraints, we also check the possibility of probing these scenarios in future GW and axion detection experiments. While most part of the currently allowed parameter space can be probed at a variety of axion experiments, some of the scenarios can also be probed at future GW experiments. In particular, vacuum misalignment with all three types of reheating namely, bosonic, fermionic and seesaw reheating can be probed at different GW experiments for specific combinations of $T_{\rm RH}$ and $n$. The kinetic misalignment scenario is found to be more constrained and require higher values of $n$ for correct DM relic. These also invite stronger $N_{\rm eff}$ constraints from overproduction of PGW. In fact, the fermionic reheating scenario with kinetic misalignment remains disfavored even with $n=10$. Therefore, such extended reheating era with low reheating temperatures offer unique predictions for axion and GW experiments if axion constitutes the DM in the Universe. It also helps in discriminating between different axion production mechanisms as well as the reheating modes. The seesaw reheating scenario also offers additional phenomenology in the form of heavy RHNs and their roles in generating light neutrino masses and baryon asymmetry of the Universe via leptogenesis. Due to low reheating temperatures much below the sphaleron decoupling temperature, such scenarios can be tightly constrained from combined requirements of axion DM and successful leptogenesis. Apart from these new phenomenological possibilities, the predictions of our scenarios can be made more robust by considering the role of non-perturbative particle production during reheating, as studied in several earlier works \cite{Kofman:1985aw,Kofman:1997yn,Greene:1998nh, Drewes:2017fmn,Drewes:2019rxn}. We leave such detailed theoretical and phenomenological studies to future works.

%In this work, we consider a single right-handed neutrino (RHN), which becomes relativistic and subsequently non-relativistic before dominating the energy density. We focus on the case where the RHN domination begins while the RHN is still relativistic. It is also possible to consider a scenario in which the RHN is already non-relativistic when it starts to dominate, which we do not consider here. The RHN masses and Yukawa couplings can be chosen to generate the required neutrino mass scale. A more general setup with multiple RHNs is also possible, in which either one RHN or a combination of RHNs can dominate the energy density. Such a setup can additionally allow for leptogenesis, as discussed in Refs.~\cite{...}, but is beyond the scope of this work.

\acknowledgments
The work of Debasish Borah is supported by the Science and Engineering Research Board (SERB), Government of India grant CRG/2022/000603. ND would like to thank Basabendu Barman and Arindam Basu for useful discussions related to RHN reheating.  Disha Bandyopadhyay would like to thank Raymond T. Co for useful discussions on isocurvature bounds in kinetic misalignment during
Axions in Seoul workshop. She also acknowledges helpful discussions and suggestions during her visits to the Korea Institute for Advanced
Study (KIAS) and the Institute for Basic Science (IBS), and thanks Seokhoon Yun for useful discussions during her visit to Kyungpook National University.

\appendix

\section{Analytical estimates: vacuum misalignment}
\label{appen1}

The axion energy density is generally given by Eq.~\eqref{energyden}. During the analytical treatment of vacuum misalignment, the axion field remains approximately frozen at $\theta_i$ until the onset of oscillations at $T_{\rm osc}$. Thus, its kinetic contribution can be neglected before the onset, and the energy density is dominated by the potential term. For small $\theta_i$, the axion potential can be expanded around its minimum, giving
\begin{equation}
\rho_a(T_{\rm osc})
\simeq
\frac{1}{2}m_a^2(T_{\rm osc})f_a^2\theta_i^2 .
\label{eq:rho_a_osc_reheat}
\end{equation}
The axion energy density at a later temperature can then be obtained by red-shifting the axion number density after the onset of oscillations. In particular, the present-day energy density is
\begin{equation}
\rho_a(T_0)
= \rho_a(T_{\rm osc})
\frac{m_a}{m_a(T_{\rm osc})}
\frac{s(T_0)}{s(T_{\rm osc})}
\begin{cases}
\displaystyle
\frac{\mathcal{S}(T_{\rm osc})}{\mathcal{S}(T_{\rm RH})},
&
T_{\rm osc}>T_{\rm RH},
\\[1.2em]
\displaystyle 1,
&
T_{\rm osc}<T_{\rm RH}.
\end{cases}
\label{eq:rho_a_today_general}
\end{equation}
If oscillations begin after reheating, $T_{\rm osc}<T_{\rm RH}$, the axion abundance follows the standard radiation-dominated result. For $\theta_i=1$, the observed dark matter abundance is obtained at approximately $f_a\simeq 1.4\times10^{12}\,\mathrm{GeV}$. We therefore focus below on oscillations before the completion of reheating, $T_{\rm osc}>T_{\rm RH}$. In this case, both the modified oscillation temperature and the subsequent entropy dilution affect the relic abundance. We examine these effects for direct and seesaw reheating separately.
\subsection{Direct Reheating}
\label{Sec:Direct Reheating}
The entropy dilution factor associated with reheating  for bosonic and fermionic inflaton decay is given by:
 \begin{equation}
\frac{\mathcal{S}(T)}{\mathcal{S}(T_{\rm RH})}
=
\left(\frac{g_{\star s}(T)}{g_{\star s}(T_{\rm RH})}\right)
\left(\frac{T}{T_{\rm RH}}\right)^3
\left(\frac{a(T)}{a_{\rm RH}}\right)^3
=
\left(\frac{g_{\star s}(T)}{g_{\star s}(T_{\rm RH})}\right)
\left(\frac{T}{T_{\rm RH}}\right)^{\frac{3\alpha-3}{\alpha}}.
\label{eq:entropy_dilution_general}
\end{equation}
which can also be written as,
\begin{equation}
\frac{\mathcal{S}(T_{\rm osc})}{\mathcal{S}(T_{\rm RH})} = \left( \frac{g_{\star s}(T_{\rm osc})}{g_{\star s}(T_{\rm RH})} \right) \begin{cases} \displaystyle \left( \frac{T_{\rm RH}}{T_{\rm osc}} \right)^{2n+1}, & \text{bosonic decay},\\[12pt] \displaystyle \left( \frac{T_{\rm RH}}{T_{\rm osc}} \right)^{\frac{7-n}{n-1}}, & \text{fermionic decay},\quad n<7. 
\end{cases}
\label{eq:entropy_dilution_reheat}
\end{equation}
The relic abundance can then be evaluated separately according to whether the axion starts oscillating before or after the QCD phase transition. We first consider the case
\subsubsection{Oscillations Above the QCD Scale: $T_{\rm osc}>T_{\rm QCD}$}
In this regime, the temperature-dependent axion mass is given by Eq.~\eqref{axion_mass}. Taking $g_{*\rho}\simeq g_{*s}$ at the relevant temperatures $T_{\rm RH}$ and $T_{\rm osc}$, and using the analytical expression for the oscillation temperature in Eq.~\eqref{eq:Tosc_above_QCD} together with the entropy dilution factor in Eq.~\eqref{eq:entropy_dilution_reheat}, the general expression in Eq.~\eqref{eq:rho_a_today_general} gives the axion relic abundance for the two inflaton decay scenarios as follows.\\
\noindent\textbf{1. Bosonic decay:}\\
\begin{equation}
\Omega_a h^2 \simeq 0.12
\left(\frac{\theta_i}{1}\right)^2
\left(\frac{g_{*\rho}(T_{\rm osc})}{60}\right)^{\frac{n+4}{2(n+2)}}
\left(\frac{17.25}{g_{*s}(T_{\rm RH})}\right)
\left(\frac{T_{\rm RH}}{0.1\,{\rm GeV}}\right)^{\frac{10-n}{n+2}}
\left(\frac{f_a}{f_{a,n}^{\rm ref}}\right)^{\frac{n+4}{n+2}},
\end{equation}
where $f_{a,n}^{\rm ref}$ denotes the reference axion decay
constant for each value of $n$, given by
\begin{equation}
\begin{array}{|c|c|} \hline
n & f_{a,n}^{\rm ref}\,({\rm GeV}) \\ \hline
2 & 1.17\times 10^{13} \\
4 &  2.77\times 10^{12} \\
6 &  1.17\times 10^{12} \\ \hline
\end{array}
\end{equation}
\noindent\textbf{2. Fermionic decay:} For $n<7$, the corresponding relic abundance is,
\begin{equation}
\Omega_a h^2 \simeq 0.12
\left(\frac{\theta_i}{1}\right)^2
\left(\frac{g_{*\rho}(T_{\rm osc})}{60}\right)^{\frac{3n}{2(3n-2)}}
\left(\frac{17.25}{g_{*s}(T_{\rm RH})}\right)
\left(\frac{T_{\rm RH}}{0.1\,{\rm GeV}}\right)^{\frac{14-3n}{3n-2}}
\left(\frac{f_a}{f_{a,n}^{\rm ref}}\right)^{\frac{3n}{3n-2}},
\end{equation}
where $f_{a,n}^{\rm ref}$ denotes the reference axion decay
constant for each value of $n$, given by
\begin{equation}
\begin{array}{|c|c|} \hline
n & f_{a,n}^{\rm ref}\,({\rm GeV}) \\ \hline
2 &  1.17\times 10^{13} \\
4 &  6.55\times 10^{11} \\
6 &  2.51\times 10^{11} \\ \hline
\end{array}
\end{equation}
\subsubsection{Oscillations below the QCD Scale: $T_{\rm osc}<T_{\rm QCD}$}
For axion oscillations occurring after the QCD phase transition, the axion mass is approximately constant and is given by Eq.~\eqref{axion_mass}. Using the analytical expression for the oscillation temperature in Eq.~\eqref{eq:Tosc_below_QCD} and taking $g_{*\rho}\simeq g_{*s}\simeq 10.75$ in the relevant temperature range, the axion relic abundance follows from the general expression in Eq.~\eqref{eq:rho_a_today_general}, together with the entropy dilution factor given in Eq.~\eqref{eq:entropy_dilution_reheat}. In normalized form, it can be written as

\begin{equation}
\Omega_a h^2 \simeq 0.12
\left(\frac{\theta_i}{1}\right)^2
\left(\frac{T_{\rm RH}}{4\,{\rm MeV}}\right)^{\frac{4-n}{n}}
\left(\frac{f_a}{f_{a,n}^{\rm ref}}\right)^{\frac{n+2}{n}}.
\label{eq:Omega_post_QCD}
\end{equation}
where $f_{a,n}^{\rm ref}$ denotes the reference axion decay
constant for each value of $n$, given by
\begin{equation}
\begin{array}{|c|c|} \hline
n & f_{a,n}^{\rm ref}\,({\rm GeV}) \\ \hline
2 & 7.59\times10^{14} \\
4 & 1.08\times10^{13} \\
6 & 1.28\times10^{12}\\ \hline
\end{array}
\label{eq:ma_ref_post_QCD}
\end{equation}
This result is consistent with the analytical result obtained in Ref \cite{Xu:2023lxw}.

\subsection{Seesaw Reheating}
\label{sec:seesaw-vac}
In seesaw reheating, axion oscillations can begin during inflaton
domination, relativistic RHN domination, or non-relativistic RHN
domination. Although the axion number is conserved after oscillations
begin, the number density continues to be diluted as the Universe
expands through the remaining stages of reheating. Eq.~\eqref{eq:rho_a_today_general} can be written as
\begin{equation}
\rho_a(T_0)
=
\rho_a(T_{\rm osc})
\frac{m_a}{m_a(T_{\rm osc})}\mathcal R_i,
\label{eq:rho_a_today_general1}
\end{equation}
where
\begin{equation}
\mathcal R_i =
\begin{cases}
\displaystyle
\frac{s_0}{s(T_{\rm RH})}
\left(\frac{a_{\rm osc}}{a_{\rm RH}}\right)^3,
& T_{\rm osc}>T_{\rm RH},
\\[1.2em]
\displaystyle
\frac{s_0}{s(T_{\rm osc})},
& T_{\rm osc}<T_{\rm RH}.
\end{cases}
\label{eq:redshift_general}
\end{equation}
For oscillations before reheating, the redshift factor includes
the expansion from $a_{\rm osc}$ to $a_{\rm RH}$, followed by
entropy-conserving evolution to today. For oscillations after
reheating, entropy conservation applies directly from
$T_{\rm osc}$ to the present.
Thus, the abundance depends both on the conditions at the onset of
oscillations and on the expansion between $a_{\rm osc}$ and
$a_{\rm RH}$. We evaluate these quantities in each of the three eras.
\subsubsection{Oscillations above the QCD Scale: $T_{\rm osc}>T_{\rm QCD}$}

For $T_{\rm osc}>T_{\rm QCD}$, the temperature dependence of the axion mass Eq. (\ref{axion_mass}) must be included in the oscillation condition $3\mathcal{H}(a_{\rm osc})=m_a(T_{\rm osc})$. Applying the appropriate $\mathcal{H}(a)$ and $T(a)$ relations from Eq. (\ref{eq:Hubble_RHN}) and Eq. (\ref{eq:Temperature_RHN}) respectively in each era gives the following scale factors and oscillation temperatures. The corresponding scale factor at the onset of oscillations is
\begin{equation}
a_{\rm osc} =
\begin{cases}
\displaystyle
a_I
\left[
\frac{3\mathcal{H}_I}{m_a}
\left(\frac{T_{\max}}{T_{\rm QCD}}\right)^4
\right]^{\frac{n+2}{3n}},
& \phi\text{-domination},
\\[1.5em]
\displaystyle
a_R
\left[
\frac{3\mathcal{H}_{\rm R}}{m_a}
\left(\frac{T_{\max}}{T_{\rm QCD}}\right)^4
\right]^{1/3},
& \text{relativistic RHN domination},
\\[1.5em]
\displaystyle
a_{\rm NR}
\left[
\frac{3\mathcal{H}_{\rm NR}}{m_a}
\left(\frac{T_{\rm NR}}{T_{\rm QCD}}\right)^4
\right]^{1/3},
& \text{non-relativistic RHN domination}.
\end{cases}
\label{eq:aosc_preQCD}
\end{equation}
The corresponding oscillation temperature can thus be obtained by using Eq. (\ref{eq:Temperature_RHN})
\begin{equation}
T_{\rm osc} =
\begin{cases}
\displaystyle
T_{\max},
& \phi\text{-domination},
\\[1.2em]
\displaystyle
T_{\max}
\left[
\frac{m_a}{3\mathcal{H}_{\rm R}}
\left(\frac{T_{\rm QCD}}{T_{\max}}\right)^4
\right]^{1/12},
& \text{relativistic RHN domination},
\\[1.2em]
\displaystyle
T_{\rm NR}
\left[
\frac{m_a}{3\mathcal{H}_{\rm NR}}
\left(\frac{T_{\rm QCD}}{T_{\rm NR}}\right)^4
\right]^{1/8},
& \text{non-relativistic RHN domination}.
\end{cases}
\label{eq:Tosc_preQCD}
\end{equation}
Substituting the values of $\mathcal{H}_{\rm R}$ and $ \mathcal{H}_{\rm NR}$ from Eq. (\ref{eq:Hubble_RHN}) the oscillation temperature thus obtained is 
\begin{equation}
T_{\rm osc} =
\begin{cases}
\displaystyle
T_{\max},
& \phi\text{-domination},
\\[1.2em]
\displaystyle
\left(
\frac{m_a M_{\rm P}T_{\rm NR}^4T_{\rm RH}^2 T_{\rm QCD}^4}
{g_{*s}^{1/2}({\rm T_{RH}})}
\right)^{1/12},
& \text{relativistic RHN domination},
\\[1.2em]
\displaystyle
\left(
\frac{m_a M_{\rm P}T_{\rm RH}^2 T_{\rm QCD}^4}
{g_{*s}^{1/2}({\rm T_{RH}})}
\right)^{1/8},
& \text{non-relativistic RHN domination}.
\end{cases}
\label{eq:Tosc_preQCD1}
\end{equation}
By substituting Eq. (\ref{eq:aosc_preQCD}) in Eq. (\ref{eq:rho_a_today_general1}) one can compute the redshift factor from the oscillation epoch to today as,
\begin{equation}
\mathcal{R}_i \approx
2.25,s_0
\begin{cases}
\displaystyle
g_{*s}^{\frac{2-n}{2n}}({\rm T_{RH}})
\frac{
T_{\max}^{\frac{24}{n}}
T_{\rm NR}^{-\frac{8}{n}}
T_{\rm RH}^{3-\frac{4}{n}}
}{
m_a^{\frac{n+2}{n}}
M_{\rm P}^{\frac{n+2}{n}}
T_{\rm QCD}^{\frac{4(n+2)}{n}}
},
& \phi\text{-dominated},
\\[2.0em]
\displaystyle
\frac{
T_{\rm RH}^{3}
}{
g_{*s}^{1/2}({\rm T_{RH}})
m_a M_{\rm P}
T_{\rm QCD}^{4}
},
& \text{relativistic RHN},
\\[2.0em]
\displaystyle
\frac{
T_{\rm RH}^{3}
}{
g_{*s}^{1/2}({\rm T_{RH}})
m_a M_{\rm P}
T_{\rm QCD}^{4}
},
& \text{non-relativistic RHN}.
\end{cases}
\end{equation}
The corresponding energy density Eq. (\ref{eq:rho_a_today_general1}) and relic abundance thus obtained is
\begin{equation}
\begin{aligned}
\frac{\Omega_a h^2}{0.12} \simeq{}&
\left(\frac{\theta_i}{1}\right)^2
\begin{cases}
\displaystyle
\begin{aligned}
&\phi\text{-domination}:\\[0.3em]
&\left(\frac{f_a}{f_{a,n}^{\rm ref}}\right)^{\frac{n+2}{n}}
 \left(\frac{g_{*s}(T_{\rm RH})}{17.25}\right)^{\frac{2-n}{2n}}
 \left(\frac{T_{\max}}{10\,{\rm GeV}}\right)^{\frac{24}{n}-4}
 \left(\frac{T_{\rm NR}}{1\,{\rm GeV}}\right)^{\frac{-8}{n}}
 \left(\frac{T_{\rm RH}}{0.1\,{\rm GeV}}\right)^{3-\frac{4}{n}},
\end{aligned}
\\[3 em]
\displaystyle
\begin{aligned}
&\text{relativistic RHN domination}:\\[0.3em]
&\left(\frac{f_a}{5.16\times10^{13}\,{\rm GeV}}\right)^{\frac{4}{3}}
 \left(\frac{g_{*s}(T_{\rm RH})}{17.25}\right)^{-\frac{1}{3}}
 \left(\frac{T_{\rm NR}}{1\,{\rm GeV}}\right)^{-\frac{4}{3}}
 \left(\frac{T_{\rm RH}}{0.1\,{\rm GeV}}\right)^{\frac{7}{3}},
\end{aligned}
\\[3 em]
\displaystyle
\begin{aligned}
&\text{non-relativistic RHN domination}:\\[0.3em]
&\left(\frac{f_a}{2.18\times10^{13}\,{\rm GeV}}\right)^{\frac{3}{2}}
 \left(\frac{g_{*s}(T_{\rm RH})}{17.25}\right)^{-\frac{1}{4}}
 \left(\frac{T_{\rm RH}}{0.1\,{\rm GeV}}\right)^2.
\end{aligned}
\end{cases}
\end{aligned}
\label{eq:preQCD_abundance}
\end{equation}
% \begin{equation}
% \Omega_a h^2 \simeq 0.12 \, \left(\frac{\theta_i}{1}\right)^2
% \begin{cases}
% \displaystyle
% \left(\frac{f_a}{f_{a,n}^{\rm ref}}\right)^{\frac{n+2}{n}}
% \left(\frac{g_{*s}({\rm T_{RH}})}{17.25}\right)^{\frac{2-n}{2n}}
% \left(\frac{T_{\max}}{10\,{\rm GeV}}\right)^{\frac{24}{n}-4}
% \left(\frac{T_{\rm NR}}{1\,{\rm GeV}}\right)^{-\frac{8}{n}}
% \left(\frac{T_{\rm RH}}{0.1\,{\rm GeV}}\right)^{3-\frac{4}{n}},
% & \phi\text{-domination},
% \\[2.0em]
% \displaystyle
% \left(\frac{f_a}{5.16\times10^{13}\,{\rm GeV}}\right)^{\frac{4}{3}}
% \left(\frac{g_{*s}({\rm T_{RH}})}{17.25}\right)^{-\frac{1}{3}}
% \left(\frac{T_{\rm NR}}{1\,{\rm GeV}}\right)^{-\frac{4}{3}}
% \left(\frac{T_{\rm RH}}{0.1\,{\rm GeV}}\right)^{\frac{7}{3}},
% & \text{relativistic RHN domination},
% \\[2.0em]
% \displaystyle
% \left(\frac{f_a}{2.18\times10^{13}\,{\rm GeV}}\right)^{\frac{3}{2}}
% \left(\frac{g_{*s}({\rm T_{RH}})}{17.25}\right)^{-\frac{1}{4}}
% \left(\frac{T_{\rm RH}}{0.1\,{\rm GeV}}\right)^2,
% & \text{non-relativistic RHN domination}.
% \end{cases}
% \label{eq:preQCD_abundance}
% \end{equation}
For $\phi$-domination, the reference value of $f_a$ depends on the equation-of-state parameter $n$ and is denoted by $f_{a,n}^{\rm ref}$. For the cases considered in this work,
\begin{equation} \begin{array}{|c|c|} \hline
n & f_{a,n}^{\rm ref}\;({\rm GeV}) \\ \hline 2  & 3.89\times10^{8} \\
4 & 1.01\times10^{12} \\
6 & 5.16\times10^{13}\\ \hline
\end{array}
\label{tab:fa_ref_preQCD}
\end{equation}
Alternatively for $\phi$-domination, the axion relic abundance for the three values
$n=2,4,6$ considered here can be written as
\begin{equation}
\frac{\Omega_a h^2}{0.12} \simeq  \left(\frac{\theta_i}{1}\right)^2 
\begin{cases}
\displaystyle
\left(\frac{f_a}{f_{a,2}^{\rm ref}}\right)^2
\left(\frac{T_{\max}}{10\,{\rm GeV}}\right)^8
\left(\frac{T_{\rm NR}}{1\,{\rm GeV}}\right)^{-4}
\left(\frac{T_{\rm RH}}{0.1\,{\rm GeV}}\right),
& n=2,
\\[1.5em]
\displaystyle
\left(\frac{f_a}{f_{a,4}^{\rm ref}}\right)^{\frac{3}{2}}
\left(\frac{g_{*s}({\rm T_{RH}})}{17.25}\right)^{-\frac{1}{4}}
\left(\frac{T_{\max}}{10\,{\rm GeV}}\right)^2
\left(\frac{T_{\rm NR}}{1\,{\rm GeV}}\right)^{-2}
\left(\frac{T_{\rm RH}}{0.1\,{\rm GeV}}\right)^2,
& n=4,
\\[1.5em]
\displaystyle
\left(\frac{f_a}{f_{a,6}^{\rm ref}}\right)^{\frac{4}{3}}
\left(\frac{g_{*s}({\rm T_{RH}})}{17.25}\right)^{-\frac{1}{3}}
\left(\frac{T_{\rm NR}}{1\,{\rm GeV}}\right)^{-\frac{4}{3}}
\left(\frac{T_{\rm RH}}{0.1\,{\rm GeV}}\right)^{\frac{7}{3}},
& n=6.
\end{cases}
\label{eq:phi_abundance_preQCD}
\end{equation}
For pre-QCD oscillations, the $n=6$ inflaton-dominated result has the same parameter scaling as the relativistic RHN-dominated result, despite their different equations of state. This equality follows from the combined dependence of the oscillation condition, the temperature-dependent axion mass, and the subsequent dilution factor; it does not imply identical expansion histories. The common expression for $\mathcal R_i$ in the two RHN-dominated regimes likewise does not give the same abundance, because $m_a(T_{\rm osc})$ depends on their different oscillation temperatures.

\subsubsection{Oscillations below the QCD Scale: $T_{\rm osc}<T_{\rm QCD}$}

When oscillations begin below $T_{\rm QCD}$, we can take the axion mass to be approximately constant, $m_a(T_{\rm osc})\simeq m_a$. The oscillation condition $3\mathcal{H}(a_{\rm osc})=m_a$ then gives a different $a_{\rm osc}$ in each of the three eras:
\begin{equation}
a_{\rm osc} =
\begin{cases}
\displaystyle
a_I
\left[
\frac{3\mathcal{H}_I}{m_a}
\right]^{\frac{n+2}{3n}},
& \phi\text{-domination},
\\[1.5em]
\displaystyle
a_R
\left[
\frac{3\mathcal{H}_{\rm R}}{m_a}
\right]^{1/2},
& \text{relativistic RHN domination},
\\[1.5em]
\displaystyle
a_{\rm NR}
\left[
\frac{3\mathcal{H}_{\rm NR}}{m_a}
\right]^{2/3},
& \text{non-relativistic RHN domination}.
\end{cases}
\label{eq:aosc_postQCD}
\end{equation}
The corresponding oscillation temperature can thus be obtained by using Eq. (\ref{eq:Temperature_RHN}) and Eq. (\ref{eq:Hubble_RHN})

\begin{equation}
T_{\rm osc}=
\begin{cases}
T_{\max}, & \phi\text{-dominated}, \\[1.0em]
\left(
m_a M_{\rm P} T_{\rm NR}^4 T_{\rm RH}^2
g_{*s}^{-1/2}({\rm T_{RH}})
\right)^{1/8},
& \text{relativistic RHN}, \\[1.2em]
\left(
m_a M_{\rm P} T_{\rm RH}^2
g_{*s}^{-1/2}({\rm T_{RH}})
\right)^{1/4},
& \text{non-relativistic RHN}.
\end{cases}
\label{Tosc:seesaw-vac}
\end{equation}
By substituting Eq. (\ref{eq:aosc_postQCD}) in Eq. (\ref{eq:rho_a_today_general1}) one can compute the redshift factor from the oscillation epoch to today as,
\begin{equation}
\mathcal{R}_i \approx
2.25 \,s_0
\begin{cases}
\displaystyle
g_{*s}^{\frac{2-n}{2n}}({\rm T_{RH}})
\frac{
T_{\max}^{\frac{16}{n}-4}
T_{\rm NR}^{-\frac{8}{n}}
T_{\rm RH}^{3-\frac{4}{n}}
}{
m_a^{\frac{n+2}{n}}
M_{\rm P}^{\frac{n+2}{n}}
},
& \phi\text{-dominated}, \\[2.0em]
\displaystyle
\,
\frac{
T_{\rm RH}^2
}{
g_{*s}^{1/4}({\rm T_{RH}})
m_a^{3/2}
M_{\rm P}^{3/2}
T_{\rm NR}^2
},
& \text{relativistic RHN}, \\[2.0em]
\displaystyle
\frac{T_{\rm RH}}
{m_a^2M_{\rm P}^2},
& \text{non-relativistic RHN}.
\end{cases}
\end{equation}
Substituting this factor into Eq.~\eqref{eq:rho_a_today_general1}, we find the following relic abundances. We have normalized them to show their dependence on the reheating parameters and on $f_a$:
\begin{equation}
\begin{aligned}
\frac{\Omega_a h^2}{0.12} \simeq{}&
\left(\frac{\theta_i}{1}\right)^2
\begin{cases}
\displaystyle
\begin{aligned}
&\phi\text{-domination}:\\[0.3em]
&\left(\frac{f_a}{f_{a,n}^{\rm ref}}\right)^{\frac{n+2}{n}}
 \left(\frac{g_{*s}(T_{\rm RH})}{17.25}\right)^{\frac{2-n}{2n}}
 \left(\frac{T_{\max}}{10\,{\rm GeV}}\right)^{\frac{16}{n}-4}
 \left(\frac{T_{\rm NR}}{1\,{\rm GeV}}\right)^{-\frac{8}{n}}
 \left(\frac{T_{\rm RH}}{0.1\,{\rm GeV}}\right)^{3-\frac{4}{n}},
\end{aligned}
\\[3em]
\displaystyle
\begin{aligned}
&\text{relativistic RHN domination}:\\[0.3em]
&\left(\frac{f_a}{2.51\times10^{14}\,{\rm GeV}}\right)^{3/2}
 \left(\frac{g_{*s}(T_{\rm RH})}{17.25}\right)^{-1/4}
 \left(\frac{T_{\rm NR}}{1\,{\rm GeV}}\right)^{-2}
 \left(\frac{T_{\rm RH}}{0.1\,{\rm GeV}}\right)^2,
\end{aligned}
\\[3em]
\displaystyle
\begin{aligned}
&\text{non-relativistic RHN domination}:\\[0.3em]
&\left(\frac{f_a}{1.52\times10^{14}\,{\rm GeV}}\right)^2
 \left(\frac{T_{\rm RH}}{0.1\,{\rm GeV}}\right).
\end{aligned}
\end{cases}
\end{aligned}
\label{eq:postQCD_abundance_seesaw}
\end{equation}
For $\phi$-domination, the reference value of $f_a$ depends on the equation-of-state parameter $n$ and is denoted by $f_{a,n}^{\rm ref}$. For the cases considered in this work,
\begin{equation} \begin{array}{|c|c|} \hline n & f_{a,n}^{\rm ref}\;({\rm GeV}) \\ \hline 2 & 1.52\times10^{12} \\ 4 & 2.51\times10^{14} \\ 6 & 3.22\times10^{15} \\\hline\end{array} \label{tab:fa_ref} \end{equation}
Alternatively for $\phi$-domination, the axion relic abundance for the three values
$n=2,4,6$ considered here can be written as
\begin{equation}
\frac{\Omega_a h^2}{0.12} \simeq   \left(\frac{\theta_i}{1}\right)^2
\begin{cases}
\displaystyle
\left(\frac{f_a}{f_{a,2}^{\rm ref}}\right)^2
\left(\frac{T_{\max}}{10\,{\rm GeV}}\right)^4
\left(\frac{T_{\rm NR}}{1\,{\rm GeV}}\right)^{-4}
\left(\frac{T_{\rm RH}}{0.1\,{\rm GeV}}\right),
\hspace{3.25 cm} n=2,
\\[1.5em]
\displaystyle
\left(\frac{f_a}{f_{a,4}^{\rm ref}}\right)^{\frac{3}{2}}
\left(\frac{g_{*s}({\rm T_{RH}})}{17.25}\right)^{-\frac{1}{4}}
\left(\frac{T_{\rm NR}}{1\,{\rm GeV}}\right)^{-2}
\left(\frac{T_{\rm RH}}{0.1\,{\rm GeV}}\right)^2,
\hspace{2.5 cm} n=4,
\\[1.5em]
\displaystyle
\left(\frac{f_a}{f_{a,6}^{\rm ref}}\right)^{\frac{4}{3}}
\left(\frac{g_{*s}({\rm T_{RH}})}{17.25}\right)^{-\frac{1}{3}}
\left(\frac{T_{\max}}{10\,{\rm GeV}}\right)^{-\frac{4}{3}}
\left(\frac{T_{\rm NR}}{1\,{\rm GeV}}\right)^{-\frac{4}{3}}
\left(\frac{T_{\rm RH}}{0.1\,{\rm GeV}}\right)^{\frac{7}{3}},
\hspace{0.0 cm} n=6.
\end{cases}
\label{eq:phi_abundance}
\end{equation}
We can see that the $n=4$ $\phi$-dominated case yields the same abundance scaling as relativistic RHN domination, since both have a radiation-like equation of state and the same relevant $\mathcal{H}(a)$ scaling. In contrast, although the $n=2$ case and non-relativistic RHN domination are both matter-like, their different $T(a)$ histories lead to different relic-abundance dependence due to the presence of intermediate relativistic RHN phase in case of axion oscillating during inflaton domination.

%\section{Analytical estimate of axion relic abundance for Kinetic misalignment mechanism}
\section{Analytical estimates: kinetic misalignment}
\label{appen2}

In kinetic misalignment, the axion has an initial angular velocity
$\dot{\theta}_{i}$ at $a=a_I$. For a sufficiently large initial
velocity, the field continues to rotate over the potential barriers
before becoming trapped and oscillating around a minimum. We consider
a fixed radial field value $f_a$ and assume that the axion energy
density remains subdominant.

While the potential force can be neglected, the axion equation of
motion gives
\begin{equation}
\ddot{\theta}+3\mathcal{H}\dot{\theta}\simeq 0,
\qquad
\dot{\theta}(a)
=
\dot{\theta}_{i}
\left(\frac{a_I}{a}\right)^3.
\label{eq:theta_velocity_KM}
\end{equation}
We estimate the trapping time by equating the kinetic energy to the
height of the axion potential barrier,
\begin{equation}
\frac{1}{2}f_a^2\dot{\theta}^{\,2}(a_{\rm osc})
\simeq
2m_a^2(T_{\rm osc})f_a^2 \implies
|\dot{\theta}(a_{\rm osc})|
\simeq 2m_a(T_{\rm osc}).
\label{eq:trapping_KM}
\end{equation}
Here $a_{\rm osc}$ denotes the onset of oscillations. These estimates apply when the initial velocity is large
enough to delay oscillation beyond the onset expected from vacuum misalignment.\\
The angular motion carries a charge density
$n_\theta=f_a^2\dot{\theta}$, whose comoving value is approximately
conserved during the rotation. We parameterize its conversion into
the axion number density by $\mathcal C_{\rm KM}$,
\begin{equation}
n_a(a_{\rm osc})
\simeq
\mathcal C_{\rm KM} f_a^2
|\dot{\theta}(a_{\rm osc})|.
\end{equation}
Using Eq.~\eqref{eq:theta_velocity_KM} and redshifting the axion
number density to today gives
\begin{equation}
\begin{aligned}
\Omega_a h^2 \simeq{}&
\mathcal C_{\rm KM}
\frac{m_a f_a^2|\dot{\theta}_{i}|s_0}{\rho_c/h^2}
\begin{cases}
\displaystyle
\frac{1}{s(T_{\rm RH})}
\left(\frac{a_I}{a_{\rm RH}}\right)^3,
& T_{\rm osc}>T_{\rm RH},
\\[1.2em]
\displaystyle
\frac{1}{s(T_{\rm osc})}
\left(\frac{a_I}{a_{\rm osc}}\right)^3,
& T_{\rm osc}<T_{\rm RH}.
\end{cases}
\end{aligned}
\label{eq:Omega_KM_cases}
\end{equation}
For oscillations before reheating, the factors of $a_{\rm osc}$
cancel directly. For oscillations after reheating, entropy
conservation gives
\begin{equation}
s(T_{\rm osc})a_{\rm osc}^3
=
s(T_{\rm RH})a_{\rm RH}^3.
\end{equation}
Both cases therefore reduce to
\begin{equation}
\Omega_a h^2
\simeq
\mathcal C_{\rm KM}
\frac{m_a f_a^2|\dot{\theta}_{i}|s_0}
{(\rho_c/h^2)s(T_{\rm RH})}
\left(\frac{a_I}{a_{\rm RH}}\right)^3.
\label{eq:Omega_KM_general}
\end{equation}
Thus, within this approximation, the explicit dependence on
$a_{\rm osc}$ cancels in both cases.

% We can see that the explicit dependence on the oscillation time cancels:
% the velocity decreases as $a^{-3}$ before trapping, and the axion
% number density follows the same scaling afterwards. The trapping
% condition is still needed to determine which regime applies and
% whether kinetic misalignment is realized.
\subsection{Direct reheating}
\label{sec:direct-kin}
For direct reheating, the expansion during inflaton domination
determines the dilution of the initial axion velocity. The scale
factor at the end of reheating is related to its initial value is given by Eq. (\ref{hub_case1})
The bosonic and fermionic decay channels have different temperature
evolutions, which affect the time at which the axion oscillates. Using the velocity evolution in
Eq.~\eqref{eq:theta_velocity_KM}, the oscillation condition can be
written as
\begin{equation}
\dot{\theta}_i
\left(\frac{a_I}{a_{\rm osc}}\right)^3
\simeq 2m_a(T_{\rm osc}).
\label{eq:trapping_KM_direct}
\end{equation}
Here, $\dot{\theta}_i$ denotes the initial angular velocity at
$a_I$, and $a_{\rm osc}$ denotes the onset of oscillations
following trapping. We evaluate this condition separately
above and below the QCD scale.

\subsubsection{Oscillations above the QCD scale:
$T_{\rm osc}>T_{\rm QCD}$}

For $T_{\rm osc}>T_{\rm QCD}$, the axion mass is
$m_a(T_{\rm osc})=m_a(T_{\rm QCD}/T_{\rm osc})^4$.
The onset of oscillations is therefore determined by the
decrease in the angular velocity together with the growth
of the axion mass as the bath cools. Substituting the
temperature evolution from Eq. (\ref{temp_case1}) for each decay channel into Eq.~\eqref{eq:trapping_KM_direct} gives
\begin{equation}
    a_{\rm osc}=
\begin{cases}
\displaystyle
a_I
\left(\frac{\dot\theta_{\rm inf}}{2m_a}\right)^{\frac{n+2}{3(n+4)}}
\left(\frac{T_{\rm RH}}{T_{\rm QCD}}\right)^{\frac{4(n+2)}{3(n+4)}}
\left(\frac{\mathcal{H}_I}{\mathcal{H}_{\rm RH}}\right)^{\frac{2(n+2)}{3n(n+4)}},
&
\text{bosonic},
\\[1.3em]
\displaystyle
a_I
\left(\frac{\dot\theta_{\rm inf}}{2m_a}\right)^{\frac{n+2}{9n}}
\left(\frac{T_{\rm RH}}{T_{\rm QCD}}\right)^{\frac{4(n+2)}{9n}}
\left(\frac{\mathcal{H}_I}{\mathcal{H}_{\rm RH}}\right)^{\frac{2(n-1)(n+2)}{9n^2}},
&
\text{fermionic},\ n<7,
\\[1.3em]
\displaystyle
a_I
\left(\frac{\dot\theta_{\rm inf}}{2m_a}\right)^{1/7}
\left(\frac{T_{\rm RH}}{T_{\rm QCD}}\right)^{4/7}
\left(\frac{\mathcal{H}_I}{\mathcal{H}_{\rm RH}}\right)^{\frac{4(n+2)}{21n}},
&
\text{fermionic},\ n>7,
\\[1.3em]
\displaystyle
a_I
\left(\frac{\dot\theta_{\rm inf}}{2m_a}\right)^{1/7}
\left(\frac{T_{\rm RH}}{T_{\rm QCD}}\right)^{4/7}
\left(\frac{\mathcal{H}_I}{\mathcal{H}_{\rm RH}}\right)^{\frac{4(n+2)}{21n}},
&
\text{radiation dominated}.
\end{cases}
\label{eq:aosc_KM_direct_above}
\end{equation}
Evaluating the temperature from Eq. (\ref{temp_case1}) at the scale factors in
Eq.~\eqref{eq:aosc_KM_direct_above}, we obtain the oscillation temperature as
\begin{equation}
    T_{\rm osc}= \begin{cases} \displaystyle T_{\rm RH} \left(\frac{\mathcal{H}_I}{\mathcal{H}_{\rm RH}}\right)^{\frac{n+2}{2n(n+4)}} \left(\frac{2m_a}{\dot\theta_{\rm inf}}\right)^{\frac{1}{2(n+4)}} \left(\frac{T_{\rm QCD}}{T_{\rm RH}}\right)^{\frac{2}{n+4}}, & \text{bosonic}, \\[1.3em] \displaystyle T_{\rm RH} \left(\frac{\mathcal{H}_I}{\mathcal{H}_{\rm RH}}\right)^{\frac{(n-1)(n+2)}{6n^2}} \left(\frac{2m_a}{\dot\theta_{\rm inf}}\right)^{\frac{n-1}{6n}} \left(\frac{T_{\rm QCD}}{T_{\rm RH}}\right)^{\frac{2(n-1)}{3n}}, & \text{fermionic},\ n<7, \\[1.3em] \displaystyle T_{\rm RH} \left(\frac{\mathcal{H}_I}{\mathcal{H}_{\rm RH}}\right)^{\frac{n+2}{7n}} \left(\frac{2m_a}{\dot\theta_{\rm inf}}\right)^{1/7} \left(\frac{T_{\rm QCD}}{T_{\rm RH}}\right)^{4/7}, & \text{fermionic},\ n>7, \\[1.3em] \displaystyle T_{\rm RH} \left(\frac{\mathcal{H}_I}{\mathcal{H}_{\rm RH}}\right)^{\frac{n+2}{7n}} \left(\frac{2m_a}{\dot\theta_{\rm inf}}\right)^{1/7} \left(\frac{T_{\rm QCD}}{T_{\rm RH}}\right)^{4/7}, & \text{radiation dominated}. \end{cases}
    \label{eq:Tosc_KM_direct_above}
\end{equation}
The first three branches of
Eqs.~\eqref{eq:aosc_KM_direct_above}
and~\eqref{eq:Tosc_KM_direct_above}
apply to trapping before the completion of reheating,
$a_{\rm osc}<a_{\rm RH}$, while the radiation-dominated
branch applies for $a_{\rm osc}>a_{\rm RH}$. The fermionic $n>7$ and radiation-dominated expressions coincide because both use $T\propto a^{-1}$ in the analytical temperature evolution. In these temperature scalings, we neglect changes in the effective relativistic degrees of freedom when estimating
$a_{\rm osc}$ and $T_{\rm osc}$ from Eq. (\ref{temp_case1}). We evaluate the subsequent
redshifting using
\begin{equation}
\mathcal R_i \simeq
\frac{s_0}{s(T_{\rm RH})}
\left(\frac{a_{\rm osc}}{a_{\rm RH}}\right)^3,
\label{eq:R_KM_scale_factor}
\end{equation}
assuming entropy conservation after reheating. This factor
relates the axion number density at trapping to its value
today. Substituting Eq.~\eqref{eq:aosc_KM_direct_above} gives

\begin{equation}
\mathcal R_i=
\frac{s_0}{s(T_{\rm RH})}
\begin{cases}
\displaystyle
\left(\frac{\dot\theta_{\rm inf}}{2m_a}\right)^{\frac{n+2}{n+4}}
\left(\frac{T_{\rm RH}}{T_{\rm QCD}}\right)^{\frac{4(n+2)}{n+4}}
\left(\frac{\mathcal{H}_{\rm RH}}{\mathcal{H}_I}\right)^{\frac{(n+2)^2}{n(n+4)}},
& \text{bosonic decay},
\\[1.5em]
\displaystyle
\left(\frac{\dot\theta_{\rm inf}}{2m_a}\right)^{\frac{n+2}{3n}}
\left(\frac{T_{\rm RH}}{T_{\rm QCD}}\right)^{\frac{4(n+2)}{3n}}
\left(\frac{\mathcal{H}_{\rm RH}}{\mathcal{H}_I}\right)^{\frac{(n+2)^2}{3n^2}},
& \text{fermionic decay}\, (n<7),
\\[1.5em]
\displaystyle
\left(\frac{\dot\theta_{\rm inf}}{2m_a}\right)^{3/7}
\left(\frac{T_{\rm RH}}{T_{\rm QCD}}\right)^{12/7}
\left(\frac{\mathcal{H}_{\rm RH}}{\mathcal{H}_I}\right)^{\frac{3(n+2)}{7n}},
& \text{fermionic decay}\, (n>7),
\\[1.5em]
\displaystyle
\left(\frac{\dot\theta_{\rm inf}}{2m_a}\right)^{3/7}
\left(\frac{T_{\rm RH}}{T_{\rm QCD}}\right)^{12/7}
\left(\frac{\mathcal{H}_{\rm RH}}{\mathcal{H}_I}\right)^{\frac{3(n+2)}{7n}},
& \text{radiation dominated}.
\end{cases}
\label{eq:R_KM_direct_above}
\end{equation}Using the axion number density at trapping,
$n_a(T_{\rm osc})\simeq
\mathcal C_{\rm KM}f_a^2\dot{\theta}(a_{\rm osc})$,
and substituting the corresponding redshift factors from
Eq.~\eqref{eq:R_KM_direct_above}, we obtain
\begin{equation}
\begin{aligned}
\rho_a(T_0)
&\simeq
m_a n_a(T_{\rm osc})\mathcal R_i
\\
&\simeq
\mathcal C_{\rm KM}m_a f_a^2\dot{\theta}_i
\left(\frac{a_I}{a_{\rm osc}}\right)^3\mathcal R_i
\end{aligned}
\label{eq:rho_KM_direct}
\end{equation}
Then the relic abundance becomes
\begin{equation}
\Omega_a h^2
\simeq
3.56\times10^{6}\,
\mathcal C_{\rm KM}\,
f_a\dot{\theta}_{i}
\left(
\frac{\pi}
{\sqrt{90}\,M_P \mathcal{H}_I}
\right)^{\frac{n+2}{n}}
T_{\rm RH}^{\frac{4-n}{n}} \,
g_{*s}^{\frac{2-n}{2n}}(T_{\rm RH}).
\label{eq:Omega_KM_direct}
\end{equation}
The same result is obtained for all branches. This is expected
from Eq.~\eqref{eq:Omega_KM_general}, where the explicit
dependence on $a_{\rm osc}$ cancels. Substituting the individual
expressions for the number density and redshift factor
therefore confirms this cancellation for each regime.
\\
For $n=10$, the abundance can be
written in the normalized form
\begin{equation}
\begin{aligned}
\Omega_a h^2 \simeq {}&
0.12
\left(\frac{\mathcal C_{\rm KM}}{2}\right)
\left(\frac{f_a}{1.6\times10^9\,{\rm GeV}}\right)
\left(\frac{\dot{\theta}_{i}}{10^{16}\,{\rm GeV}}\right)
\left(\frac{\mathcal{H}_I}{1.6\times10^8\,{\rm GeV}}\right)^{-\frac{6}{5}} \\ 
& \times 
\left(\frac{T_{\rm RH}}{1\,{\rm GeV}}\right)^{-\frac{3}{5}}
\left(\frac{g_{*s}(T_{\rm RH})}{17.25}\right)^{-\frac{2}{5}}
\end{aligned}
\label{eq:Omega_KM_direct_n10}
\end{equation}
At fixed initial conditions and relativistic degrees of freedom, the $n=10$ abundance scales as $T_{\rm RH}^{-3/5}$.
The reheating history thus affects the abundance even when
oscillation occurs after reheating, because the initial velocity is specified at $a_I$ and is diluted throughout the preceding
expansion.
\subsubsection{Oscillations below the QCD scale:
$T_{\rm osc}<T_{\rm QCD}$}

For $T_{\rm osc}<T_{\rm QCD}$, the axion mass is constant, $m_a(T_{\rm osc}) = m_a$. The trapping condition
in Eq.~\eqref{eq:trapping_KM_direct} then depends only on the
dilution of the initial angular velocity. The trapping scale
factor is therefore common to all the decay channels:
\begin{equation}
a_{\rm osc}
=
a_I
\left(\frac{\dot{\theta}_i}{2m_a}\right)^{1/3}.
\label{eq:aosc_KM_direct_below}
\end{equation}
The corresponding temperature still depends on the thermal
history. Evaluating the temperature evolution at
$a_{\rm osc}$ from Eq. (\ref{temp_case1}) gives
\begin{equation}
T_{\rm osc}=
\begin{cases}
\displaystyle
T_{\rm RH}
\left(\frac{\mathcal{H}_I}{\mathcal{H}_{\rm RH}}\right)^{\frac{1}{2n}}
\left(\frac{2m_a}{\dot\theta_{\rm inf}}\right)^{\frac{1}{2(n+2)}},
& \text{bosonic decay},\quad T_{\rm osc}>T_{\rm RH},
\\[1.2em]
\displaystyle
T_{\rm RH}
\left(\frac{\mathcal{H}_I}{\mathcal{H}_{\rm RH}}\right)^{\frac{n-1}{2n}}
\left(\frac{2m_a}{\dot\theta_{\rm inf}}\right)^{\frac{n-1}{2(n+2)}},
& \text{fermionic decay},\quad n<7,\quad T_{\rm osc}>T_{\rm RH},
\\[1.2em]
\displaystyle
T_{\rm RH}
\left(\frac{\mathcal{H}_I}{\mathcal{H}_{\rm RH}}\right)^{\frac{n+2}{3n}}
\left(\frac{2m_a}{\dot\theta_{\rm inf}}\right)^{1/3},
& \text{fermionic decay},\quad n>7,\quad T_{\rm osc}>T_{\rm RH},
\\[1.2em]
\displaystyle
T_{\rm RH}
\left(\frac{\mathcal{H}_I}{\mathcal{H}_{\rm RH}}\right)^{\frac{n+2}{3n}}
\left(\frac{2m_a}{\dot\theta_{\rm inf}}\right)^{1/3},
& \text{radiation dominated},\quad T_{\rm osc}<T_{\rm RH}.
\end{cases}
\label{eq:Tosc_KM_direct_below}
\end{equation}
Substituting Eq.~\eqref{eq:aosc_KM_direct_below} into
Eq.~\eqref{eq:R_KM_scale_factor}, gives
\begin{equation}
    \mathcal R_i
=
\frac{45s_0}{4\pi^2}
\frac{\dot\theta_{\rm inf}}{m_a}
\left[
\frac{\sqrt{\pi^2g_\rho(T_{\rm RH})/90}}
{M_P\mathcal{H}_I}
\right]^{\frac{n+2}{n}}
T_{\rm RH}^{\frac{4-n}{n}}
\dfrac{1}{g_{*s}(T_{\rm RH})}
\label{eq:R_KM_direct_below}
\end{equation}
Combining Eq.~\eqref{eq:R_KM_direct_below} with the axion number density at the onset of oscillation reproduces Eq.~\eqref{eq:Omega_KM_direct}, including the $n=10$ expression in Eq.~\eqref{eq:Omega_KM_direct_n10}.
The abundance therefore has the same form for $T_{\rm osc}$ above and below the QCD scale. This follows from the $a^{-3}$ scaling for the evolution of $\dot{\theta}$
and of the axion number density afterwards, with the conversion described by the same $\mathcal C_{\rm KM}$. The oscillation temperatures remain necessary for determining the applicable regime and verifying that the initial velocity is sufficient for kinetic misalignment.

% \begin{equation}
% \Omega_a h^2
% \simeq
% 3.56\times10^{6}\,
% \mathcal C_{\rm KM}\,
% f_a\dot{\theta}_{i}
% \left(
% \frac{\pi}
% {\sqrt{90}\,M_P \mathcal{H}_I}
% \right)^{\frac{n+2}{n}}
% T_{\rm RH}^{\frac{4-n}{n}} \,
% g_{*s}(T_{\rm RH})^{\frac{2-n}{2n}}.
% \end{equation}

% for n=10;
% \begin{equation}
% \begin{aligned}
% \Omega_a h^2 \simeq {}&
% 0.12
% \left(\frac{\mathcal C_{\rm KM}}{2}\right)
% \left(\frac{f_a}{1.6\times10^9\,{\rm GeV}}\right)
% \left(\frac{\dot{\theta}_{i}}{10^{16}\,{\rm GeV}}\right)
% \left(\frac{\mathcal{H}_I}{1.6\times10^8\,{\rm GeV}}\right)^{-\frac{6}{5}}
% \left(\frac{T_{\rm RH}}{1\,{\rm GeV}}\right)^{-\frac{3}{5}}
% \left(\frac{g_{*s}(T_{\rm RH})}{17.25}\right)^{-\frac{2}{5}}
% \end{aligned}
% \end{equation}
\subsection{Seesaw Reheating}
\label{sec:seesaw-kin}
In seesaw reheating, the expansion proceeds through inflaton domination, relativistic RHN domination, and non-relativistic RHN domination before the radiation-dominated era begins. Axion oscillations can start during any of these stages. We determine the onset using the condition in
Eq.~\eqref{eq:trapping_KM_direct}, together with the temperature evolution in Eq.~\eqref{eq:Temperature_RHN}.
The characteristic scale factors are related by Eq.~\eqref{scale_fac}.
\subsubsection{Oscillations above the QCD scale: $T_{\rm osc}>T_{\rm QCD}$}
For oscillations above the QCD scale, we use
$m_a(T)=m_a(T_{\rm QCD}/T)^4$ in the oscillation condition given in Eq.~\eqref{eq:trapping_KM_direct}.
The different temperature scalings in
Eq.~\eqref{eq:Temperature_RHN} then give the following scale factors at the onset of oscillations:
\begin{equation}
a_{\rm osc}=
\begin{cases}
\displaystyle
a_I
\left(\frac{\dot{\theta}_{i}}{2m_a}\right)^{1/3}
\left(\frac{T_{\max}}{T_{\rm QCD}}\right)^{4/3},
& \phi\text{-dominated},
\\[1.2em]
\displaystyle
\left[
a_I^3\,
\frac{\dot{\theta}_{i}}{2m_a}
\left(\frac{T_{\max}}{T_{\rm QCD}}\right)^4
a_{\rm R}
\right]^{1/4},
& \text{relativistic RHN},
\\[1.2em]
\displaystyle
\left[
a_I^3\,
\frac{\dot{\theta}_{i}}{2m_a}
\left(\frac{T_{\rm NR}}{T_{\rm QCD}}\right)^4
a_{\rm NR}^{3/2}
\right]^{2/9},
& \text{non-relativistic RHN},
\\[1.2em]
\displaystyle
\left[
a_I^3\,
\frac{\dot{\theta}_{i}}{2m_a}
\left(\frac{T_{\rm RH}}{T_{\rm QCD}}\right)^4
a_{\rm RH}^{4}
\right]^{1/7},
& \text{radiation dominated}.
\label{eq:aosc_KM_seesaw_above}
\end{cases}
\end{equation}
To express the oscillation temperatures in terms of the
reheating parameters, we define
\begin{equation}
X\equiv\frac{\mathcal{H}_I}{\mathcal{H}_{\rm R}}
=
\frac{3\sqrt{10}\,\mathcal{H}_I M_P T_{\rm NR}^4T_{\rm RH}^2}
{\pi\,g_{*s}^{1/2}(T_{\rm RH})T_{\max}^8},
\label{eq:X_KM_seesaw}
\end{equation}
where we have used
$g_{*\rho}(T_{\rm RH})\simeq g_{*s}(T_{\rm RH})$.
The relations in Eq.~\eqref{scale_fac} become
\begin{equation}
\begin{aligned}
a_{\rm R}
&=a_I X^{\frac{n+2}{3n}},\\
a_{\rm NR}
&=a_I X^{\frac{n+2}{3n}}
\left(\frac{T_{\max}}{T_{\rm NR}}\right)^4,\\
a_{\rm RH}
&=a_I X^{\frac{n+2}{3n}}
\left(\frac{T_{\max}}{T_{\rm NR}}\right)^4
\left(\frac{T_{\rm NR}}{T_{\rm RH}}\right)^{8/3}.
\end{aligned}
\label{eq:scales_KM_seesaw}
\end{equation}
Evaluating Eq.~\eqref{eq:Temperature_RHN} at the scale factors
in Eq.~\eqref{eq:aosc_KM_seesaw_above}, we obtain
\begin{equation}
T_{\rm osc}=
\begin{cases}
T_{\max},
& \phi\text{-dominated},
\\[0.8em]
T_{\max}
\left(\dfrac{T_{\rm QCD}}{T_{\max}}\right)^{\frac{1}{4}}
X^{\frac{n+2}{16n}}
\left(\dfrac{2m_a}{\dot\theta_{\rm inf}}\right)^{\frac{1}{16}},
& \text{relativistic RHN},
\\[1em]
T_{\max}
\left(\dfrac{T_{\rm QCD}}{T_{\rm NR}}\right)^{\frac{1}{3}}
X^{\frac{n+2}{12n}}
\left(\dfrac{2m_a}{\dot\theta_{\rm inf}}\right)^{\frac{1}{12}},
& \text{non-relativistic RHN},
\\[1em]
T_{\rm RH}
\left(\dfrac{T_{\rm QCD}}{T_{\rm RH}}\right)^{\frac{4}{7}}
X^{\frac{n+2}{7n}}
\left(\dfrac{T_{\max}}{T_{\rm NR}}\right)^{\frac{12}{7}}
\left(\dfrac{T_{\rm NR}}{T_{\rm RH}}\right)^{\frac{8}{7}}
\left(\dfrac{2m_a}{\dot\theta_{\rm inf}}\right)^{\frac{1}{7}},
& \text{radiation dominated}.
\end{cases}
\label{eq:Tosc_KM_seesaw_above}
\end{equation}
$T_{\rm osc}=T_{\max}$ in the inflaton-dominated branch follows from the approximately constant temperature
assumed during that stage. Similar to the direct-reheating calculation, we neglect changes in the effective relativistic degrees of freedom when estimating the oscillation temperature and scale factor. We evaluate the redshift factor using Eq.~\eqref{eq:R_KM_scale_factor}, assuming entropy conservation after reheating. Substituting Eqs.~\eqref{eq:aosc_KM_seesaw_above}
and~\eqref{eq:scales_KM_seesaw} gives
% \begin{equation}
% \begin{aligned}
% \mathcal R_i=
% \frac{45s_0}{2\pi^2 g_{*s}({\rm T_{RH}})} \\ 
% \times \begin{cases}
% \displaystyle
% \left(\frac{\dot\theta_{\rm inf}}{2m_a}\right)
% T_{\rm QCD}^{-4}
% \left[
% \frac{\sqrt{\pi^2g_{*s}({\rm T_{RH}})/90}}
% {M_P\mathcal{H}_I}
% \right]^{\frac{n+2}{n}}
% T_{\max}^{\frac{16}{n}}
% T_{\rm NR}^{-\frac8n}
% T_{\rm RH}^{\frac{3n-4}{n}},
% & \phi\text{-dominated},
% \\[1.5em]
% \displaystyle 
% \left(\frac{\dot\theta_{\rm inf}}{2m_a}\right)^{3/4}
% T_{\rm QCD}^{-3}
% \left[
% \frac{\sqrt{\pi^2 g_{*s}({\rm T_{RH}})/90}}
% {M_P\mathcal{H}_I}
% \right]^{\frac{3(n+2)}{4n}}
% T_{\max}^{\frac{3(4-n)}{n}}
% T_{\rm NR}^{\frac{n-6}{n}}
% T_{\rm RH}^{\frac{7n-6}{2n}},
% & \text{relativistic RHN},
% \\[1.5em]
% \displaystyle 
% \left(\frac{\dot\theta_{\rm inf}}{2m_a}\right)^{2/3}
% T_{\rm QCD}^{-8/3}
% \left[
% \frac{\sqrt{\pi^2g_{*s}({\rm T_{RH}})/90}}
% {M_P\mathcal{H}_I}
% \right]^{\frac{2(n+2)}{3n}}
% T_{\max}^{\frac{8(4-n)}{3n}}
% T_{\rm NR}^{-\frac{16}{3n}}
% T_{\rm RH}^{\frac{11n-8}{3n}},
% & \text{non-relativistic RHN},
% \\[1.5em]
% \displaystyle 
% \left(\frac{\dot\theta_{\rm inf}}{2m_a}\right)^{3/7}
% T_{\rm QCD}^{-12/7}
% \left[
% \frac{\sqrt{\pi^2 g_{*s}({\rm T_{RH}})/90}}
% {M_P\mathcal{H}_I}
% \right]^{\frac{3(n+2)}{7n}}
% T_{\max}^{\frac{12(4-n)}{7n}}
% T_{\rm NR}^{-\frac{24}{7n}}
% T_{\rm RH}^{\frac{3(3n-4)}{7n}},
% & \text{radiation dominated}.
% \end{cases}
% \end{aligned}
% \label{eq:R_KM_seesaw_above}
% \end{equation}
\begin{equation}
\begin{aligned}
&\mathcal R_i=
\frac{45s_0}{2\pi^2 g_{*s}(T_{\rm RH})}
\\[0.5em]
&\quad\times
\begin{cases}
\displaystyle
\left(\frac{\dot{\theta}_i}{2m_a}\right)
T_{\rm QCD}^{-4}
\left[
\frac{\sqrt{\frac{\pi^2g_{*s}(T_{\rm RH})}{90}}}
{M_P \mathcal{H}_I}
\right]^{\frac{n+2}{n}}
T_{\max}^{\frac{16}{n}}
T_{\rm NR}^{-\frac{8}{n}}
T_{\rm RH}^{\frac{3n-4}{n}},
& \phi\text{-dominated},
\\[1.5em]
\displaystyle
\left(\frac{\dot{\theta}_i}{2m_a}\right)^{\frac{3}{4}}
T_{\rm QCD}^{-3}
\left[
\frac{\sqrt{\frac{\pi^2g_{*s}(T_{\rm RH})}{90}}}
{M_P \mathcal{H}_I}
\right]^{\frac{3(n+2)}{4n}}
T_{\max}^{\frac{3(4-n)}{n}}
T_{\rm NR}^{\frac{n-6}{n}}
T_{\rm RH}^{\frac{7n-6}{2n}},
& \text{relativistic RHN},
\\[1.5em]
\displaystyle
\left(\frac{\dot{\theta}_i}{2m_a}\right)^{\frac{2}{3}}
T_{\rm QCD}^{-\frac{8}{3}}
\left[
\frac{\sqrt{\frac{\pi^2g_{*s}(T_{\rm RH})}{90}}}
{M_P \mathcal{H}_I}
\right]^{\frac{2(n+2)}{3n}}
T_{\max}^{\frac{8(4-n)}{3n}}
T_{\rm NR}^{-\frac{16}{3n}}
T_{\rm RH}^{\frac{11n-8}{3n}},
& \text{non-relativistic RHN},
\\[1.5em]
\displaystyle
\left(\frac{\dot{\theta}_i}{2m_a}\right)^{\frac{3}{7}}
T_{\rm QCD}^{-\frac{12}{7}}
\left[
\frac{\sqrt{\frac{\pi^2g_{*s}(T_{\rm RH})}{90}}}
{M_P \mathcal{H}_I}
\right]^{\frac{3(n+2)}{7n}}
T_{\max}^{\frac{12(4-n)}{7n}}
T_{\rm NR}^{-\frac{24}{7n}}
T_{\rm RH}^{\frac{3(3n-4)}{7n}},
& \text{radiation dominated}.
\end{cases}
\end{aligned}
\label{eq:R_KM_seesaw_above}
\end{equation}
Using the axion number density at the onset of oscillation,
$n_a(T_{\rm osc})\simeq
\mathcal C_{\rm KM}f_a^2\dot{\theta}(a_{\rm osc})$,
and substituting the corresponding redshift factors from
Eq.~\eqref{eq:R_KM_seesaw_above}, we obtain
\begin{equation}
\begin{aligned}
\rho_a(T_0)
&\simeq m_a n_a(T_{\rm osc})\mathcal R_i\\
&\simeq
\mathcal C_{\rm KM}m_a f_a^2\dot{\theta}_i
\left(\frac{a_I}{a_{\rm osc}}\right)^3\mathcal R_i.
\end{aligned}
\label{eq:rho_KM_seesaw}
\end{equation}
Then the relic abundance becomes
\begin{equation}
\begin{aligned}
\Omega_a h^2 \simeq{}&
3.56\times10^{6}\,
\mathcal C_{\rm KM}\,
f_a\dot{\theta}_i
\left(
\frac{\pi}{\sqrt{90}\,M_P \mathcal{H}_I}
\right)^{\frac{n+2}{n}}
g_{*s}^{\frac{2-n}{2n}}(T_{\rm RH})
T_{\max}^{\frac{4(4-n)}{n}}
T_{\rm NR}^{-\frac8n}
T_{\rm RH}^{\frac{3n-4}{n}}.
\end{aligned}
\label{eq:Omega_KM_seesaw}
\end{equation}
The dimensionful quantities in this numerical expression are expressed in GeV. The same result is obtained for all four branches.
As expected from Eq.~\eqref{eq:Omega_KM_general}, the
dependence on $a_{\rm osc}$ cancels when the number density at trapping is combined with its subsequent redshift factor. Substituting the individual expressions therefore confirms this cancellation for each regime. The remaining dependence on $T_{\max}$ and $T_{\rm NR}$ arises through the total expansion between $a_I$ and $a_{\rm RH}$ in Eq.~\eqref{eq:scales_KM_seesaw}.
% \begin{equation}
% \Omega_a h^2 \simeq 0.12
% \left(\frac{\mathcal C_{\rm KM}}{2}\right)
% \left(\frac{f_a}{4.7\times10^9\,{\rm GeV}}\right)
% \left(\frac{\dot{\theta}_{i}}{10^{17}\,{\rm GeV}}\right)
% \left(\frac{g_{*s}({\rm T_{RH}})}{17.25}\right)^{-2/5}
% \left(\frac{\mathcal{H}_I}{4.7\times10^8\,{\rm GeV}}\right)^{-6/5}
% \left(\frac{T_{\max}}{0.2\,{\rm GeV}}\right)^{-12/5}
% \left(\frac{T_{\rm NR}}{0.1\,{\rm GeV}}\right)^{-4/5}
% \left(\frac{T_{\rm RH}}{0.05\,{\rm GeV}}\right)^{13/5}.
% \end{equation}
For $n=10$, the abundance can be written in the normalized form
\begin{equation}
\begin{aligned}
\Omega_a h^2 \simeq{}&
0.12
\left(\frac{\mathcal C_{\rm KM}}{2}\right)
\left(\frac{f_a}{4.7\times10^9\,{\rm GeV}}\right)
\left(\frac{\dot{\theta}_i}{10^{17}\,{\rm GeV}}\right)
\left(\frac{\mathcal{H}_I}{4.7\times10^8\,{\rm GeV}}\right)^{-\frac{6}{5}} \\
&\times
\left(\frac{g_{*s}(T_{\rm RH})}{17.25}\right)^{-\frac{2}{5}}
\left(\frac{T_{\max}}{0.2\,{\rm GeV}}\right)^{-\frac{12}{5}}
\left(\frac{T_{\rm NR}}{0.1\,{\rm GeV}}\right)^{-\frac{4}{5}}
\left(\frac{T_{\rm RH}}{0.05\,{\rm GeV}}\right)^{\frac{13}{5}}.
\end{aligned}
\label{eq:Omega_KM_seesaw_n10}
\end{equation}
The abundance retains a dependence on the earlier reheating
stages even when oscillations begin during radiation
domination. This follows because the initial angular
velocity is specified at $a_I$ and is diluted throughout
the expansion preceding the onset of oscillations.

\subsubsection{Oscillations below the QCD scale:
$T_{\rm osc}<T_{\rm QCD}$}

Below the QCD scale, the axion mass is
constant, $m_a(T_{\rm osc})= m_a$. The oscillation  condition in Eq.~\eqref{eq:trapping_KM_direct} therefore
gives the same scale factor irrespective of the era:
\begin{equation}
a_{\rm osc}
=
a_I\left(\frac{\dot{\theta}_i}{2m_a}\right)^{1/3}.
\label{eq:aosc_KM_seesaw_below}
\end{equation}
The corresponding temperature depends on the stage in
which this scale factor is reached. Using
Eqs.~\eqref{eq:Temperature_RHN}
and~\eqref{eq:scales_KM_seesaw}, we obtain
\begin{equation}
T_{\rm osc}=
\begin{cases}
T_{\max},
& \phi\text{-dominated},
\\[1.0em]
T_{\max}
\left(
\frac{
3\sqrt{10}\,\mathcal{H}_I M_P T_{\rm NR}^4 T_{\rm RH}^2
}{
\pi\,g_{*s}^{1/2}({\rm T_{RH}}) T_{\max}^8
}
\right)^{\frac{n+2}{12n}}
\left(
\frac{2 m_a}{\dot{\theta}_{i}}
\right)^{1/12},
& \text{relativistic RHN},
\\[1.2em]
T_{\rm NR}
\left(
\frac{
3\sqrt{10}\,\mathcal{H}_I M_P T_{\rm NR}^4 T_{\rm RH}^2
}{
\pi\,g_{*s}^{1/2} ({\rm T_{RH}})T_{\max}^8
}
\right)^{\frac{n+2}{8n}}
\left(
\frac{T_{\max}}{T_{\rm NR}}
\right)^{3/2}
\left(
\frac{2 m_a}{\dot{\theta}_{i}}
\right)^{1/8},
& \text{non-relativistic RHN},
\\[1.2em]
T_{\rm RH}
\left(
\frac{
3\sqrt{10}\,\mathcal{H}_I M_P T_{\rm NR}^4 T_{\rm RH}^2
}{
\pi\,g_{*s}^{1/2} ({\rm T_{RH}}) T_{\max}^8
}
\right)^{\frac{n+2}{3n}}
\left(
\frac{T_{\max}}{T_{\rm NR}}
\right)^4
\left(
\frac{T_{\rm NR}}{T_{\rm RH}}
\right)^{8/3}
\left(
\frac{2 m_a}{\dot{\theta}_{i}}
\right)^{1/3},
& \text{radiation dominated}.
\end{cases}
\label{eq:Tosc_KM_seesaw_below}
\end{equation}
For the inflaton-dominated branch, the latter condition requires $T_{\max}<T_{\rm QCD}$ within the adopted temperature approximation.\\
Substituting Eq.~\eqref{eq:aosc_KM_seesaw_below} into
Eq.~\eqref{eq:R_KM_scale_factor}, and using
Eq.~\eqref{eq:scales_KM_seesaw}, gives
\begin{equation}
\begin{aligned}
\mathcal R_i \simeq{}&
\frac{45s_0}{4\pi^2g_{*s}(T_{\rm RH})}
\frac{\dot{\theta}_i}{m_a}
\left[
\frac{\sqrt{\pi^2g_{*s}(T_{\rm RH})/90}}
{M_P \mathcal{H}_I}
\right]^{\frac{n+2}{n}}
T_{\max}^{\frac{4(4-n)}{n}}
T_{\rm NR}^{-\frac8n}
T_{\rm RH}^{\frac{3n-4}{n}}.
\end{aligned}
\label{eq:R_KM_seesaw_below}
\end{equation}
Combining Eq.~\eqref{eq:R_KM_seesaw_below} with the axion
number density at the onset of oscillations reproduces
Eq.~\eqref{eq:Omega_KM_seesaw}, including its $n=10$ form
in Eq.~\eqref{eq:Omega_KM_seesaw_n10}.
As in direct reheating, the abundance has the same form for oscillations above and below the QCD scale. The oscillation temperatures are still needed to identify
the applicable era and verify that the initial velocity is sufficient for kinetic misalignment.
%\bibliographystyle{JHEP}
%\bibliography{ref1} 

\providecommand{\href}[2]{#2}\begingroup\raggedright\endgroup

\end{document}